\documentclass[transmag]{IEEEtran}
\usepackage{latexsym}
\usepackage{color}
\usepackage{colortbl}
\usepackage{graphicx}
\usepackage{subfigure}
\usepackage{amsmath}
\usepackage{amssymb}
\usepackage{amsthm}
\usepackage{afterpage}
\usepackage{verbatim}
\usepackage{psfrag}
\usepackage{color}
\usepackage[table,xcdraw]{xcolor}

\usepackage{cite}
\usepackage{setspace}
\usepackage{epsfig}
\usepackage{footmisc}
\usepackage{fmtcount}
\usepackage{stackrel}
\usepackage{float}
\usepackage{breqn}
\usepackage{enumerate}
\usepackage{rotating}
\usepackage{url}
\usepackage{breakurl}
\usepackage{tabularx}
\usepackage{tikz}

\usepackage[normalem]{ulem}
\usepackage{makecell}

\usepackage[nolist]{acronym}
\usepackage{esint} % various fancy integral symbols
\usepackage{multirow}

\usepackage[per-mode=symbol]{siunitx}
\definecolor{Linen}{rgb}{0.9803,0.9411,0.9019}
\definecolor{White}{rgb}{1,1,1}
\definecolor{Lightred}{rgb}{1,0.3803,0.3803}
\definecolor{Coral}{rgb}{1,0.4980,0.3137}
\definecolor{Grayblue}{rgb}{0.9411,0.9411,0.9803}
\definecolor{DarkLinen}{rgb}{0.729,0.7176,0.635}

\usetikzlibrary{shapes,backgrounds,positioning-plus,node-families,calc}

\tikzstyle{dummy} = [rectangle, text width=0.1em, draw=white, white,
                      minimum width=0.1em, minimum height=3em, opacity=0.0]
                      
  \tikzstyle{survey} = [rectangle, draw=black, rounded corners, text centered, text=black, text width=8cm, minimum height=2cm]
  \tikzstyle{survey2} = [rectangle, draw=black, rounded corners, text centered, text=black, text width=8cm, minimum height=2cm,fill=Lightred]
  
  \tikzstyle{section} = [rectangle, draw=black, rounded corners, text centered, text=black, text width=4cm, minimum height=2cm]
  \tikzstyle{section2} = [rectangle, draw=black, rounded corners, fill=Grayblue, text centered, text=black, text width=4cm, minimum height=2cm]

  \tikzstyle{subsection} = [rectangle, draw=black, rounded corners, text centered, text=black, text width=4cm]
  \tikzstyle{subsection2} = [rectangle, draw=black, rounded corners, fill=Linen, text centered, text=black, text width=4cm]

\begin{document}
% 	\newpage
% 	\begin{small}
% 		\tableofcontents
% 	\end{small}
%
	\begin{acronym}

\acro{5G-NR}{5G New Radio}
\acro{3GPP}{3rd Generation Partnership Project}
\acro{AC}{address coding}
\acro{ACF}{autocorrelation function}
\acro{ACR}{autocorrelation receiver}
\acro{ADC}{analog-to-digital converter}
\acrodef{aic}[AIC]{Analog-to-Information Converter}     
\acro{AIC}[AIC]{Akaike information criterion}
\acro{aric}[ARIC]{asymmetric restricted isometry constant}
\acro{arip}[ARIP]{asymmetric restricted isometry property}

\acro{ARQ}{Automatic Repeat Request}
\acro{AUB}{asymptotic union bound}
\acrodef{awgn}[AWGN]{Additive White Gaussian Noise}     
\acro{AWGN}{additive white Gaussian noise}

\acro{APSK}[PSK]{asymmetric PSK} 

\acro{waric}[AWRICs]{asymmetric weak restricted isometry constants}
\acro{warip}[AWRIP]{asymmetric weak restricted isometry property}
\acro{BCH}{Bose, Chaudhuri, and Hocquenghem}        
\acro{BCHC}[BCHSC]{BCH based source coding}
\acro{BEP}{bit error probability}
\acro{BFC}{block fading channel}
\acro{BG}[BG]{Bernoulli-Gaussian}
\acro{BGG}{Bernoulli-Generalized Gaussian}
\acro{BPAM}{binary pulse amplitude modulation}
\acro{BPDN}{Basis Pursuit Denoising}
\acro{BPPM}{binary pulse position modulation}
\acro{BPSK}{Binary Phase Shift Keying}
\acro{BPZF}{bandpass zonal filter}
\acro{BSC}{binary symmetric channels}              
\acro{BU}[BU]{Bernoulli-uniform}
\acro{BER}{bit error rate}
\acro{BS}{base station}

\acro{CP}{Cyclic Prefix}
\acrodef{cdf}[CDF]{cumulative distribution function}   
\acro{CDF}{Cumulative Distribution Function}
\acrodef{c.d.f.}[CDF]{cumulative distribution function}
\acro{CCDF}{complementary cumulative distribution function}
\acrodef{ccdf}[CCDF]{complementary CDF}               
\acrodef{c.c.d.f.}[CCDF]{complementary cumulative distribution function}
\acro{CD}{cooperative diversity}

\acro{CDMA}{Code Division Multiple Access}
\acro{ch.f.}{characteristic function}
\acro{CIR}{channel impulse response}
\acro{cosamp}[CoSaMP]{compressive sampling matching pursuit}
\acro{CR}{cognitive radio}
\acro{cs}[CS]{compressed sensing}                   
\acrodef{cscapital}[CS]{Compressed sensing} %will not include it in the list
\acrodef{CS}[CS]{compressed sensing}
\acro{CSI}{channel state information}
\acro{CCSDS}{consultative committee for space data systems}
\acro{CC}{convolutional coding}
\acro{Covid19}[COVID-19]{Coronavirus disease}

\acro{DAA}{detect and avoid}
\acro{DAB}{digital audio broadcasting}
\acro{DCT}{discrete cosine transform}
\acro{dft}[DFT]{discrete Fourier transform}
\acro{DR}{distortion-rate}
\acro{DS}{direct sequence}
\acro{DS-SS}{direct-sequence spread-spectrum}
\acro{DTR}{differential transmitted-reference}
\acro{DVB-H}{digital video broadcasting\,--\,handheld}
\acro{DVB-T}{digital video broadcasting\,--\,terrestrial}
\acro{DL}{DownLink}
\acro{DSSS}{Direct Sequence Spread Spectrum}
\acro{DFT-s-OFDM}{Discrete Fourier Transform-spread-Orthogonal Frequency Division Multiplexing}
\acro{DAS}{Distributed Antenna System}
\acro{DNA}{DeoxyriboNucleic Acid}

\acro{EC}{European Commission}
\acro{EED}[EED]{exact eigenvalues distribution}
\acro{EIRP}{Equivalent Isotropically Radiated Power}
\acro{ELP}{equivalent low-pass}
\acro{eMBB}{Enhanced Mobile Broadband}
\acro{EMF}{ElectroMagnetic Field}
\acro{EU}{European union}

\acro{FC}[FC]{fusion center}
\acro{FCC}{Federal Communications Commission}
\acro{FEC}{forward error correction}
\acro{FFT}{fast Fourier transform}
\acro{FH}{frequency-hopping}
\acro{FH-SS}{frequency-hopping spread-spectrum}
\acrodef{FS}{Frame synchronization}
\acro{FSsmall}[FS]{frame synchronization}  
\acro{FDMA}{Frequency Division Multiple Access}

\acro{GA}{Gaussian approximation}
\acro{GF}{Galois field }
\acro{GG}{Generalized-Gaussian}
\acro{GIC}[GIC]{generalized information criterion}
\acro{GLRT}{generalized likelihood ratio test}
\acro{GPS}{Global Positioning System}
\acro{GMSK}{Gaussian Minimum Shift Keying}
\acro{GSMA}{Global System for Mobile communications Association}

\acro{HAP}{high altitude platform}

\acro{IDR}{information distortion-rate}
\acro{IFFT}{inverse fast Fourier transform}
\acro{iht}[IHT]{iterative hard thresholding}
\acro{i.i.d.}{independent, identically distributed}
\acro{IoT}{Internet of Things}                      
\acro{IR}{impulse radio}
\acro{lric}[LRIC]{lower restricted isometry constant}
\acro{lrict}[LRICt]{lower restricted isometry constant threshold}
\acro{ISI}{intersymbol interference}
\acro{ITU}{International Telecommunication Union}
\acro{ICNIRP}{International Commission on Non-Ionizing Radiation Protection}
\acro{IEEE}{Institute of Electrical and Electronics Engineers}
\acro{ICES}{IEEE international committee on electromagnetic safety}
\acro{IEC}{International Electrotechnical Commission}
\acro{IARC}{International Agency on Research on Cancer}
\acro{IS-95}{Interim Standard 95}

\acro{LEO}{low earth orbit}
\acro{LF}{likelihood function}
\acro{LLF}{log-likelihood function}
\acro{LLR}{log-likelihood ratio}
\acro{LLRT}{log-likelihood ratio test}
\acro{LOS}{Line-of-Sight}
\acro{LRT}{likelihood ratio test}
\acro{wlric}[LWRIC]{lower weak restricted isometry constant}
\acro{wlrict}[LWRICt]{LWRIC threshold}
\acro{LPWAN}{Low Power Wide Area Network}
\acro{LoRaWAN}{Low power long Range Wide Area Network}
\acro{NLOS}{Non-Line-of-Sight}
\acro{LiFi}[Li-Fi]{light-fidelity}
 \acro{LED}{light emitting diode}

\acro{MB}{multiband}
\acro{MC}{multicarrier}
\acro{MDS}{mixed distributed source}
\acro{MEC}{Mobile Edge Computing}
\acro{MF}{matched filter}
\acro{m.g.f.}{moment generating function}
\acro{MI}{mutual information}
\acro{MIMO}{Multiple-Input Multiple-Output}
\acro{MISO}{multiple-input single-output}
\acrodef{maxs}[MJSO]{maximum joint support cardinality}                       
\acro{ML}[ML]{maximum likelihood}
\acro{MMSE}{minimum mean-square error}
\acro{MMV}{multiple measurement vectors}
\acrodef{MOS}{model order selection}
\acro{M-PSK}[${M}$-PSK]{$M$-ary phase shift keying}                       
\acro{M-APSK}[${M}$-PSK]{$M$-ary asymmetric PSK} 

\acro{M-QAM}[$M$-QAM]{$M$-ary quadrature amplitude modulation}
\acro{MRC}{maximal ratio combiner}                  
\acro{maxs}[MSO]{maximum sparsity order}                                      
\acro{M2M}{Machine-to-Machine}                                                
\acro{MUI}{multi-user interference}
\acro{mMTC}{massive Machine Type Communications}      
\acro{mm-Wave}{millimeter-wave}
\acro{MP}{mobile phone}
\acro{MPE}{maximum permissible exposure}
\acro{MAC}{media access control}
\acro{NB}{narrowband}
\acro{NBI}{narrowband interference}
\acro{NLA}{nonlinear sparse approximation}
\acro{NLOS}{Non-Line of Sight}
\acro{NTIA}{National Telecommunications and Information Administration}
\acro{NTP}{National Toxicology Program}
\acro{NHS}{National Health Service}

\acro{OC}{optimum combining}                             
\acro{OC}{optimum combining}
\acro{ODE}{operational distortion-energy}
\acro{ODR}{operational distortion-rate}
\acro{OFDM}{Orthogonal Frequency-Division Multiplexing}
\acro{omp}[OMP]{orthogonal matching pursuit}
\acro{OSMP}[OSMP]{orthogonal subspace matching pursuit}
\acro{OQAM}{offset quadrature amplitude modulation}
\acro{OQPSK}{offset QPSK}
\acro{OFDMA}{Orthogonal Frequency-division Multiple Access}
\acro{OPEX}{Operating Expenditures}
\acro{OQPSK/PM}{OQPSK with phase modulation}

\acro{PAM}{pulse amplitude modulation}
\acro{PAR}{peak-to-average ratio}
\acrodef{pdf}[PDF]{probability density function}                      
\acro{PDF}{probability density function}
\acrodef{p.d.f.}[PDF]{probability distribution function}
\acro{PDP}{power dispersion profile}
\acro{PMF}{probability mass function}                             
\acrodef{p.m.f.}[PMF]{probability mass function}
\acro{PN}{pseudo-noise}
\acro{PPM}{pulse position modulation}
\acro{PRake}{Partial Rake}
\acro{PSD}{power spectral density}
\acro{PSEP}{pairwise synchronization error probability}
\acro{PSK}{phase shift keying}
\acro{PD}{power density}
\acro{8-PSK}[$8$-PSK]{$8$-phase shift keying}

\acro{FSK}{Frequency Shift Keying}

\acro{QAM}{Quadrature Amplitude Modulation}
\acro{QPSK}{Quadrature Phase Shift Keying}
\acro{OQPSK/PM}{OQPSK with phase modulator }

\acro{RD}[RD]{raw data}
\acro{RDL}{"random data limit"}
\acro{ric}[RIC]{restricted isometry constant}
\acro{rict}[RICt]{restricted isometry constant threshold}
\acro{rip}[RIP]{restricted isometry property}
\acro{ROC}{receiver operating characteristic}
\acro{rq}[RQ]{Raleigh quotient}
\acro{RS}[RS]{Reed-Solomon}
\acro{RSC}[RSSC]{RS based source coding}
\acro{r.v.}{random variable}                               
\acro{R.V.}{random vector}
\acro{RMS}{root mean square}
\acro{RFR}{radiofrequency radiation}
\acro{RIS}{Reconfigurable Intelligent Surface}
\acro{RNA}{RiboNucleic Acid}

\acro{SA}[SA-Music]{subspace-augmented MUSIC with OSMP}
\acro{SCBSES}[SCBSES]{Source Compression Based Syndrome Encoding Scheme}
\acro{SCM}{sample covariance matrix}
\acro{SEP}{symbol error probability}
\acro{SG}[SG]{sparse-land Gaussian model}
\acro{SIMO}{single-input multiple-output}
\acro{SINR}{signal-to-interference plus noise ratio}
\acro{SIR}{signal-to-interference ratio}
\acro{SISO}{Single-Input Single-Output}
\acro{SMV}{single measurement vector}
\acro{SNR}[\textrm{SNR}]{signal-to-noise ratio} 
\acro{sp}[SP]{subspace pursuit}
\acro{SS}{spread spectrum}
\acro{SW}{sync word}
\acro{SAR}{specific absorption rate}
\acro{SSB}{synchronization signal block}

\acro{TH}{time-hopping}
\acro{ToA}{time-of-arrival}
\acro{TR}{transmitted-reference}
\acro{TW}{Tracy-Widom}
\acro{TWDT}{TW Distribution Tail}
\acro{TCM}{trellis coded modulation}
\acro{TDD}{Time-Division Duplexing}
\acro{TDMA}{Time Division Multiple Access}

\acro{UAV}{unmanned aerial vehicle}
\acro{uric}[URIC]{upper restricted isometry constant}
\acro{urict}[URICt]{upper restricted isometry constant threshold}
\acro{UWB}{ultrawide band}
\acro{UWBcap}[UWB]{Ultrawide band}   
\acro{URLLC}{Ultra Reliable Low Latency Communications}
         
\acro{wuric}[UWRIC]{upper weak restricted isometry constant}
\acro{wurict}[UWRICt]{UWRIC threshold}                
\acro{UE}{User Equipment}
\acro{UL}{UpLink}

\acro{WiM}[WiM]{weigh-in-motion}
\acro{WLAN}{wireless local area network}
\acro{wm}[WM]{Wishart matrix}                               
\acroplural{wm}[WM]{Wishart matrices}
\acro{WMAN}{wireless metropolitan area network}
\acro{WPAN}{wireless personal area network}
\acro{wric}[WRIC]{weak restricted isometry constant}
\acro{wrict}[WRICt]{weak restricted isometry constant thresholds}
\acro{wrip}[WRIP]{weak restricted isometry property}
\acro{WSN}{wireless sensor network}                        
\acro{WSS}{Wide-Sense Stationary}
\acro{WHO}{World Health Organization}
\acro{Wi-Fi}{Wireless Fidelity}

\acro{sss}[SpaSoSEnc]{sparse source syndrome encoding}

\acro{VLC}{Visible Light Communication}
\acro{VPN}{Virtual Private Network} 
\acro{RF}{Radio Frequency}
\acro{FSO}{Free Space Optics}
\acro{IoST}{Internet of Space Things}

\acro{GSM}{Global System for Mobile Communications}
\acro{2G}{Second-generation cellular network}
\acro{3G}{Third-generation cellular network}
\acro{4G}{Fourth-generation cellular network}
\acro{5G}{Fifth-generation cellular network}	
\acro{gNB}{next-generation Node-B Base Station}
\acro{NR}{New Radio}
\acro{UMTS}{Universal Mobile Telecommunications Service}
\acro{LTE}{Long Term Evolution}

\acro{QoS}{Quality of Service}
\end{acronym}
	
	%% EMF definitions
\newcommand{\SAR} {\mathrm{SAR}}
\newcommand{\WBSAR} {\mathrm{SAR}_{\mathsf{WB}}}
\newcommand{\gSAR} {\mathrm{SAR}_{10\si{\gram}}}
\newcommand{\Sab} {S_{\mathsf{ab}}}
\newcommand{\Eavg} {E_{\mathsf{avg}}}
\newcommand{\ft}{f_{\textsf{th}}}
\newcommand{\alphatf}{\alpha_{24}}

	%\doublespace
	%\title{5G Technology: Between Efficient Communication and Health Concerns}
	%\title{Recent Advances in Wireless Communications and Their Implication on Human Health: An Engineering Perspective}
	\title{Health Risks Associated with 5G Exposure: \\A View from the Communications Engineering Perspective}
	\author{Luca Chiaraviglio\IEEEauthorrefmark{1},~\IEEEmembership{Senior Member,~IEEE}, Ahmed Elzanaty\IEEEauthorrefmark{1},~\IEEEmembership{Member,~IEEE}, and \\ Mohamed-Slim Alouini,~\IEEEmembership{Fellow,~IEEE}
\thanks{\IEEEauthorrefmark{1}L. Chiaraviglio and  A. Elzanaty contributed equally to this work.}
\thanks{L. Chiaraviglio is with the Electrical Engineering (EE) Department, University of Rome Tor Vergata, 00133 Rome, Italy, and also with the Consorzio Nazionale Interuniversitario per le Telecomunicazioni, 43124 Parma, Italy (e-mail: luca.chiaraviglio@uniroma2.it).}
\thanks{A. Elzanaty and M.-S. Alouini are with the Computer Electrical and Mathematical Sciences \& Engineering (CEMSE) Division, King Abdullah University of Science and Technology (KAUST), Thuwal, Makkah Province, Kingdom of Saudi Arabia, 23955-6900  (e-mail: \{ahmed.elzanaty,slim.alouini\}@kaust.edu.sa).}
% 			\author{Luca Chiaraviglio,${^{(1,2)}}$~\IEEEmembership{Senior Member,~IEEE}, Ahmed Elzanaty,${^{(3)}}$ ~\IEEEmembership{Member,~IEEE}, \\ Mohamed-Slim Alouini,${^{(3)}}$~\IEEEmembership{Fellow,~IEEE}
% 		\\
% 		1) University of Rome Tor Vergata, Italy, luca.chiaraviglio@uniroma2.it\\
% 	         2) Consorzio Nazionale Interuniversitario per le Telecomunicazioni (CNIT), Italy\\
% 		3) King Abdullah University of Science and Technology (KAUST), Kingdom of Saudi Arabia, \{ahmed.elzanaty,slim.alouini\}@kaust.edu.sa\\
	}

	%%%%%%%%%%%%%%%%%%%%%%%%%
\IEEEtitleabstractindextext{
	\begin{abstract}
The deployment of the fifth-generation (5G) wireless communication services requires the installation of 5G next-generation Node-B Base Stations (gNBs) over the territory and the wide adoption of 5G User Equipment (UE). In this context, the population is concerned about the potential health risks associated with the Radio Frequency (RF) emissions from 5G equipment, with several communities actively working toward stopping the 5G deployment. To face these concerns, in this work, we analyze the health risks associated with 5G exposure by adopting a new and comprehensive viewpoint, based on the communications engineering perspective. By exploiting our background, we {\color{black} investigate} the alleged health effects of 5G exposure and critically review the latest works that are often referenced to support the health concerns from 5G. We then precisely examine the up-to-date metrics, regulations, and assessment of compliance procedures for 5G exposure, by evaluating the latest guidelines from the Institute of Electrical and Electronics Engineers (IEEE), the International Commission on Non-Ionizing Radiation Protection (ICNIRP), the International Telecommunication Union (ITU), the International Electrotechnical Commission (IEC), and the United States Federal Communications Commission (FCC), as well as the national regulations in more than 220 countries. We also thoroughly analyze the main health risks that are frequently associated with specific 5G features (e.g., multiple-input multiple-output (MIMO), beamforming, cell densification, adoption of millimeter waves, and connection of millions of devices). Finally, we examine the risk mitigation techniques based on communications engineering that can be implemented to reduce the exposure from 5G gNB and UE. Overall, we argue that the widely perceived health risks that are attributed to 5G are not supported by scientific evidence from communications engineering. In addition, we explain how the solutions to minimize the health risks from 5G (including currently unknown effects) are already mature and ready to be implemented. Finally, future works, e.g., aimed at evaluating long-term impacts of 5G exposure, as well as innovative solutions to further reduce the RF emissions, are suggested.
	\end{abstract}
	\IEEEpeerreviewmaketitle
% 	\begin{IEEEkeywords} 
% 		5G health risks, 5G health effects, \acs{EMF} exposure of 5G devices, 5G \acs{EMF} regulations, 5G \acs{EMF} metrics, compliance assessment of 5G exposure, health risks of 5G features, 5G risk mitigation techniques
% 	\end{IEEEkeywords}
}
	\maketitle
	%%%%%%%%%%%%%%%%%%%%%%%%%%%%%%%%%%%%%%%%%%%%%%%%%%%%%%%%%%%%%%%%%%%%%%%%%%%%%%%%%%%%%%%%%%%%%%%%%%%%%%%%%%%%%%%%%%%%%%%%%%%%%%%%%%%%%%%%%%%%%%%%%%%%%%%%%%%%%%%%%%%%%%%%%%%%%%%%%%%%%%%%%%%%%%%%%%%%%%%%%%%%%%%%%%%%%%%%%%%%%%%
		\acresetall
	%%%%%%%%%%%%%%%%%%%%%%%%%%%%%%%%%%%%%%%%%%%%%%%%%%%%%%%%%%%%%%%%%%%%%%%%%%%%%%%%%%%%%%%%%%%%%%%%%%%%%%%%%%%%%%%%%%%%%%%%%%%%%%%%%%%%%%%%%%%%%%%%%%%%%%%%%%%%%%%%%%%%%%%%%%%%%%%%%%%%%%%%%%%%%%%%%%%%%%%%%%%%%%%%%%%%%%%%%%%%%%%
  \section{Introduction}
\label{sec:intro}
\begin{figure}[t]
	\centering
	\includegraphics[width=8cm]{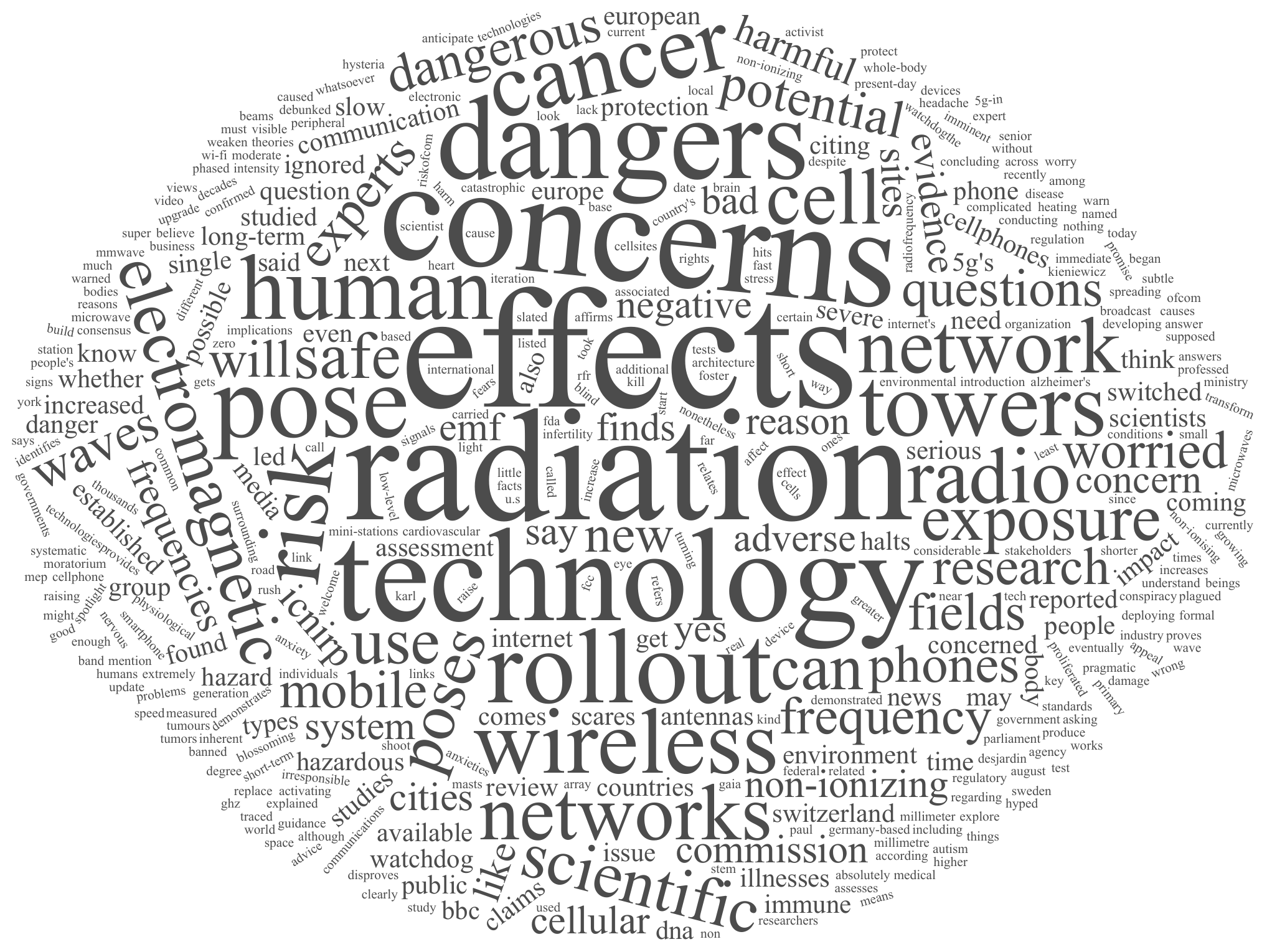}
	\caption{Word cloud of the first five pages of Google results for the search terms ``5G health risks'' (excluding from the word cloud the search terms).}
	\label{fig:tag-cloud}
	%\vspace{-4mm}
\end{figure}

The rolling out of \ac{5G} networks is a fundamental step to enable the variegate set of services offered by 5G across the world. The deployment of 5G networks requires installing new 5G \acp{gNB} over the territory, as well as the diffusion of 5G \ac{UE} among the users. Historically, the large-scale adoption of each new technology has always been accompanied by a mixture of positive and negative feelings by the population \cite{electricityfears}. Nowadays, a similar controversy involves the 5G technology, i.e., a non-negligible number of people firmly convinced that 5G constitutes a real danger for human health \cite{NybergHardel:17}. As a consequence, the words ``5G'' and ``risks'' are often associated together, with a negative impact on the perception of 5G among the population. For example, Google retrieves more than 88 million results when searching the terms ``5G health risks''. As graphically shown in Fig.~\ref{fig:tag-cloud}, the words appearing in the search results (excluding the search terms) often include negative nuances and expressions of concerns. Fuelled by the social media, the sentiment of fear against 5G {is spreading} across the world ({not among the whole set of citizens, but at least in part of the population}), {leading} {some} communities/municipalities {to ban the deployment of 5G sites in their territory} \cite{switzerland, png, jamaica}, as well as {driving several} sabotages of towers that host 5G {(and pre-5G)} equipment \cite{derby5gfire,dutch5Gfire,liverpool5Gfire}.

\begin{figure}[t]
	\centering
	\includegraphics[width=8cm]{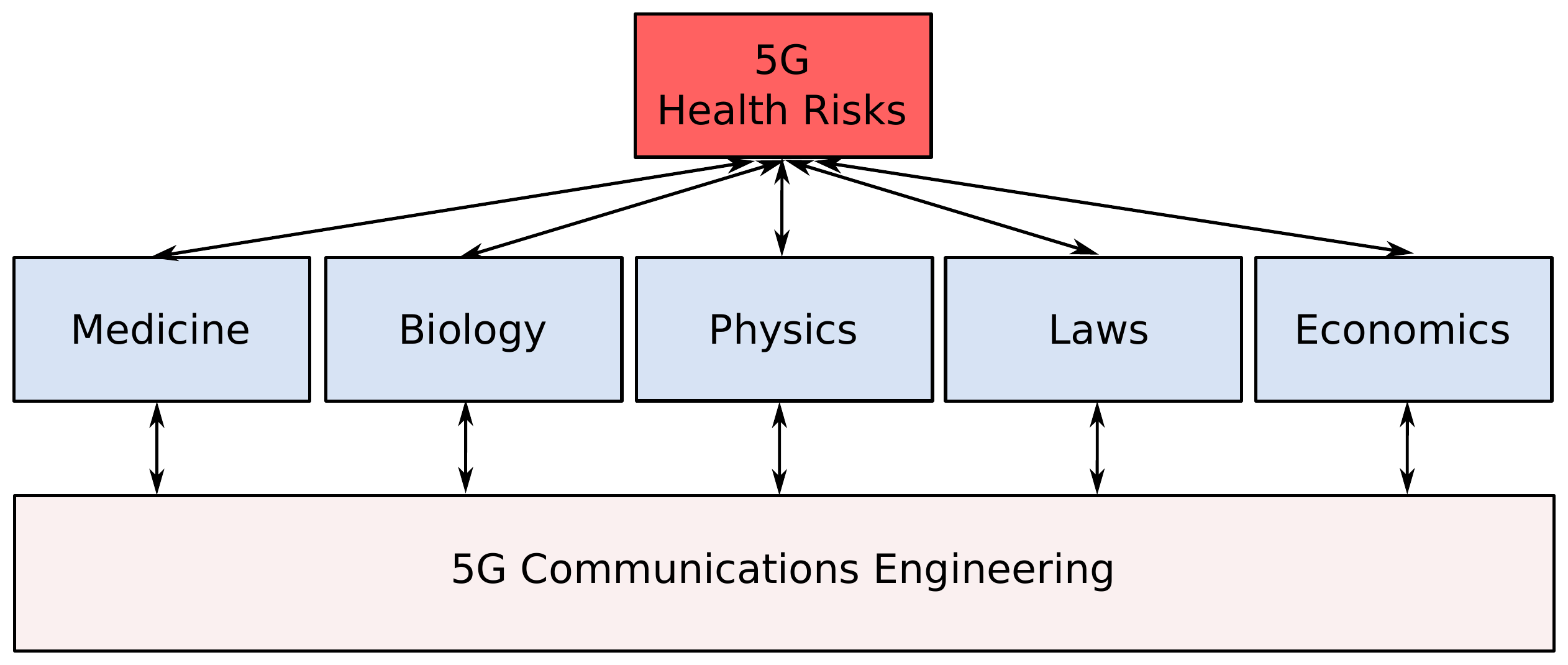}
	\caption{5G communications engineering is the glue to analyze the different disciplines involved in the assessment of health risks from 5G exposure.}
	\label{fig:general_scheme}
	%\vspace{-4mm}
\end{figure}

The fear of 5G technology is mainly due to a biased feeling among the population, which is often driven by weak theories (a.k.a. pseudoscience), developed without solid scientific evidence. {Clearly, such theories can be easily debunked when considering 5G frequencies below 6~GHz. However, there is currently a lack of well done scientific studies focused on the assessment of (potential) health effects from 5G devices operating in the mm-Wave band} \cite{karipidis20215g}, {thus fuelling the argument that not enough research has been done to demonstrate the safety of 5G}. {Not surprisingly, part of the population is convinced that} exposure to \acp{EMF} generated by 5G \acp{gNB} and 5G \ac{UE}  {is dangerous for health} \cite{ETNO2021}.

Although the research community well knows that{, at present time,} there are no proven health effects from an \ac{EMF} exposure kept below the maximum limits enforced by law ({see e.g.,} \cite{bushberg2020ieee}), the health risks associated with 5G are overly perceived by the general public.\footnote{In line with the recommendations of international organizations (such as the \ac{WHO} and the \ac{ITU}), we also advocate the need of continuing to investigate possible - yet still unknown at present - health effects due to 5G exposure, {especially at higher frequencies}.} This is (likely) due to multiple reasons, which include both rational and irrational aspects. In general, we observe: \textit{i}) a widespread fragmentation of research across the different disciplines that are involved in the health risks assessment of 5G, \textit{ii}) a diffuse feeling of a suspect against the institutions that are supposed to control the health risks of 5G, and \textit{iii}) a continuous fabrication of fake news (misinformation), which generally convey the message of severe health risks triggered by 5G exposure in an immediate and catching way compared to the scientific community.  For example,  the misinformation or ``infodemic'' related to the connection between \ac{EMF} exposure from 5G \ac{gNB} and the infection of \ac{Covid19} disease \cite{Caulfield:20} is currently very widespread in non-scientific communities.\footnote{The ``infodemic'' expression has been used by the \ac{WHO} to describe the excessive amount of misinformation regarding \ac{Covid19}  pandemic.}

\begin{figure*}[t]
	\centering
	\resizebox{\textwidth}{!}{
		\begin{tikzpicture}[node distance = 0.5cm, thick, nodes = {align = center}, 
		>=latex] 
		\small
	
		\definecolor{lightred}{rgb}{255,97,97};
		\definecolor{linen}{RGB}{250,240,230};
		\definecolor{coral}{RGB}{255,127,80};

	    \node[survey2] (survey2) {\large \textbf{Health Risks Associated with 5G Exposure}}; 
		\node[survey] (survey) {\large \textbf{Health Risks Associated with 5G Exposure}}; 
		
		\node[dummy, below = of survey] (dummy2) {};
		
		\node[dummy, below = of dummy2] (dummy) {};
		\node[section2, left = of dummy] (sec3) {\large Sec.~\ref{sec:metrics_regulations}\\5G Exposure: Regulations and Compliance Assessments};
		\node[section, left = of dummy] (sec3) {\large Sec.~\ref{sec:metrics_regulations}\\5G Exposure: Regulations and Compliance Assessments};
		
		\node[section, below = of sec3] (sec3a) {\large A. International Guidelines on 5G EMF Exposure};
		\node[section, below = of sec3a] (sec3b) {\large B. Impact of National Regulations};
		\node[section, below = of sec3b] (sec3c) {\large C. Compliance Assessment of 5G Exposure};
		
	    \draw[-,-,->, thick,] (sec3.south) -- (sec3a.north);  
        \draw[-,-,->, thick,] (sec3a.south) -- (sec3b.north);  
        \draw[-,-,->, thick,] (sec3b.south) -- (sec3c.north);

		\node[section2, left = of sec3] (sec2) {\large Sec.~\ref{sec:medical_evidence} \\Health Effects from 5G Exposure};
		\node[section, left = of sec3] (sec2) {\large Sec.~\ref{sec:medical_evidence} \\Health Effects from 5G Exposure};
		
	%	\node[section2, below = of sec2] (sec2a) {\large Basic Principles of \\5G Exposure};
		\node[section, below = of sec2] (sec2a) {\large A. Basic Principles of 5G Exposure};
		\node[section, below = of sec2a] (sec2b) {\large B. 5G Exposure Metrics};
		\node[section, below = of sec2b] (sec2c) {\large C. Alleged Health Effects of 5G Exposure};
		\node[section, below = of sec2c] (sec2d) {\large D. Relevant Medical Studies};
		\node[section, below = of sec2d] (sec2e) {\large E. Review of the Studies from Comm. Eng. Perspective};
		
		 \draw[-,-,->, thick,] (sec2.south) -- (sec2a.north);  
        \draw[-,-,->, thick,] (sec2a.south) -- (sec2b.north);  
        \draw[-,-,->, thick,] (sec2b.south) -- (sec2c.north);
		\draw[-,-,->, thick,] (sec2c.south) -- (sec2d.north);
		\draw[-,-,->, thick,] (sec2d.south) -- (sec2e.north);

		\node[section2, left = of sec2] (sec1) {\large Sec.~\ref{sec:intro}\\Introduction};
		\node[section, left = of sec2] (sec1) {\large Sec.~\ref{sec:intro}\\Introduction};

		\node[section2, right = of dummy] (sec4) {\large Sec.~\ref{sec:health_risks_5G_features}\\Health Risks Associated with 5G Features};
		\node[section, right = of dummy] (sec4) {\large Sec.~\ref{sec:health_risks_5G_features}\\Health Risks Associated with 5G Features};
		
			\node[section, below = of sec4] (sec4a) {\large A. Extensive Adoption of Massive MIMO and Beamforming};
			
			\node[section, below = of sec4a] (sec4b) {\large B. Densification of 5G Sites Over the Territory};
			
			\node[section, below = of sec4b] (sec4c) {\large B. Adoption of Frequencies in the mm-Wave bands};
			
			\node[section, below = of sec4c] (sec4d) {\large C. Connection of Millions of IoT Devices};
			
			\node[section, below = of sec4d] (sec4e) {\large D. Coexistence of 5G with Legacy Technologies};
			
			 \draw[-,-,->, thick,] (sec4.south) -- (sec4a.north);  
        \draw[-,-,->, thick,] (sec4a.south) -- (sec4b.north);  
        \draw[-,-,->, thick,] (sec4b.south) -- (sec4c.north);
		\draw[-,-,->, thick,] (sec4c.south) -- (sec4d.north);
		\draw[-,-,->, thick,] (sec4d.south) -- (sec4e.north);
		
		\node[section2, right = of sec4] (sec5) {\large Sec.~\ref{sec:risk_mitigation_techniques}\\Risk Mitigation Techniques for 5G Exposure};
		\node[section, right = of sec4] (sec5) {\large Sec.~\ref{sec:risk_mitigation_techniques}\\Risk Mitigation Techniques for 5G Exposure};

		\node[section, below = of sec5] (sec5a) {\large A. Device-based Approaches};
			
		\node[section, below = of sec5a] (sec5b) {\large B. Architectural-based Approaches};
		
		\node[section, below = of sec5b] (sec5c) {\large C. Network-based Approaches};
		
		\node[section, below = of sec5c] (sec5d) {\large D. Regulation-based Approaches};
		
	    \draw[-,-,->, thick,] (sec5.south) -- (sec5a.north);  
        \draw[-,-,->, thick,] (sec5a.south) -- (sec5b.north);  
        \draw[-,-,->, thick,] (sec5b.south) -- (sec5c.north);
		\draw[-,-,->, thick,] (sec5c.south) -- (sec5d.north);
		
		\node[section2, right = of sec5] (sec6) {\large Sec.~\ref{sec:summary_concl}\\Summary and Conclusions};
		\node[section, right = of sec5] (sec6) {\large Sec.~\ref{sec:summary_concl}\\Summary and Conclusions};

		\draw[-,-,->, thick,] (survey.south) -- (sec1.north);  
		\draw[-,-,->, thick,] (survey.south) -- (sec2.north);
		\draw[-,-,->, thick,] (survey.south) -- (sec3.north);  
		\draw[-,-,->, thick,] (survey.south) -- (sec4.north);  
		\draw[-,-,->, thick,] (survey.south) -- (sec5.north);  
		\draw[-,-,->, thick,] (survey.south) -- (sec6.north);  
	%	\draw[-,-,->, thick,] (sec1.east) -- (sec2.west);  
%		\draw[-,-,->, thick,] (sec2.east) -- (sec3.west);  
%		\draw[-,-,->, thick,] (sec3.east) -- (sec4.west);  
%		\draw[-,-,->, thick,] (sec4.east) -- (sec5.west);  
%		\draw[-,-,->, thick,] (sec5.east) -- (sec6.west);  

		%\draw[-,-,->, thick,] (riskmittech.south) -- (archbased.north);   
		%\draw[-,-,->, thick,] (riskmittech.south) -- (netbased.north);    
		%\draw[-,-,->, thick,] (riskmittech.south) -- (regbased.north);    

		%\draw[|-,-|,-, thick,] (dismission.west) -- +(-0.5em,0) |- (regbased.west); 
		%\draw[|-,-|,-, thick,] (harmonization.west) -- +(-0.5em,0) |- (regbased.west); 
		%\draw[|-,-|,-, thick,] (reducingexposure.west) -- +(-0.5em,0) |- (regbased.west);  
		%\draw[|-,-|,-, thick,] (emfmeasureport.west) -- +(-0.5em,0) |- (regbased.west);  
		
		%\draw[|-,-|,-, thick,] (sarawaredesign.west) -- +(-0.5em,0) |- (devicebased.west); 
		%\draw[|-,-|,-, thick,] (emfwaregnb.west) -- +(-0.5em,0) |- (devicebased.west); 

		\end{tikzpicture} 
	}
	\caption{{Organization of our work.}}
	\label{fig:survey_organization}
\end{figure*}
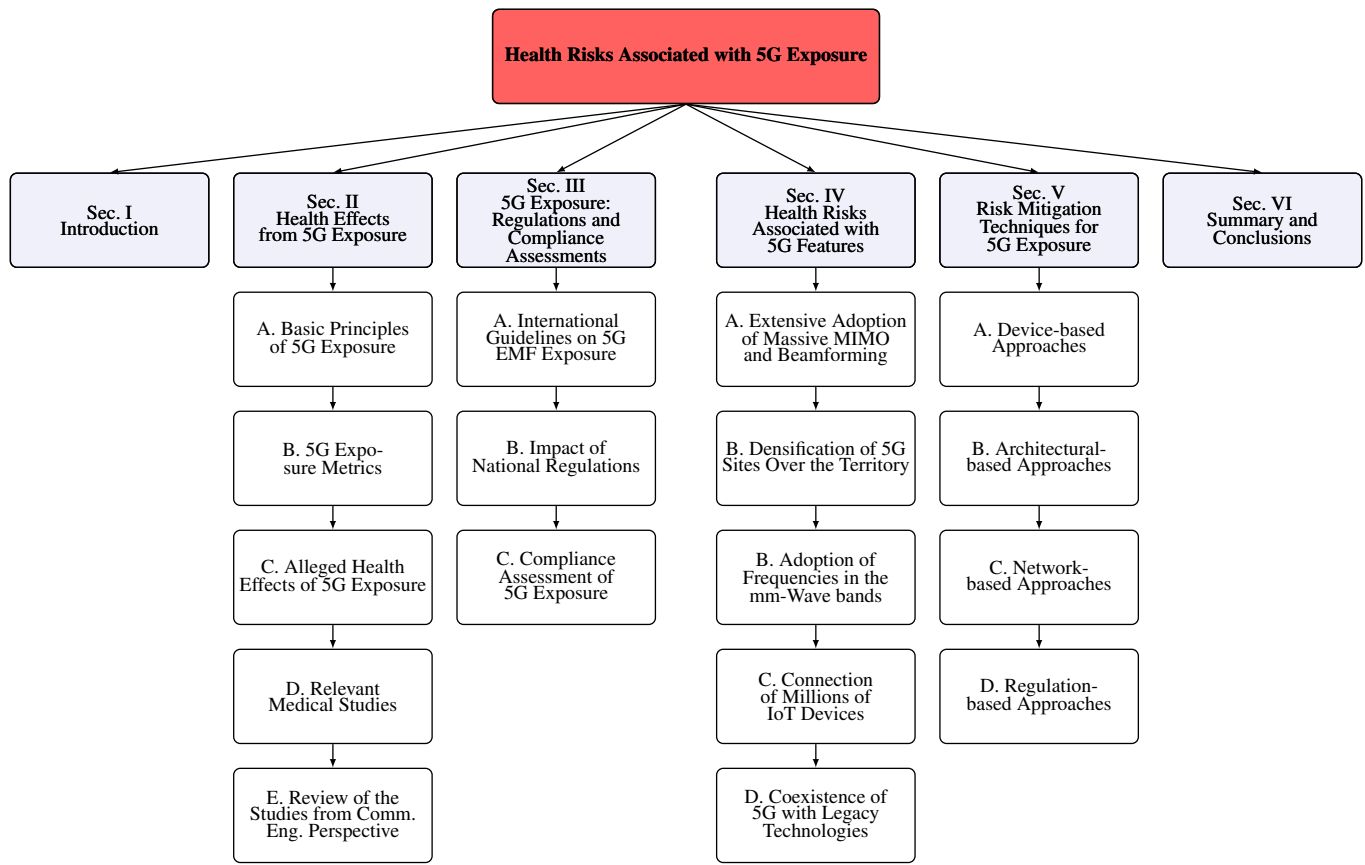

In this scenario, analyzing the scientific literature targeting the health risks of 5G is a fundamental task on one side and a challenging (and multi-faceted) problem. Indeed, the health risks assessment of 5G covers several disciplines, which include (to cite a few): medicine, biology, physics, economics, and laws. Although we recognize the relevance of each of the previous fields, the scientific research about health risks associated with 5G is frequently polarized towards a single aspect of the whole picture, with little attention to the other areas. For example, medical studies are often focused on assessing the health diseases triggered by 5G exposure (including legacy mobile generations), with little emphasis on the meaningfulness of the adopted test conditions. Also, the conditions of the experiments are often very conservative and pretty far from the real settings of the radio equipment under operation in a deployed network. Since it is challenging to achieve a unique view of health risks across all the involved disciplines, the population tends to believe in the large number of fake theories/allegations claiming severe health risks triggered by 5G. Apparently, this issue also severely increases the sense of suspect against the institutions devoted to controlling health risks.

Given this background, a key question naturally emerges: Is it possible to scientifically analyze the health risks associated with 5G through holistic work spanning across the different disciplines that are involved in the problem? Our ambitious goal is to provide an answer to this intriguing question. More concretely, we adopt a 5G communications engineering perspective as the glue that links the research works from the different fields into a unique big picture. {Clearly, our goal is not to compete with the communications efforts done by health agencies on the theme, but rather to add another (important) voice in the wide topic of health risk assessment of 5G exposure.}

As sketched in Fig.~\ref{fig:general_scheme}, communications engineering is a common denominator for all the disciplines involved in assessing the health risks associated with 5G. For example, communications engineering can provide insights about realistic patterns of power radiated by 5G equipment, allowing a realistic assessment of 5G exposure. On the other hand, communications engineering can drive the design of new 5G equipment and protocols tailored to the minimization of the \acp{EMF} and, consequently, of the health risks. In addition, the communications engineering can provide indications about the effectiveness of the laws that regulate the 5G exposure, e.g., to assess if some laws are too conservative or too relaxed compared to the real conditions at which 5G devices operate. In a nutshell, communications engineering is the \textit{passe-partout} to analyze the health risks of 5G. 

%Consequently, in this survey, we analyze all the aspects involved in health risks from 5G exposure under this new and rather unexplored perspective. 
Our key contributions include: 
\begin{enumerate}
	\item the analysis of the medical research focused on long-term EMF exposure, by exploiting the 5G communications engineering knowledge;
	\item the evaluation of the EMF metrics and the EMF regulations across all the countries in the world from the perspective of 5G communications engineering;
	\item the overview of the methods to assess the exposure compliance w.r.t. the maximum limits when considering 5G equipment;
	\item the analysis of the main 5G and beyond 5G technology features and their potential impact on the health risks; 
	\item the discussion of the mitigation techniques based on communications engineering that can be implemented to reduce the health risks of 5G.
\end{enumerate} 

\subsection{Paper Positioning}

\renewcommand{\arraystretch}{1}
\begin{table*}
	\footnotesize
	\centering
	\caption{Positioning of this work against other relevant papers analyzing the health risks of 5G technology.}
	\label{tab:comp_surveys}
	\begin{tabular}{|p{0.5cm}|p{0.5cm}|p{4cm}|p{4.2cm}|p{3.6cm}|p{3.4cm}|}
		\hline
		\rowcolor{Lightred} \multirow{2}{*}{\textbf{Work}} & \multirow{2}{*}{\textbf{Year}}  & \multirow{2}{*}{\textbf{Health Effects from 5G Exposure}} & \textbf{5G Exposure Metrics, Regulations and Compliance Assessment}  & \multirow{2}{*}{\textbf{Health Risks of 5G Features}} & \multirow{2}{*}{\textbf{5G Risks Mitigation}} \\
		%\rowcolor{Lightred} \multirow{-2}{*}{\textbf{Work}} & \textbf{5G Exposure} & \textbf{Compliance Assessment} & \textbf{of 5G Features} & \textbf{Techniques} \\
		%\cite{ZhaChaSau:11}
		\hline
		& & & & & \\[-0.8em]
		\begin{minipage}{0.4cm}\vspace{-1.5mm}\cite{Ciaula:18}\end{minipage}  &  2018 & \begin{minipage}{3.7cm}Partially covered:  authors mainly focused on works investigating the biological effects of pre-5G technologies  (including generic mm-Waves).\end{minipage} & Not covered & Not covered & Not covered \\[-0.9em]
		& & & & & \\
		\hline
		\rowcolor{Linen} & & & & & \\[-0.8em]
		\rowcolor{Linen} \begin{minipage}{0.4cm}\vspace{-1.5mm}\cite{Russell:18}\end{minipage} & 2018 & \begin{minipage}{3.7cm}Partially covered: \textit{i}) brief overview of works investigating the health risks of pre-5G technologies, \textit{ii}) review of the works investigating health effects from generic mm-Waves (not radiated by 5G antennas)\vspace{0.8mm}.\end{minipage}& \begin{minipage}{3.7cm}\vspace{-0.8mm}Partially covered: \textit{i}) brief overview of the FCC regulations, \textit{ii}) no discussion about other international {guidelines}, \textit{iii}) compliance assessment procedure only briefly mentioned.\end{minipage} & \begin{minipage}{3.7cm}Partially covered: \textit{i}) possible effect of 5G frequencies (only mm-Waves are mentioned, while sub-GHz and sub-6GHz frequencies are not reported at all), \textit{ii}) impact of \ac{gNB} densification only briefly analyzed.\vspace{0.8mm}\end{minipage}& Not covered. \\%[-0.9em]
		%    \cellcolor{Linen} & \cellcolor{Linen} & \cellcolor{Linen} & \cellcolor{Linen} & \cellcolor{Linen} & \cellcolor{Linen} \\
		\hline
		& & & & & \\[-0.8em]
		\begin{minipage}{0.4cm}\vspace{-1.5mm}\cite{chiaraviglio2018planning}\end{minipage} & 2018 &Not covered & \begin{minipage}{3.7cm}Partially covered: \textit{i}) only incident field \ac{EMF} for \ac{gNB} and no metric for \ac{UE}, \textit{ii}) impact of national regulations on \ac{gNB} planning in a single country, \textit{iii}) \ac{ICNIRP} regulations briefly introduced, \textit{iv}) compliance assessment procedure tailored to a single country.\end{minipage} & Brief discussion, no comprehensive overview of the related works. & Partially covered: network based solutions for \ac{gNB}. \\[-0.9em]
		& & & & & \\  
		\hline
		\rowcolor{Linen} & & & & & \\[-0.8em]
		\rowcolor{Linen} \begin{minipage}{0.4cm}\vspace{-1.5mm}\cite{SimMat:19}\end{minipage} & 2019  & \begin{minipage}{3.7cm}Partially covered: focus is on the research works investigating biological effects due to exposure from generic RF sources (not tailored to 5G emissions).\end{minipage} & \begin{minipage}{3.7cm}Partially covered: \textit{i}) exposure metrics briefly mentioned, \textit{ii}) incident \ac{EMF} strength taken into account, \textit{iii}) international {guidelines} only briefly mentioned.\vspace{0.8mm}\end{minipage} & Not covered & Not covered  \\
		\hline
		& & & & & \\[-0.8em]
		\begin{minipage}{0.4cm}\vspace{-1.5mm}\cite{PawKraZur:19}\end{minipage} & 2019  &Not covered & \begin{minipage}{3.7cm}Partially covered: \textit{i}) Review of exposure metrics for \ac{gNB} (and not for \ac{UE}), \textit{ii}) Brief overview of one international {guideline} and a set of national regulations (with a focus on Poland) for limiting the maximum \ac{EMF} strength, \text{iii}) overview of a generic procedure for compliance assessment of exposure, \textit{iv}) international compliance assessment procedures only briefly mentioned.\end{minipage} &  \begin{minipage}{3.7cm}Partially covered: \textit{i}) brief discussion on the impact of \ac{MIMO}, mm-Waves and densification without reviewing the literature.\end{minipage} & \begin{minipage}{3.5cm}Partially covered: authors briefly discussed the impact of strict regulations and the monitoring activities based on measurements.\end{minipage} \\[-0.9em]
		& & & & & \\   
		\hline
		\rowcolor{Linen} \begin{minipage}{0.4cm}\vspace{-1.5mm}\cite{JamHelBro:19}\end{minipage} & 2020  &\begin{minipage}{3.7cm}Partially covered: \textit{i}) brief overview of the alleged health effects from RF exposure, \textit{ii}) brief summary of medical studies investigating the impact of RF exposure on health.\end{minipage} & \begin{minipage}{3.7cm}\vspace{0.8mm} Partially covered: \textit{i}) detailed overview of exposure metrics, \textit{ii}) brief overview of the international guidelines (with a focus on UE), \textit{iii}) no overview of national regulations stricter than international guidelines, \textit{iv}) assessment of compliance only introduced.\vspace{0.8mm} \end{minipage} & \begin{minipage}{3.7cm}Covered in terms of basic 5G features (MIMO, densification, mm-Waves)\end{minipage} & \begin{minipage}{3.5cm}Partially covered in terms of network based and regulation based solutions (with a focus on \ac{EMF} mitigation).\end{minipage}\\  
		\hline
		\begin{minipage}{0.4cm}\vspace{-1.5mm}\cite{imran2019low}\end{minipage} & 2019 & Not covered   & \begin{minipage}{3.7cm}\vspace{0.8mm}Partially covered: \textit{i}) exposure metrics for \ac{gNB} and \ac{UE}, \textit{ii}) brief overview of international exposure regulations, \textit{iii}) no detailed analysis of country-specific exposure regulations, \textit{iv}) limited analysis of compliance assessment procedures.\vspace{0.8mm}\end{minipage} &  Not covered  & \begin{minipage}{3.5cm}Partially covered: \textit{i}) network based (limited to resource allocation), \textit{ii}) device based.\end{minipage} \\
		\hline
		\rowcolor{Linen}  \begin{minipage}{0.4cm}\begin{sideways}\textbf{This work}\end{sideways}\end{minipage} & 2020 &\begin{minipage}{3.7cm}Full coverage of: \textit{i}) basic principles of \ac{RF} exposure, \textit{ii}) overview of the main allegations against 5G exposure (updated on 2020), \textit{iii}) analysis of the animal-based and the population-based studies relevant to 5G.\end{minipage} & \begin{minipage}{3.7cm}\vspace{0.8mm}In-depth review with main contributions: \textit{i}) coverage of 5G exposure metrics for \ac{UE} and \ac{gNB}, \textit{ii}) analysis of international  regulations (updated on 2020), \textit{iii}) analysis of local regulations and their impact on 5G deployment (data from more than 225 countries), \textit{iv}) analysis of the state-of-the-art compliance assessment procedures (updated on 2020)\vspace{0.8mm}\end{minipage} & \begin{minipage}{3.7cm} Comprehensive analysis of the impact from: \textit{i}) MIMO and beamforming, \textit{ii}) \ac{gNB} densification, \textit{iii}) mm-Waves, \textit{iv}) connection of millions of devices, \textit{v}) coexistence with legacy technologies (2G/3G/4G, radio and TV broadcasting, weather satellites)\end{minipage} &  \begin{minipage}{3.5cm}Comprehensive overview of the solutions for \ac{gNB} and \ac{UE}: \textit{i}) device based, \textit{ii}) architectural based, \textit{iii}) network based, \textit{iv}) regulation based.\end{minipage}\\
		\hline
		\rowcolor{Lightred} \multirow{2}{*}{\textbf{Work}} & \multirow{2}{*}{\textbf{Year}}  & \multirow{2}{*}{\textbf{Health Effects from 5G Exposure}} & \textbf{5G Exposure Metrics, Regulations and Compliance Assessment}  & \multirow{2}{*}{\textbf{Health Risks of 5G Features}} & \multirow{2}{*}{\textbf{5G Risks Mitigation}} \\
		\hline
	\end{tabular}
\end{table*}
\renewcommand{\arraystretch}{1.0}

Tab.~\ref{tab:comp_surveys} reports the positioning of our work w.r.t. the relevant papers \cite{Ciaula:18,Russell:18,chiaraviglio2018planning,SimMat:19,PawKraZur:19,JamHelBro:19,imran2019low} already published in the literature. Although we recognize the importance of such previous works, 
to the best of our knowledge, this is the first paper targeting the analysis of health risks due to 5G by adopting a communications engineering perspective in a comprehensive and in-depth manner. Specific aspects of our work (and not covered by previous papers) include: 
\textit{i}) a comprehensive approach that covers the health risks associated with exposure from both 5G \acp{gNB} and 5G \ac{UE};
\textit{ii}) a detailed analysis of the different 5G \ac{EMF} metrics and the different 5G \ac{EMF} regulations defined by international organizations (\ac{ICNIRP}, \ac{IEEE}), federal commissions (\ac{FCC}) and even single nations (by extracting data from more than 225 countries);
\textit{iii}) the review of the latest guidelines from \ac{IEEE}, \ac{IEC} and \ac{ITU} to perform the compliance assessment of 5G exposure;
\textit{iv}) the risk analysis of the set of 5G features that are associated with health issues by the population;
and \textit{v}) the review of the main risk mitigation techniques at a device, architectural, network, and regulation levels.
	
	% Previous surveys in the literature are focused on specific aspects, such as: \textbf{TBD}

	\subsection{Paper Organization}
	
	The rest of the paper is organized by following the scheme reported in Fig.~\ref{fig:survey_organization}. {We initially analyze the health effects of 5G exposure in } Sec.~\ref{sec:medical_evidence}. {In particular, we briefly summarize the basic principles of 5G exposure. We then provide a concise overview of the exposure metrics that are relevant to 5G. In the following step, we overview the main health effects (particularly the negative ones) that are associated with 5G exposure. We then provide an overview of the main medical studies that are relevant to 5G exposure. Finally, we review the main medical studies from the perspective of communications engineering, e.g., by considering the differences between the test conditions of such studies against the real settings at which 5G equipment operate}.
	
	%analyzes the main health effects of \ac{RF} exposure under the light of 5G communications. This section is also tailored to a critical review of the (recent) studies aimed at finding connections between the emergence of tumors and exposure to \ac{RF} devices from the perspective of the real settings at which 5G equipment will operate. 

	Sec.~\ref{sec:metrics_regulations} {moves one step further in the risk assessment by reporting an overview of the international guidelines governing 5G exposure. In addition, the section focuses on the differences introduced by national regulations w.r.t. international guidelines, and on the impact that such regulations have on the perceived health risks of 5G. Moreover, we review the main procedures of the assessment of compliance of 5G exposure against the maximum limits defined by law.}

	%quantifies the health risks by considering 5G exposure. To this aim, we shed light on exposure metrics, exposure regulations, and compliance assessment procedures relevant to the context of 5G. 
	
	{Sec.~\ref{sec:health_risks_5G_features} is devoted to a review of the main allegations that are raised against specific 5G features. In particular, we tackle the impact of massive \ac{MIMO} and beamforming on the perceived health risks. We then move our attention to the densification of cell sites over the territory, and its associated claims about a dramatic increase of exposure. In the following step, we consider the impact of frequencies in the mm-Wave bands on the health risks. Eventually, we tackle the issue of connecting millions of \ac{IoT} devices per cell. In the following step, we discuss how 5G can coexists with other technologies, and how this feature will affect the health risks.}
	
	{Sec.~\ref{sec:risk_mitigation_techniques} focuses on the techniques that can be put into place to mitigate the risks of 5G exposure. In particular, we survey the works targeting the reduction of exposure at the device, architectural, network and regulation levels.}
	
	%We then examine in Sec.~\ref{sec:health_risks_5G_features} the main allegations of health risks associated with 5G features (e.g., adoption of mm-Waves, the proliferation of 5G antennas, the large adoption of \ac{MIMO} and beamforming, and the connection of millions of devices), by analyzing the relevant literature from the communications engineering perspective. We then discuss in Sec.~\ref{sec:risk_mitigation_techniques} the main techniques that can be put into place to minimize the (potential) risks of 5G exposure at the device, architectural, network and regulation levels.  
	
	Finally, Sec.~\ref{sec:summary_concl} concludes our work.
	%%%%%%%%%%%%%%%%%%%%%%%%%%%%%%%%%%%%%%%%%%%%%%%%%%%%%%%%%%%%%%%%%%%%%%%%%%%%%%%%%%%%%%%%%%%%%%%%%%%%%%%%%%%%%%%%%%%%%%%%%%%%%%%%%%%%%%%%%%%%%%%%%%%%%%%%%%%%%%%%%%%%%%%%%%%%%%%%%%%%%%%%%%%%%%%%%%%%%%%%%%%%%%%%%%%%%%%%%%%%%%%
	%%%%%%%%%%%%%%%%%%%%%%%%%%%%%%%%%%%%%%%%%%%%%%%%%%%%%%%%%%%%%%%%%%%%%%%%%%%%%%%%%%%%%%%%%%%%%%%%%%%%%%%%%%%%%%%%%%%%%%%%%%%%%%%%%%%%%%%%%%%%%%%%%%%%%%%%%%%%%%%%%%%%%%%%%%%%%%%%%%%%%%%%%%%%%%%%%%%%%%%%%%%%%%%%%%%%%%%%%%%%%%%
		\section{Health Effects from 5G Exposure}
	\label{sec:medical_evidence}
	
	We perform our analysis under the following avenues: \textit{i}) basic principles of \ac{RF} exposure, \textit{ii}) summary of the alleged health effects from \ac{RF} exposure, \textit{iii}) overview of the relevant medical studies in the context of 5G communications, \textit{iv}) critical review of these medical studies from the perspective of 5G communications engineering.

	\subsection{Basic Principles of \ac{RF} Exposure}
	
	The exposure from \ac{EMF} can be categorized according to the effects on the cells generated by the electromagnetic waves. In particular, we distinguish between ionizing radiations and non-ionizing radiations. The former category includes the waves that have enough energy to remove the electrons from the atoms in the living cells, causing the atom to become ionized. For example, X-rays with frequencies in the range $3\times10^{16}$~[\si{\hertz}] - $3\times10^{19}$~[\si{\hertz}] and gamma-rays  with frequencies larger than $3\times10^{19}$~[\si{\hertz}] fall within the ionizing radiation. {Depending on the dose level, the} cells exposed to ionizing radiation {may die or} become cancerous, thus posing a risk for the health effects. On the other hand, \acp{EMF} belonging to  the non-ionizing radiation group are composed of waves that do not have enough energy to ionize the cells, thus (likely) avoiding cancer and death for the exposed cells. However, the waves may have enough energy to vibrate the molecules, causing a possible health issue. 
	
In this scenario, exposure from \ac{RF} communications equipment falls within the non-ionizing radiation category. 	More specifically, the biological effects of \ac{RF} radiation can be further classified into thermal effects and non-thermal effects. Focusing on the thermal effects, this group is characterized by an \ac{RF} exposure that can produce a heating of the exposed tissues. An example of \ac{EMF} source introducing thermal effects is the micro-wave oven (although this device is not intended to be used for \ac{RF} communications). In this context, the mechanism that triggers the raising of the temperature in the exposed tissues is well understood and deeply analyzed in the literature, since the massive adoption of radio equipment for broadcast transmission \cite{tell1972broadcast}. To face this issue, regulatory authorities (e.g., the \ac{EC} in Europe and the \ac{FCC} in the USA), international commissions (e.g., \ac{ICNIRP}) and international organizations (e.g., \ac{IEEE}) define maximum \ac{RF} exposure limits that allow preventing the heating effects on the exposed tissues. % such that a \ac{RF} intensity below the limit is not able to increase the temperature of the exposed tissue by more than \num{1}\si{\degreeCelsius}. 
	
	Regarding the non-thermal effects, the majority of the literature and reports of international organizations state that there is not a clear causal correlation between \ac{EMF} exposure levels generated by \ac{RF} sources operating below maximum limits defined by law and emergence of biological effects, see, e.g., the  Swedish radiation safety authority report  \cite{Danker:19}, \ac{WHO} and \ac{ITU} statements \cite{whoemf1,whoemf2,itu}, and recent \ac{ICNIRP} guidelines \cite{ICNIRPGuidelines:20}. However, since the mechanism by which the \ac{RF} exposure may cause non-thermal effects is still not entirely known (if there is any), it is essential to continue the research in this field.
	
	\begin{figure*}[t]
		\centering
		\includegraphics[width=0.99\linewidth,clip]{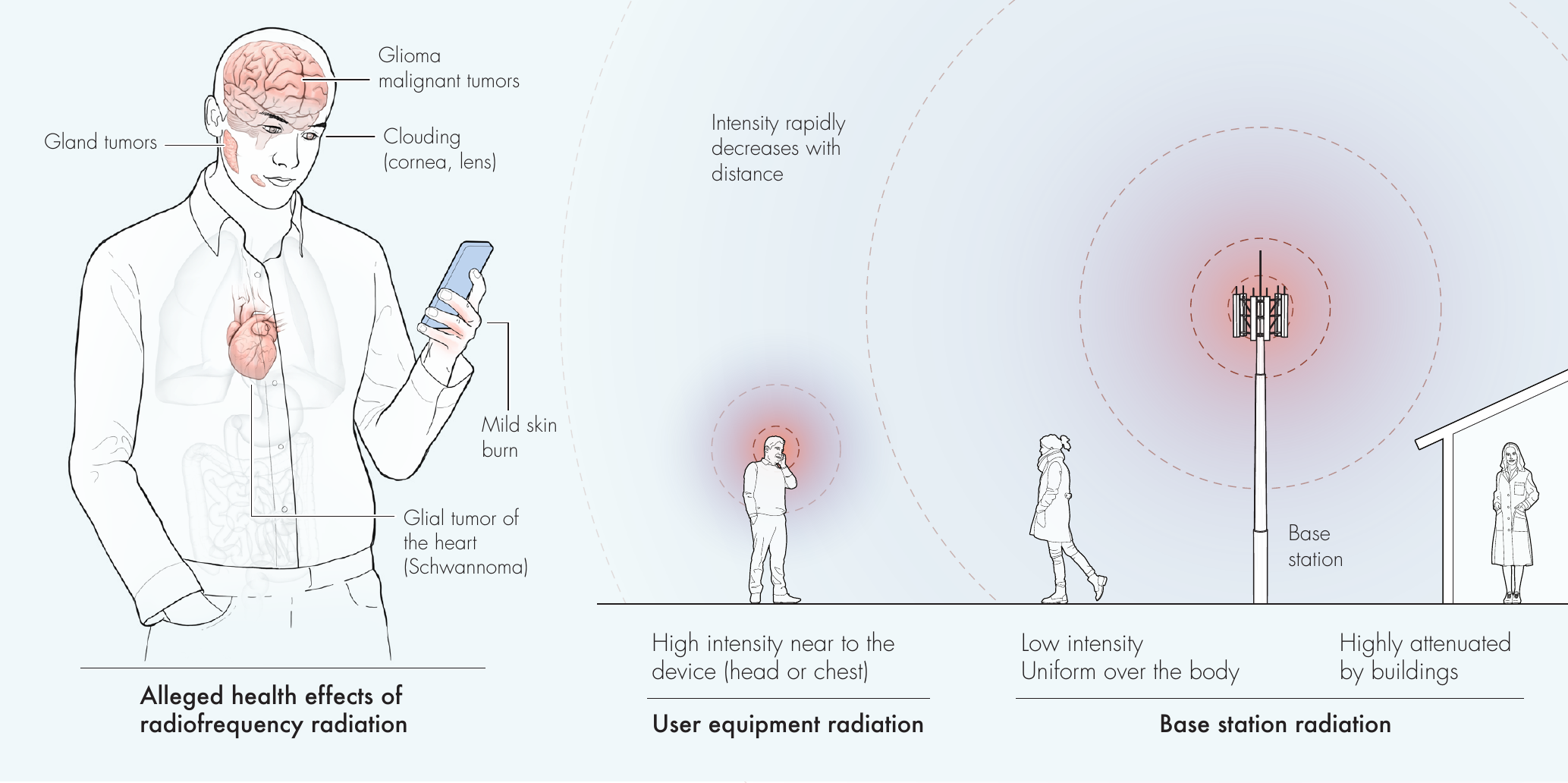}
		\caption{Left part: main health effects {that are alleged as a consequence of} exposure from \ac{RF} devices (including 5G equipment). Glioma and Schwannoma tumors have been observed only in animals exposed to high levels of \ac{EMF}. Right part: the primary sources of \ac{EMF} exposure when considering mobile network equipment. The \ac{EMF} exposure from \ac{UE} tends to be higher and more localized on the body than the one from radio base stations. Moreover, the \ac{EMF} intensity rapidly decreases as the distance between the base station and the user is increased. Finally, buildings introduce a shielding effect that attenuates the exposure from outdoor base stations.}
		\label{fig:202006slimfacompt4fig-1}
	\end{figure*}
	
	Fig.~\ref{fig:202006slimfacompt4fig-1} on the right shows the typical conditions of \ac{EMF} exposure from \ac{RF} devices, i.e., \ac{UE} and base stations. In general, \ac{UE} radiate close to users, by generating an \ac{EMF} that is localized either on the head or chest. On the other hand, base stations radiate over the whole body and large portion of the territory compared to \ac{UE}. However, the \ac{EMF} generated by base stations tends to rapidly decrease in intensity as the distance from the \ac{RF} source increases. Moreover, a shielding effect from base station \ac{EMF} occurs inside buildings. Therefore, the exposure from base stations is, in general, lower compared to the one radiated from \ac{UE}. Despite this fact, the population associates higher health risks to base station emissions w.r.t. \ac{UE} radiation. In the following, we provide more details about the alleged health effects of \ac{RF} exposure. 
    
    \subsection{{5G Exposure Metrics}} 
    \label{subsec:exposure_metrics}
The main metrics that are used to characterize 5G (and pre-5G) exposure are: \textit{i}) \ac{EMF} strength, \textit{ii}) power density, \textit{iii}) \ac{SAR} value. In the following, we provide a concise definition of each metric. We refer the interested reader to \cite{ICNIRPGuidelines:20} and references therein to obtain more details about exposure metrics for \ac{RF} sources.

\subsubsection{Electromagnetic Field Strength}
Each \ac{RF} source generates an \ac{EMF} that is spread over the environment. The field is composed of an electric component and a magnetic one. Let us denote the electric field as $\boldsymbol {E}$, with a measurement unit in terms of Volt per meter [V/m].  Similarly, let us denote the magnetic field as $\boldsymbol {H}$, with a measurement unit in terms of {Ampere per meter [A]}. In general, both $\boldsymbol {E}$ and $\boldsymbol {H}$ are time-averaged values, i.e., they are estimated over a sufficiently long-time-interval (e.g., in the order of minutes \cite{ICNIRPGuidelines:20}). Under far-field conditions, the \ac{EMF} is characterized by solely analyzing $\boldsymbol {E}$. Otherwise, when the \ac{EMF} is evaluated under near-field conditions, both $\boldsymbol {H}$ and $\boldsymbol {E}$ are needed to fully characterize the \ac{EMF} strength.

Apart from time-averaged values, the \ac{EMF} can be computed as an average from different points in the space. For example, the spatially averaged electric field strength $\Eavg$ over volume $V$ is computed by applying a root mean square operation. More formally, we have: 
\begin{equation}
\label{eq:eavg}
\Eavg= \sqrt{{\frac  {1}{V}} \int_{\displaystyle \scriptstyle V}{{|{\boldsymbol  {E}}|^{2}}} \mathrm{d} v}  \quad \text[\si{\volt\per\meter}].
\end{equation} 
%In the previous equation, $\Eavg$ is computed by applying a root mean square operator.

\subsubsection{Power Density}
A second metric used to assess the level of exposure is the \ac{PD}, which can be either the absorbed power density $S_{\text{ab}}$ or the incident power density $S_{\text{inc}}$.  More formally, the absorbed power density $S_{\text{ab}}$ is  expressed as:
\begin{equation} 
\label{eq:sab}
S_{\text{ab}}= \iint_A  \frac{1}{A} \mathrm {Re}[\boldsymbol {E} \times \boldsymbol {H}^{*}] \,\mathrm{d} s, \quad [\text{W/m}^2],
 \end{equation}
 where the body surface is at position $0$~[cm], $A$~[cm$^2$] is the x-y integral area,  $\boldsymbol {E}$ is the electric field, $\boldsymbol {H}$ is the magnetic field, $\mathrm{d} s$ is the integral variable vector whose direction is orthogonal w.r.t. $A$,  while  $\mathrm {Re}(\cdot)$ and  $(\cdot)^*$ denote the real part and the complex conjugate, respectively.
 
The incident power density $S_{\text{inc}}$ is defined as the modulus of the complex Poynting vector. More formally, $S_{\text{inc}}$ is expressed as:
\begin{equation} 
\label{eq:sinc}
S_{\text{inc}}= |\boldsymbol {E} \times \boldsymbol {H}^{*}|, \quad [\text{W/m}^2].
 \end{equation}
 
Under far-field conditions or  transverse electromagnetic plane wave, Eq.~(\ref{eq:sinc}) is simplified as:
\begin{equation}
\label{eq:sincff}
S_{\text{inc}}= \frac{|\boldsymbol {E} |^2}{Z}=|\boldsymbol {H} |^2 \times Z , \quad [\text{W/m}^2]'
  \end{equation}   
where $Z=377$~[$\Omega$] is the characteristic impedance of the free space. It is important to remark that Eq.~(\ref{eq:sincff}) is also used when evaluating the equivalent power density metric (which is commonly denoted as $S_{\text{eq}}$).

Finally, the absorbed power density is related to the incident power density through the following equation:
\begin{equation}
\label{eq:sabsinc}
S_{\text{ab}}= \left( 1 - |\Gamma|^2 \right) \times S_{\text{inc}}, \quad [\text{W/m}^2],
\end{equation}
where $\Gamma$ is a reflection coefficient, which depends on multiple physical features ({e.g., the body tissue and/or the clothing above the body}). We refer the interested reader to \cite{ICNIRPGuidelines:20} for a more detailed overview of such properties. 

In general, the international {guidelines} define \ac{PD} limits that are expressed in terms of maximum $S_{\text{inc}}$ values, since the incident power density is easier to be measured compared to the absorbed power density $S_{\text{ab}}$.

\subsubsection{Specific Absorption Rate}
According to \cite{ICNIRPGuidelines:20}, \ac{SAR} is the time derivative of the energy consumed by heating that is absorbed by a mass, included in a volume of a given mass density. When considering biological tissues and/or organs, the \ac{SAR} is expressed as:
\begin{equation}
\label{eq:sardef}
\SAR= \frac{\sigma}{\rho} |\boldsymbol {E}|^{2}, \quad \text[\si{\watt\per\kilogram}]
\end{equation}
where $\sigma$~[S/m] is the electrical conductivity, $\rho$~[kg/m$^3$] is the density of the tissue/organ, and $\boldsymbol {E}$~[V/m] is the internal electric field.

Under not significant heat loss processes \cite{ICNIRPGuidelines:20}, it is possible to express the \ac{SAR} by considering the temperature rise.  More formally, we have:
\begin{equation}
\label{eq:sardefalt}
\SAR=c\,\frac{\Delta T}{\Delta t},
\end{equation}
where $c$~[J/(kg $\cdot$ Celsius)] is the tissue specific heat,  $\Delta T$~[Celsius] is the temperature rise, and $\Delta t$~[s] is the exposure duration.

In general, limits considering \ac{SAR} as exposure metric assume two distinct spatially-averaged values, namely whole body \ac{SAR} and local \ac{SAR}.
The whole body \ac{SAR} takes into account the body mass and the total energy absorbed by the body. On the other hand, the local \ac{SAR} assumes a given (small) volume with a given (small) mass.

Measuring the \ac{SAR} becomes challenging for assessing the compliance of the exposure w.r.t. the regulations for high frequencies (like the mm-Waves ones). When the frequency increases, the penetration depth of the wave decreases. Under such a condition, the temperature rise is more superficial, and the heat tends to be lost across the environment, as pointed out by \cite{ICNIRPGuidelines:20}. On the other hand, it is feasible to measure the \ac{PD} instead of the \ac{SAR} for high frequencies. In general, the majority of the regulatory standards assign a frequency threshold, denoted as $f_{\textsf{th}}$, after which the considered limits switch from \ac{SAR} to \ac{PD}. However, some regulations (like \cite{ICNIRPGuidelines:20}) additionally include \ac{SAR} limits also for frequencies larger than $f_{\textsf{th}}$, in order to apply a conservative assumption. In any case, all the regulations differentiate between whole body \ac{SAR} and local \ac{SAR} (e.g., head, chest). 
    
    \subsection{Alleged Health Effects from \ac{RF} Exposure}
    
    Fig.~\ref{fig:202006slimfacompt4fig-1} on the left sketches the (main) health diseases that are associated with \ac{RF} exposure. Although some diseases have been only observed in animals (and not in humans), the debate about possible health consequences due to \ac{RF} exposure is a hot (and controversial) topic. {For example, the impact of brain-related diseases, including brain tumors and/or sleeping disorders, is highly critical in modern society.} To shed light on {these} aspects, we briefly summarize in the following the alleged health effects (including severe and not severe ones).
    
    \textbf{Cancer.} The \ac{IARC} listed non-ionizing \ac{RF} radiation from cell phones in Group 2B as ``Possibly carcinogenic to humans'' in 2010 \cite{IARC:11,IARCMonograph:13}, {mainly based on the analysis of epidemiological studies}. %This action was taken based on different experiments that analyze the carcinogenic effect on animals, which were exposed to EMF levels generated by RF equipment \cite{SzmSzuJol:82,LarSchBle:11,interphone2011acoustic,ChoGuyKru:92}. 
More recently, {a subset of} works ({see e.g.,} \cite{NTP:18a,NTP:18b,FalBua:18,HruNeuFra:08}) have found a statistically significant increase of rare cancers (i.e., glioma malignant tumors in the brain, glial tumors of the heart, and parotid gland tumors) associated to \ac{RF} exposure in rats. 
    
    %\textbf{LC: The gland tumor reported in the figure is not analyzed here. Please include it, {\color{black} added a reference for gland tumor in rats study and added the gland tumor}}

    \textbf{Skin Effects.}  
    The \ac{RF} exposure with high power density can lead to an increase in the temperature of the exposed body tissue \cite{NelnelWal:20}. However, a modest localized heat exposure can be compensated by the human body's heat regulation system. High doses of absorbed \ac{RF} exposure can cause a sensation of warmth in the skin, causing mild skin burns \cite{Roelandts:03}.
  %  \textbf{LC: Is this effect the mild skin burn reported in the figure? {\color{black} Yes, I added causing mild skin burn}}

    \textbf{Ocular Effects.} 
    High levels of \ac{RF} exposure with sufficiently high power density may cause several ocular effects \cite{Elder:03}, including cataracts, retina damages, and cornea issues. 
    
    \textbf{Glucose metabolism.}
    \ac{RF} exposure may affect the Glucose metabolism process in human cells \cite{VolTomWong:11}. The effect can be noticed in the body organs exposed to high levels of \acp{EMF}, e.g., the brain.  
    
    \textbf{Male Fertility.}
    According to a subset of studies (see e.g., \cite{SinNathRav:18,EroOztPek:06,YanAgrMat:07}) high levels \ac{RF} exposure may be associated with negative effects on reproductive health in terms of sperm-fertilizing ability. However, the connection of such effects with \ac{RF} exposure from communications equipment is to our best knowledge scientifically not proven. 
    %\textbf{LC: What does WHO state about this aspect? Apart from the studies it would be better to include the positions of the international organizations.} {\color{black} I did not find it in WHO, I searched carefully, we can remove it if you want}
    
    \textbf{Electromagnetic Hypersensitivity.}
    Some individuals report that \ac{RF} exposure causes several sensitivity symptoms to them, e.g., headache, fatigue, stress, burning sensations, and rashes.
    However, many independent studies (see, e.g., \cite{WhoEMH:05,RubMunWes:05}) have demonstrated that such symptoms are not correlated with the levels of \ac{RF} exposure.
    
    \textbf{Spreading of the \ac{Covid19} Disease.}
    Recently, different fake theories claim that there is a connection between the \ac{RF} from 5G equipment and the spreading of the \ac{Covid19} disease\cite{BBCCoVID:20}. In particular, the fake theories include:
\begin{itemize}
\item higher infection rates for regions of territory exposed to \ac{RF} from 5G experimental trials (e.g., Wuhan region, Lombardy region) compared to those not covered by 5G \cite{Covid5GConspiracy:20};
\item   a dangerous interaction at a cell level between the \ac{DNA} and  \ac{RFR} from 5G equipment, causing a fatal inflammation of lungs;
\item a supposed interaction between the \ac{RNA} of the \ac{Covid19} virus and the mm-Waves of 5G devices .
\end{itemize}
Such fake theories are not based on any scientific evidence, although they are widespread among the population. According to the UK \ac{NHS} \cite{covi19nhs5g}, the diffusion of fake theories trying to connect \ac{Covid19} and 5G is outrageous and dangerous. %At the time of preparing this manuscript, the growing fear of connections between 5G and \ac{Covid19} disease has also lead to different sabotages of sites hosting 5G equipment \cite{derby5gfire,dutch5Gfire,liverpool5Gfire} (as already mentioned in the Introduction).

    \textbf{Oxygen Effects.} Another allegation trying to link \ac{RF} from 5G equipment and health diseases include \textit{i}) a supposed oxygen absorption of 5G equipment out of the lugs, and \textit{ii}) the increase of carbon dioxide due to the cutting of the trees to improve the signal coverage of 5G. Focusing on \textit{i}), this allegation is not based on any scientific base. Focusing on \textit{ii}), there is no plan to cut the trees to improve the signal coverage. As a result, the claimed increase in carbon dioxide emissions due to 5G is fake news.
    
    \textbf{Summary and Next Steps.} Several health effects are associated with \ac{RF} exposure, ranging from scientific-based ones to allegations based on fake theories. In the following subsection, we provide more details about the works that aim at shedding light on the connection between exposure from 5G equipment and the emergence of tumors, which is one of the most controversial aspects brought to the attention of the general public. We intentionally leave apart skin, ocular, and glucose metabolism effects, as these phenomena are observed only for \ac{EMF} levels consistently higher than the ones radiated by 5G equipment. Therefore, using 5G equipment under realistic conditions guarantees that such effects do not occur in practice. Similarly, we also skip additional analysis about male fertility and electromagnetic hypersensitivity, as their connection with 5G communications is not scientifically proven \cite{WHOEMF16,WHOEMFHyper16}. Other health effects, which are based on hoaxes and fake theories, are not further discussed.
    %{\color{black}I added two references from WHO her to support this}. The 

    \subsection{Relevant Medical Studies in the Context of 5G Communications}

    We then focus our attention on the medical studies that are relevant to the exposure from 5G communications. {Tab.~\ref{tab:medical_studies_overview} reports a high level overview of the studies considered in this work. In particular,} we divide the related works according to the type of experiment, which can be either animal-based or population-based. {Other types of studies, based e.g., on in-vitro and/or ex-vivo experiments (e.g., living tissues extracted from surgery) are intentionally not treated and left as future works.} 
    
    \begin{table}[t]
\centering
\caption{{Medical Studies Considered in This Work}}
\label{tab:medical_studies_overview}
\footnotesize
\begin{tabular}{|c|c|c|c|}
\hline
\rowcolor{Lightred} \textbf{Category} & \textbf{Name} & \textbf{Summary} & \textbf{Review}  \\
\hline
 & NTP \cite{NTP:18a,NTP:18b} & & \\
\multirow{-2}{*}{Animal} & Ramazzini Institute \cite{FalBua:18} & \multirow{-2}{*}{Sec.~\ref{subsubsec:animal_based}} & \multirow{-2}{*}{Sec.~\ref{subsubsec:aninmal_based_review}}\\
\hline
& INTERPHONE \cite{Interphone:10,interphone2011acoustic} & & \\
& Danish Cohort  \cite{johansen2001cellular,frei2011use} & & \\
& Million Women \cite{benson2013mobile} & & \\
\multirow{-4}{*}{Population} & Cefalo Case-Control  \cite{aydin2011mobile}  & \multirow{-4}{*}{Sec.~\ref{subsubsec:population_based}} & \multirow{-4}{*}{Sec.~\ref{subsubsec:population_based_review}} \\
\hline
%\ac{RF} source - NTP & 3800~[W] (65~[dBm) & \cite{capstick2017radio} \\
%\ac{RF} source - Ramazzini Institute & 100~[W] (50~[dBm]) & \cite{FalBua:18} \\
% 5G macro \ac{gNB} & 200~[W] (53~[dBm]) & \cite{air5121} \\
%5G \ac{UE} & 0.2~[W] (23~[dBm]) & \cite{etsitr} \\

\end{tabular}
\end{table}

    \subsubsection{Animal-based Studies} 
    \label{subsubsec:animal_based}

    In this category, experiments are conducted on living animals (e.g., rats and mice), exposed to \acp{EMF} to mimic the exposure from \acp{gNB} and \ac{UE}. The number of works falling in this category is vast, with hundreds of animal-based studies that analyzed the potential health effects from \ac{RF} exposure over the last four decades (see, e.g., \cite{WikahmSel:80,MagZen:97,HaeIaeYun:09,Husabdgal:16,RecDemagliaSca:19,YinHuiQin:19}). 
    %\textbf{LC: provides some references here} 
    However, the majority of works presents multiple issues, including an insufficient duration of the experiment to extract long-term indications, and/or a too-small number of animals to derive statistically significant conclusions which are not subject to large biases. To face these issues, different international organizations (such as \ac{WHO}, \ac{NTP}, and other international bodies) 
    %\textbf{LC: Is it possible to add more references to different organization guidelines? It looks that the NTP have promoted guidelines and then it has conducted a study by following (obviously) their own guidelines.}
    %
    have provided guidelines for the procedures that need to be followed by animal-based studies that investigate the emergence of severe diseases (e.g., cancer) \cite{HufJacDav:08,NTPProcedures:11,GifCalJinVan:13,OCED:09,WHOassesment:86,JosMuiFer:08}.
    %\textbf{LC: Ahmed, do you have the reference to the guidelines of WHO?}
    %
    For example, the promoted guidelines define a minimum number of animals to be used (e.g., at least 50 animals for each group), a minimum temporal duration of the experiment (e.g., 2 to 3 years), and a minimum number of \ac{EMF} intensity levels (e.g., 3) \cite{VorFalManBel:19}.

    In this scenario, the most recent (and relevant)  studies that fulfill the above requirements are the \ac{NTP} study \cite{NTP:18a,NTP:18b} and the study of the Ramazzini Institute \cite{FalBua:18}. In the following, we provide more details about each of the aforementioned research works.
    
    \textbf{NTP Study.} \ac{NTP} performed in \cite{NTP:18a,NTP:18b} one of the longest bioassay conducted so far to evaluate the impact of \ac{EMF} exposure from \ac{RF} equipment on rats and mice. {The study addressed the 2G technology, but however it is frequently cited by the opponents of 5G.} In the experiments performed by \ac{NTP}, the animals were exposed to \ac{RF} in special chambers for several hours per day until the natural death. The total duration of the experiment was set to 2 years, with an initial assessment done after the first 28 days, and a final one performed at the end of the experiment.  \ac{RF} equipment used to generate the \ac{EMF} employed frequencies in the sub-GHz band for \cite{NTP:18a} and in the mid-band (i.e., above 1~[GHz] and below 6~[GHz] bands) for \cite{NTP:18b}. The radiated power of the \ac{RF} equipment was adjusted to satisfy a given level of whole-body exposure in the chamber, with different exposure levels assigned to the chambers. In addition, the generated \ac{EMF} levels were continuously monitored in each chamber, to verify the adherence of the exposure to the \ac{EMF} level imposed during the experiment.

    Focusing on the outcomes of the studies, we refer the reader to \cite{NTP:18a,NTP:18b} for a detailed analysis, while here we report a concise summary. In brief, the study conducted over the sub-GHz frequency \cite{NTP:18a} found clear evidence of carcinogenic activity in Sprague-Dawley male rats due to malignant Schwannoma of the heart. However, the same clear evidence of heart Schwannoma incidence was not found when considering the female rats. Besides, the incidence of other tumors (e.g., malignant glioma of the brain) was also related to the \ac{RF} exposure (when considering male rats again). In general, other severe diseases were also observed, without however, a clear connection to the \ac{RF} exposure level. Focusing then on the study adopting the mid-band frequencies \cite{NTP:18b}, no clear evidence of tumors was found by considering male or female rats. Eventually, the incidence of severe diseases may have been related to \ac{RF} exposure (although the observed cases were not statistically significant). Finally, the outcomes of \cite{NTP:18a,NTP:18b} are also analyzed by \cite{SmiWydStoWit:19}, concluding that \ac{RF} exposure may be capable of causing an increase in \ac{DNA} damage.

    \textbf{Ramazzini Institute Study.} This research work evaluated the impact of \ac{RF} exposure on Sprague-Dawley rats \cite{FalBua:18}. More specifically, the rats were exposed from prenatal life until death to a \ac{EMF} generated by a \ac{RF} for several hours per day. Like the NTP studies, the rats were divided into multiple groups, each of them exposed to different \ac{EMF} levels. The study found a statistically significant increase in the occurrence of a single disease (i.e., the heart Schwannomas), which was only observed in male rats exposed to the highest \ac{EMF} level. No statistically significant increase w.r.t the exposure was found for the other diseases. Moreover, female rats did not report a statistically significant increase for any of the diseases. According to the authors, their findings corroborate the \ac{NTP} studies \cite{NTP:18a,NTP:18b} and previous epidemiological research on cellular phones, e.g.,   \cite{SzmSzuJol:82,LarSchBle:11,interphone2011acoustic,ChoGuyKru:92},
    %\textbf{LC: references to be added}
    %
    thus making necessary a revision of the \ac{IARC} classification of \ac{RF} exposure \cite{IARCMonograph:13}.

    \subsubsection{Population-based Studies}
\label{subsubsec:population_based}

The studies belonging to this category aim at investigating the relationship between people affected by severe diseases (e.g., brain tumors) and the level exposure from base stations and/or \ac{UE}. We do not intentionally focus on population-based studies tailored to base stations exposure, due to the following reasons:
\begin{enumerate}
\item base stations represent a minor source of exposure compared to \ac{UE} (as proven by previous works  e.g., \cite{joseph2010comparison,durrenberger2014emf});
\item the exposure from base stations tend to be notably reduced as the distance between the base stations, and the user is increased (see, e.g., \cite{ChiaraviglioGala:19,chiaraviglio2020safe}) and more in general when indoor conditions are experienced (see, e.g., \cite{ChiBonFioWia:19});
\item previous population-based studies (see, e.g., the note \cite{cancerbs} of the American Cancer Society and the comprehensive work of  \cite{elliott2010mobile}) did not found any causal relationship between the exposure from base stations and the increase in the risk of developing tumors.
\end{enumerate}

Focusing then on population-based studies on \ac{UE} exposure, it is well known that this \ac{RF} source represents a major source of exposure in proximity to users (see e.g. \cite{joseph2010comparison,durrenberger2014emf}). Therefore, we consider here population-based studies that aim at finding a causal correlation between emergence of tumors and \ac{UE} exposure. The main works performed in the past, which are relevant also in the context of 5G, are: \textit{i}) the \textsc{INTERPHONE} study \cite{Interphone:10,interphone2011acoustic}, \textit{ii}) the Danish cohort study \cite{johansen2001cellular,frei2011use}, \textit{iii}) the million Women study \cite{benson2013mobile} and \textit{iv}) the \textsc{CEFALO} case-control study \cite{aydin2011mobile}. In the following, we provide more details about each study.

\textbf{INTERPHONE Study.} The \textsc{INTERPHONE} Study \cite{Interphone:10,interphone2011acoustic} was {\color{black}coordinated}  by \ac{IARC}. The research, based on a very-large case-control approach, was performed across 13 countries in the world during the years 2000-2012. The project goal was to study the impact of \ac{UE} usage in people that developed severe diseases (i.e., glioma, meningioma, and acoustic neuroma), which may be connected to the usage of \ac{UE}.  The number of people involved in the study was quite important, i.e., more than 5000 patients with glioma or meningioma and 1000 patients with acoustic neuroma. Also, a similar group of people, not affected by any of the tumors mentioned above, was also monitored. The adopted methodology involved several aspects (e.g., personal interviews and validation studies) in obtaining, as much as possible, reliable data about \ac{UE} usage (e.g., duration and frequency of the calls), as well as other relevant information, e.g., \ac{UE} model, network operator, localization of the calls, user mobility and adoption of headsets or hands-free devices.

The results of the study \cite{Interphone:10,interphone2011acoustic} did not prove any connection between the usage of \ac{UE} and the risk of developing glioma, meningioma, or acoustic neuroma. Eventually, an increased risk of glioma for the largest \ac{RF} exposure level was observed. However, the presence of biases and errors in the data prevented a causal interpretation of such results. The reduction of these biases is targeted by \cite{MomSieMcbKre:17}, taking into account the \textsc{INTERPHONE} data collected in Canada during the years 2001-2004. By applying a probabilistic multiple-bias model to address the (possible) biases at the same time, the authors demonstrated that there was little evidence of an increase of tumors with the rise in \ac{UE} usage. Eventually, the importance of investigating possible long-term effects due to the heavy usage of \ac{UE} was advocated by the team involved in the \textsc{INTERPHONE} project. 

\textbf{Danish Cohort Study.} The goal of the Danish cohort study \cite{johansen2001cellular,frei2011use} was to investigate the risks of developing tumors for Danish people having a subscription with a cellular operator against the remaining of the Danish population not having any subscription. The study was updated continuously throughout the years, being the first version spanning the years 1982-1995 \cite{johansen2001cellular} and the latest one covering the 1990-2007 period \cite{frei2011use}. The number of persons taken under consideration is huge, being the number of subscribers in \cite{frei2011use} larger than 380000. The study did not show any link between the use of \ac{UE} - even for more than 13 years - and the risk of developing tumors of the central nervous system. However, the principle adopted to distinguish between exposed people and not exposed people is solely based on their subscription with a mobile operator, without going into more in-depth details like the ones taken into account by the \textsc{INTERPHONE} project.

\textbf{Million Women Study.}  A wide-scale approach is also pursued by the Million Women study \cite{benson2013mobile}. The methodology involved a postal questionnaire, which was completed by 1.3 million middle-aged women in the UK for different times during the years 1999-2009. The survey included specific questions to assess \ac{UE} exposure, which was posed two times during the considered period. The results of the study \cite{benson2013mobile} showed that \ac{UE} use was not associated with an increased incidence of glioma, meningioma, or tumors of the central nervous system. However, it is important to remark that the study is based on self-compiled questionnaires, and therefore bias and errors may have been (unintentionally) introduced by the participants.

\textbf{CEFALO Case-Control Study.}  The \textsc{CEFALO} case-control study \cite{aydin2011mobile} investigated the impact of \ac{UE} exposure on young children and adolescents (with age 7-19) that developed brain tumors between 2004 and 2008 in Denmark, Sweden, Norway, and Switzerland countries. More than 350 patients were interviewed about \ac{UE} usage (i.e., number of calls and call duration) and other relevant information, including, e.g., type of operator, number of subscriptions, starting and ending date of each subscription, adoption of hands-free devices, position of the \ac{UE} during the usage, and (eventual) changes in the \ac{UE} usage. Whenever possible, the retrieved information was also double-checked by analyzing the logs that were made available by mobile operators in a subset of countries. The outcomes were then compared against a group of other adolescents/children, not affected by brain tumors, thus acting as control subjects. Results confirmed that children/adolescents regularly using \ac{UE} were not statistically significantly more likely to have been diagnosed with brain tumors compared to subjects not using the \ac{UE}.
Also, no increased risk in developing brain tumors was observed for children/adolescents receiving the highest exposure. Eventually, the subscription duration was statistically significant w.r.t. the risk of developing a brain tumor for a small subset of the participants, whose activity information was retrieved from the logs of the mobile operators. However, as recognized by the authors of the \textsc{CEFALO} study \cite{aydin2011mobile}, this outcome might be affected by multiple factors, including \textit{i}) a small cardinality of children/adolescent considered in the subset (only 35\% of case-patients and only 34\% of control subjects), \textit{ii}) the fact that the \ac{UE} might have been used by other people in the family and/or friends (i.e., not by the considered subject), \text{iii}) the possible presence of a reverse causality effect (i.e., children/adolescents affected by brain tumors use more frequently their \ac{UE} compared to the ones not affected by the disease). Finally, the authors concluded that their work could not support a causal association between the use of \ac{UE} and brain tumors.

    \subsection{Review of the Studies from the Perspective of 5G Communications} \label{sec:enginpresp}

We now review both the animal-based studies and the population-based ones from the perspective of 5G communications.

\subsubsection{Animal-based Studies}
\label{subsubsec:aninmal_based_review}

We compare the NTP and Ramazzini Institute studies \cite{NTP:18a,NTP:18b,FalBua:18} against 5G equipment under the following key metrics: \textit{i}) operating frequencies, \textit{ii}) test chambers vs. real deployment, \textit{iii}) maximum radiated power, \textit{iv}) power management, \textit{v}) \ac{EMF} exposure levels, \textit{vi}) \ac{SAR} levels, \textit{vii}) transmission and modulation techniques.

\textbf{Operating Frequencies.}
We recall that 5G will operate in three main frequency bands:
\begin{enumerate}
\item sub-GHz band (i.e., $< 1$~[GHz]);
\item mid-band (i.e., between 1~[GHz] and 6~[GHz]);
\item mm-wave (i.e., with frequencies in the order of dozens of GHz and more).
\end{enumerate}
In this scenario, the NTP studies \cite{NTP:18a,NTP:18b} adopt frequencies belonging to the sub-GHz band and to the mid-band. More in-depth, the 900~[MHz] frequency used by \cite{NTP:18a} is very close to the one in use by 5G in the sub-GHz band. In Italy, for example, this frequency is set to 700~[MHz]. On the other hand, \cite{NTP:18b} exploits the 1900~[MHz] frequency, which is {used for 5G services in some countries of the world (e.g., USA), while other ones (like Italy) adopt different frequencies}. Focusing then on the Ramazzini Institute study \cite{FalBua:18}, the adopted frequency is equal to 1800~[MHz], which is again comparable to the 5G frequencies in the mid-band. 

Eventually, it is important to remark that none of the studies \cite{NTP:18a,NTP:18b,FalBua:18} investigate the impact of frequencies in the mm-Wave band, whose waves have very different properties (e.g., less penetration in inner tissues) compared to micro-waves. A natural question is then: Why do the studies in \cite{NTP:18a,NTP:18b,FalBua:18} not investigate mm-Wave? To answer this question, we need to remind that \cite{NTP:18a,NTP:18b,FalBua:18} assume to adopt 2G technologies (not 5G), for which the use of frequencies in the mm-Wave band was not possible. As a result, we can claim that the studies \cite{NTP:18a,NTP:18b,FalBua:18} are only partially representative of 5G frequencies.

    \begin{figure}[t]
        \centering
        \includegraphics[width=8cm]{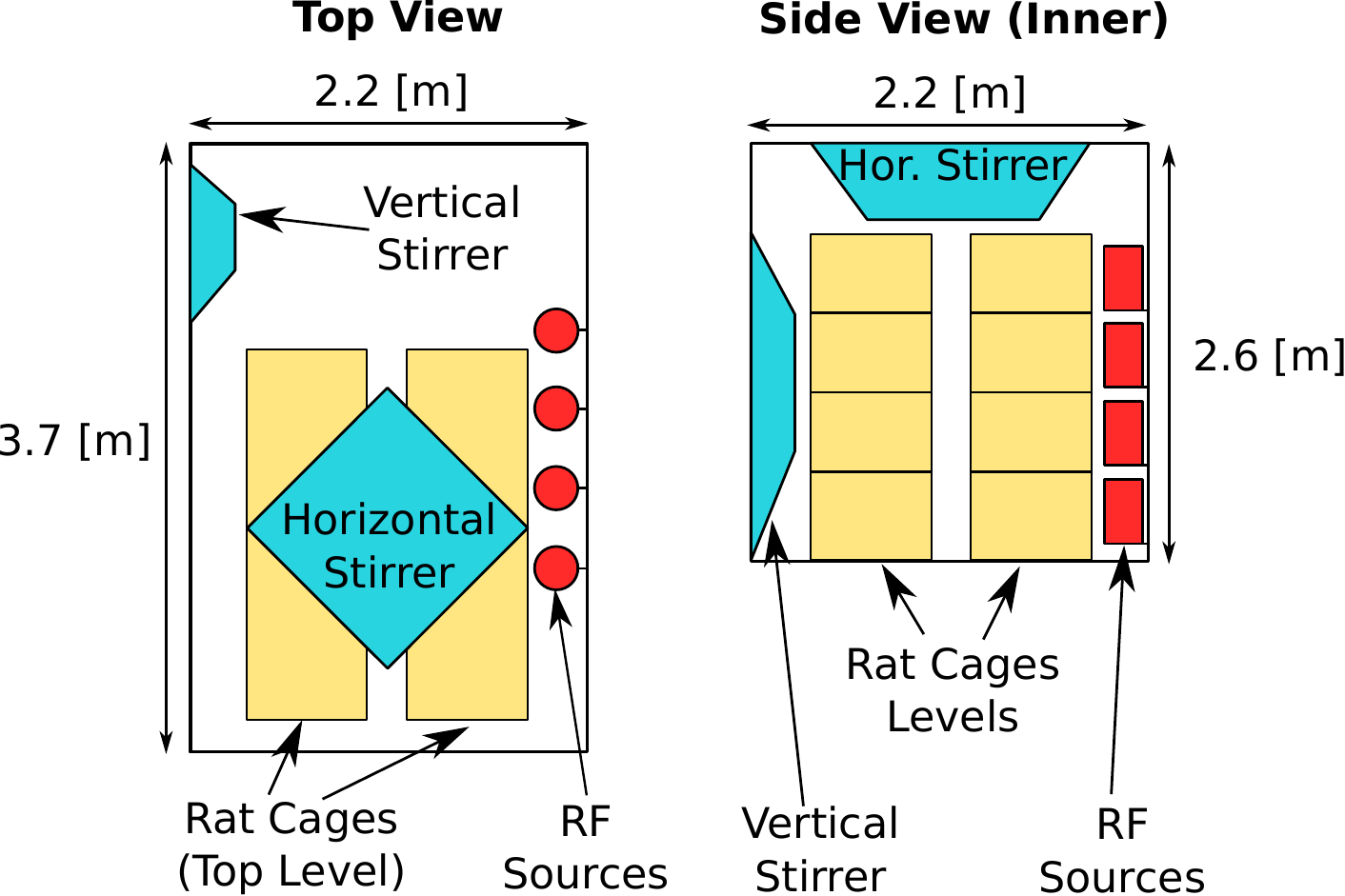}
        \caption{Test conditions adopted in the NTP experiments. Multiple \ac{RF} sources provide active exposure, while two stirrers (one horizontal and one vertical) are used as passive elements to generate an uniform exposure in the environment.}
        \label{fig:test_conditions_ntp}
    \end{figure}

\textbf{Test Chambers vs. Real Deployment.} {We then compare} the chambers used to perform the test against the real environment in which 5G equipment operates.
We initially focus on the test chambers of the NTP studies \cite{NTP:18a,NTP:18b}, which are also sketched in Fig.~\ref{fig:test_conditions_ntp}. We refer the reader to \cite{capstick2017radio} for a detailed description, while here, we report the salient features. In brief, the NTP studies employed chambers whose dimensions are comparable to a small room. In each chamber, the rat cages are positioned in the center, with different levels of cages that are vertically stacked. Inside the chamber, many standard gain antennas are placed. The exact number of deployed antennas is not provided (neither in \cite{NTP:18a,NTP:18b} or in \cite{capstick2017radio}). Besides, two elements, called stirrers, are placed on top and on the side of the chamber. Each stirrer is used as a target when setting the antenna tilting (with a subset of antennas directed towards the top stirrer, and the other ones towards the side stirrer). The stirrers are then used as passive elements to reflect the radiation and generate a uniform \ac{EMF} across the chamber. In this scenario, both the antennas and the stirrers are placed in close proximity to the exposed rats.

    \begin{figure}[t]
        \centering
        \includegraphics[width=8cm]{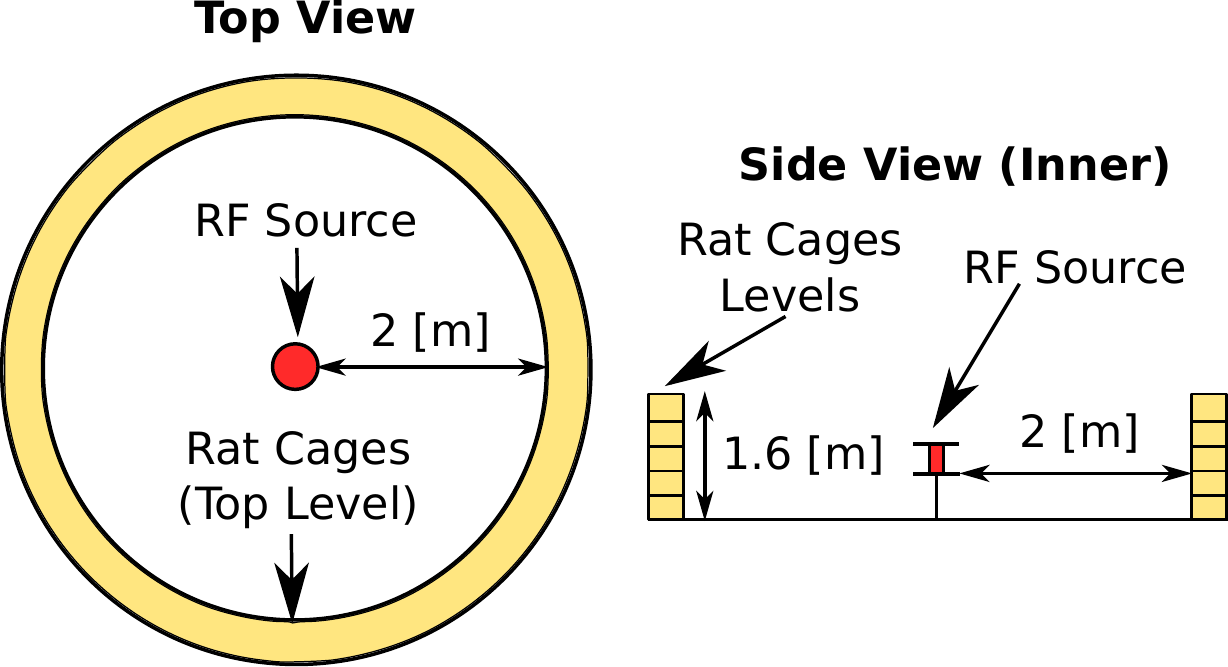}
        \caption{Test conditions adopted in the Ramazzini experiments. The \ac{RF} source is placed in close proximity to the exposed rats.}
        \label{fig:test_conditions_ramazzini}
    \end{figure}

Focusing on the test conditions adopted in the study of the Ramazzini Institute \cite{FalBua:18}, the rat cages are disposed of in a torus structure around the \ac{RF} source, as sketched in Fig.~\ref{fig:test_conditions_ramazzini}. Moreover, a minimum distance of 2~[m] is ensured between the \ac{RF} source and the rat cages, to achieve far-field conditions. Eventually, the whole structure is placed in a chamber (not shown in the figure for the sake of simplicity) that is completely shielded, in order to create a uniform \ac{EMF} in the room.

\begin{figure}[t]
        \centering
        \includegraphics[width=8cm]{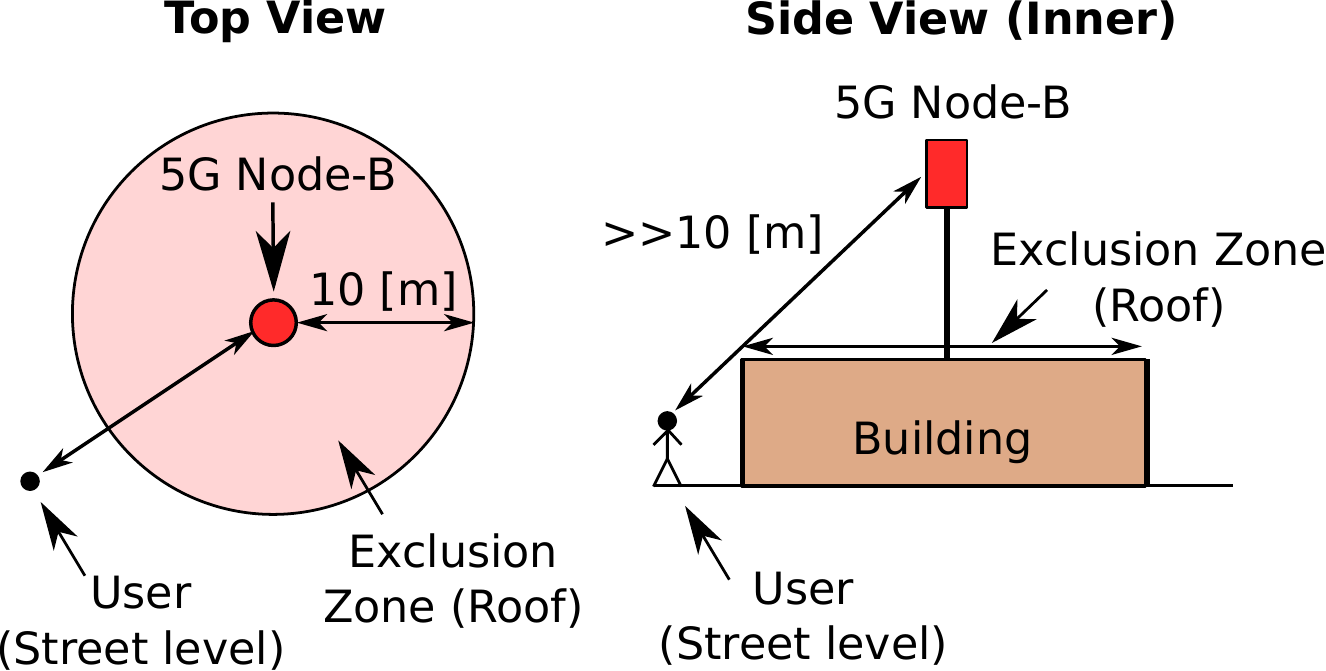}
        \caption{Realistic conditions adopted in 5G \acp{gNB} deployments (macro sites). The exposed users in Line of Sight are always at dozens of meters from the radiating 5G \ac{gNB}.}
        \label{fig:realistic_conditions}
\end{figure}

We then compare the test conditions of the studies \cite{NTP:18a,NTP:18b,FalBua:18} against a realistic 5G deployment of a macro \ac{gNB}, sketched in Fig.~\ref{fig:realistic_conditions}. More in detail, we consider a 3.5~[GHz] omni-directional \ac{gNB}, mounted on a pole, and then placed on a roof of a building. A similar deployment, exploiting a three sectorial 5G \ac{gNB}, is analyzed in \cite{5gimpact}. In this scenario, the roof of the building delimits an exclusion zone from the center of the 5G \ac{gNB}. Such a zone is intended to be accessible only by the technicians that need to perform maintenance operations on the 5G \ac{gNB}. Clearly, this zone is forbidden to the general public, which is therefore physically prevented from entering. According to \cite{5gimpact}, a minimum distance to delimit the exclusion zone in a 5G deployment is in the order of 10~[m] from 5G \acp{gNB}.
Consequently, we have imposed in Fig.~\ref{fig:realistic_conditions} an exclusion zone of 10~[m], which delimits the roof of the building. As a result, users in \ac{LOS} from the 5G \ac{gNB} tend to be pretty far from the source of radiation. By comparing the distance between the exposed users/rats and the radiating source, we can note that both the NTP study \cite{NTP:18a,NTP:18b} and the Ramazzini Institute one \cite{FalBua:18} assume a distance from the \ac{RF} source much closer than the minimum distance from a radiated user in a realistic 5G deployment. This is a second and essential outcome that obviously differentiates the laboratory studies w.r.t. the real deployment of 5G \acp{gNB}.

In the following step, we compare the test chambers of \cite{NTP:18a,NTP:18b,FalBua:18} against the real conditions at which a 5G \ac{UE} operates. First of all, it is important to remark that a 5G \ac{UE} is also used outdoor, and not only in a chamber like in \cite{NTP:18a,NTP:18b,FalBua:18}. In addition, mobility is another important aspect that strongly impacts the exposure conditions of 5G terminals, which on the other hand, is not considered by the static deployment of \cite{NTP:18a,NTP:18b,FalBua:18}. Moreover, the distance between the \ac{UE} and the exposed zone of the body is clearly lower than the one imposed in the NTP and Ramazzini Institute studies. Eventually, the exposure from a \ac{UE} is not uniform across the environment like in laboratory studies, but it tends to be localized to the closest tissues/organs. Therefore, the test conditions adopted in \cite{NTP:18a,NTP:18b,FalBua:18} are clearly far from the actual operating conditions of a 5G \ac{UE}.

\begin{table}[t]
\centering
\caption{Maximum output power $P_{\text{MAX}}$ for the different devices.}
\label{tab:max_power_comp}
\footnotesize
\begin{tabular}{|c|c|c|}
\hline
\rowcolor{Lightred} \textbf{5G Device} & \textbf{Value} & \textbf{Reference} \\
\hline
\ac{RF} source - NTP & 3800~[W] (65~[dBm) & \cite{capstick2017radio} \\
\ac{RF} source - Ramazzini Institute & 100~[W] (50~[dBm]) & \cite{FalBua:18} \\
 5G macro \ac{gNB} & 200~[W] (53~[dBm]) & \cite{air5121} \\
5G \ac{UE} & 0.2~[W] (23~[dBm]) & \cite{etsitr} \\
\hline
\end{tabular}
\end{table}

\textbf{Maximum Radiated Power.} As a third aspect, we consider the maximum radiated power $P_{\text{MAX}}$ of the \ac{RF} sources employed in \cite{NTP:18a,NTP:18b,FalBua:18}, and their comparison against real 5G communications equipment (i.e., a 5G macro \ac{gNB} and a 5G \ac{UE}). To this aim, Tab.~\ref{tab:max_power_comp} reports the comparison across the different types of devices adopted in the studies and the ones deployed in 5G networks. Two considerations hold in this case. First, the value of $P_{\text{MAX}}$ adopted in the NTP study is one order of magnitude higher than the one used in a 5G macro \ac{gNB}, and four orders of magnitude higher than the one of a 5G \ac{UE}. Although the use of enormous radiated power values is also recognized by the authors of \cite{NTP:18a,NTP:18b}, it is important to remark that such values are outside the operating range of realistic 5G \acp{gNB}. Second, the value of $P_{\text{MAX}}$ used in the Ramazzini Institute study \cite{FalBua:18} is comparable with the one of a 5G macro \ac{gNB}. However, the maximum radiated power of the \ac{RF} source in \cite{FalBua:18} is still three orders of magnitude higher than a 5G \ac{UE}. As a result, we can conclude that none of the studies \cite{NTP:18a,NTP:18b,FalBua:18} adopt realistic $P_{\text{MAX}}$ values for 5G \ac{UE}, and only \cite{FalBua:18} imposes a value of $P_{\text{MAX}}$ comparable to the one radiated by a 5G macro \ac{gNB}.

\begin{figure}[t]
\centering
\centering
    \subfigure[High Traffic Condition]
    {
    \includegraphics[width=6cm]{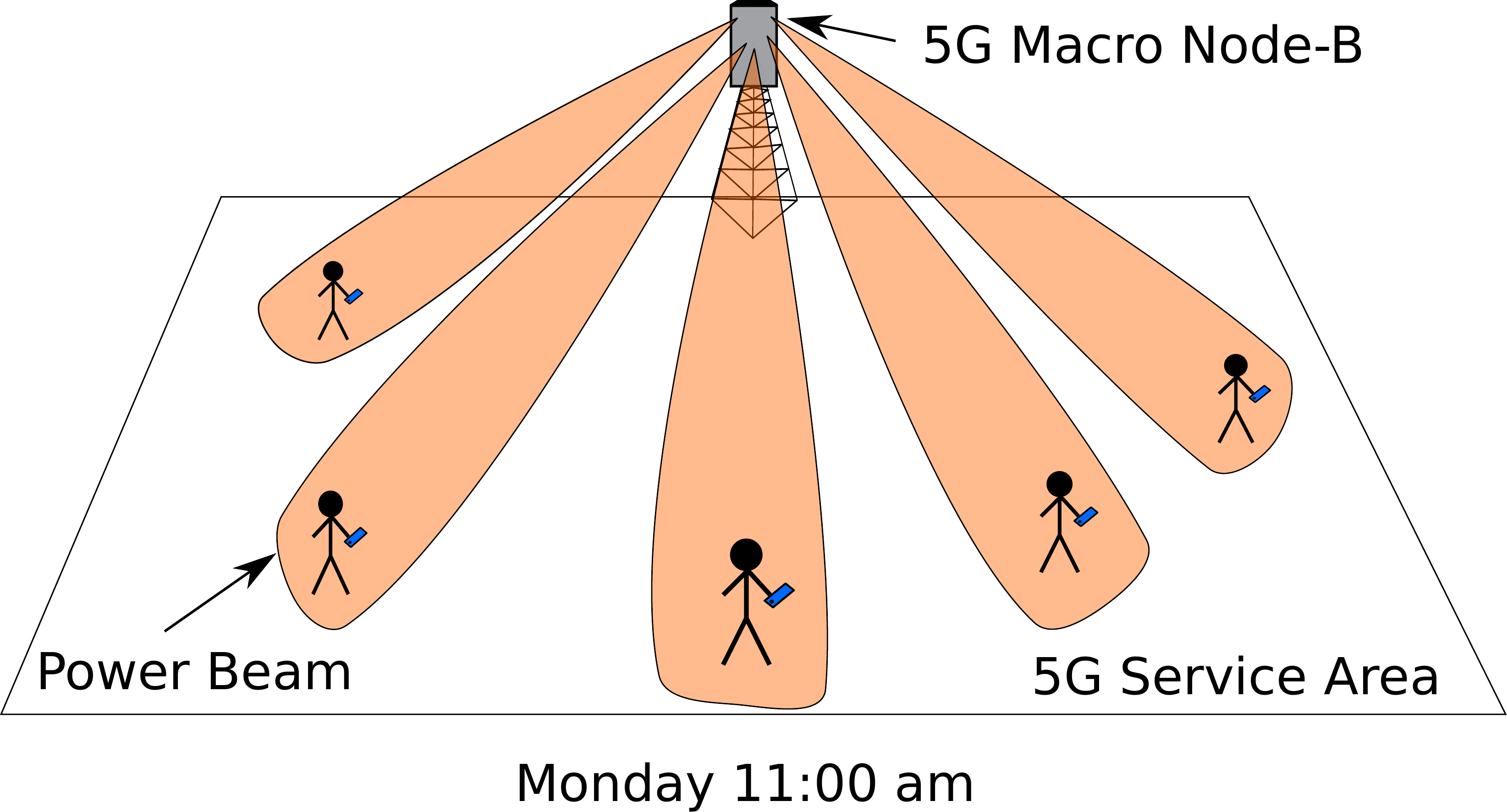}
    \label{fig:power_high_traffic}
    }

    \subfigure[High Traffic Condition - Another Day]
    {
    \includegraphics[width=6cm]{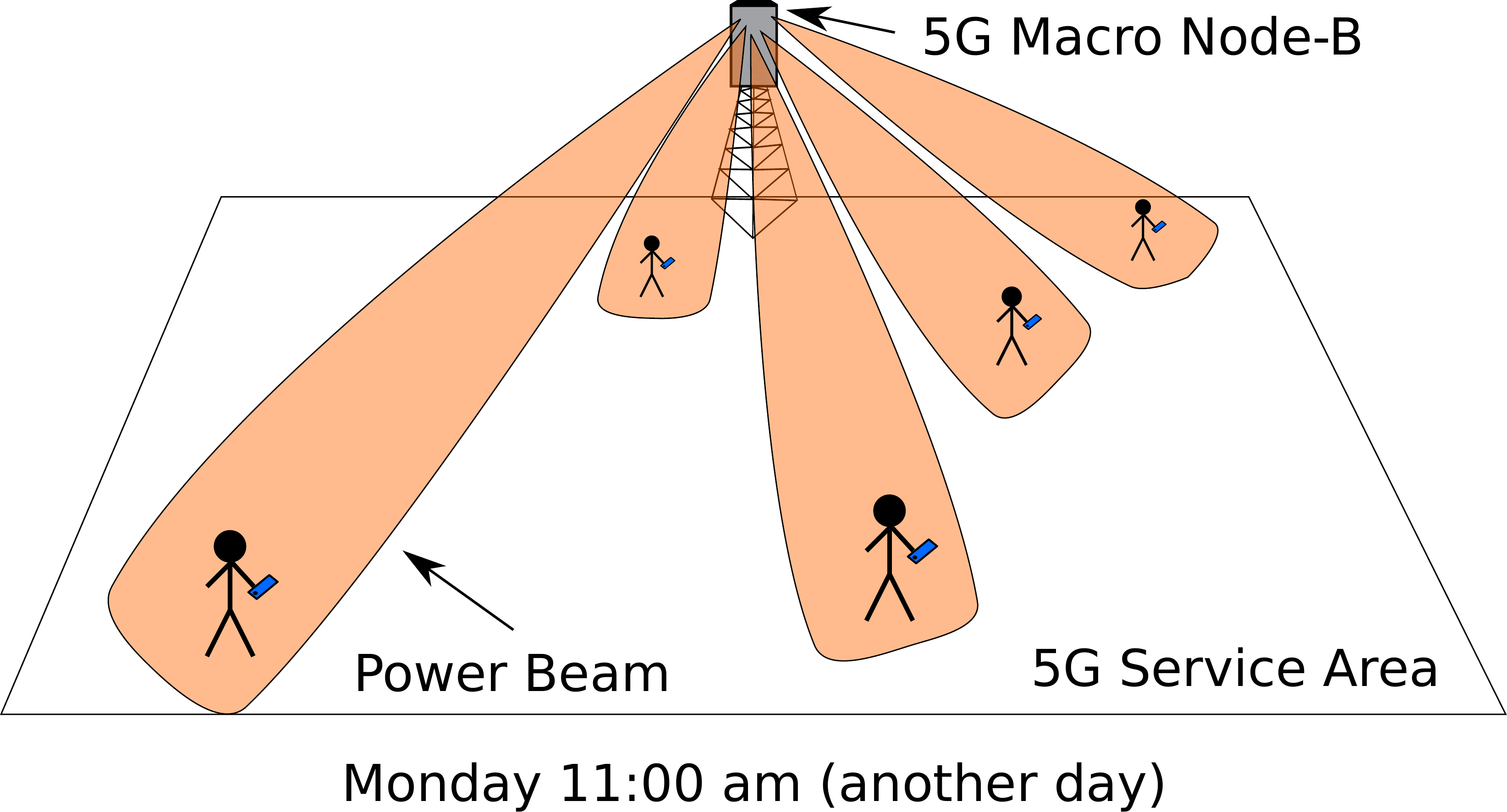}
    \label{fig:power_high_traffic_another_day}
    }
    
    \subfigure[Low Traffic Condition]
    {
    \includegraphics[width=6cm]{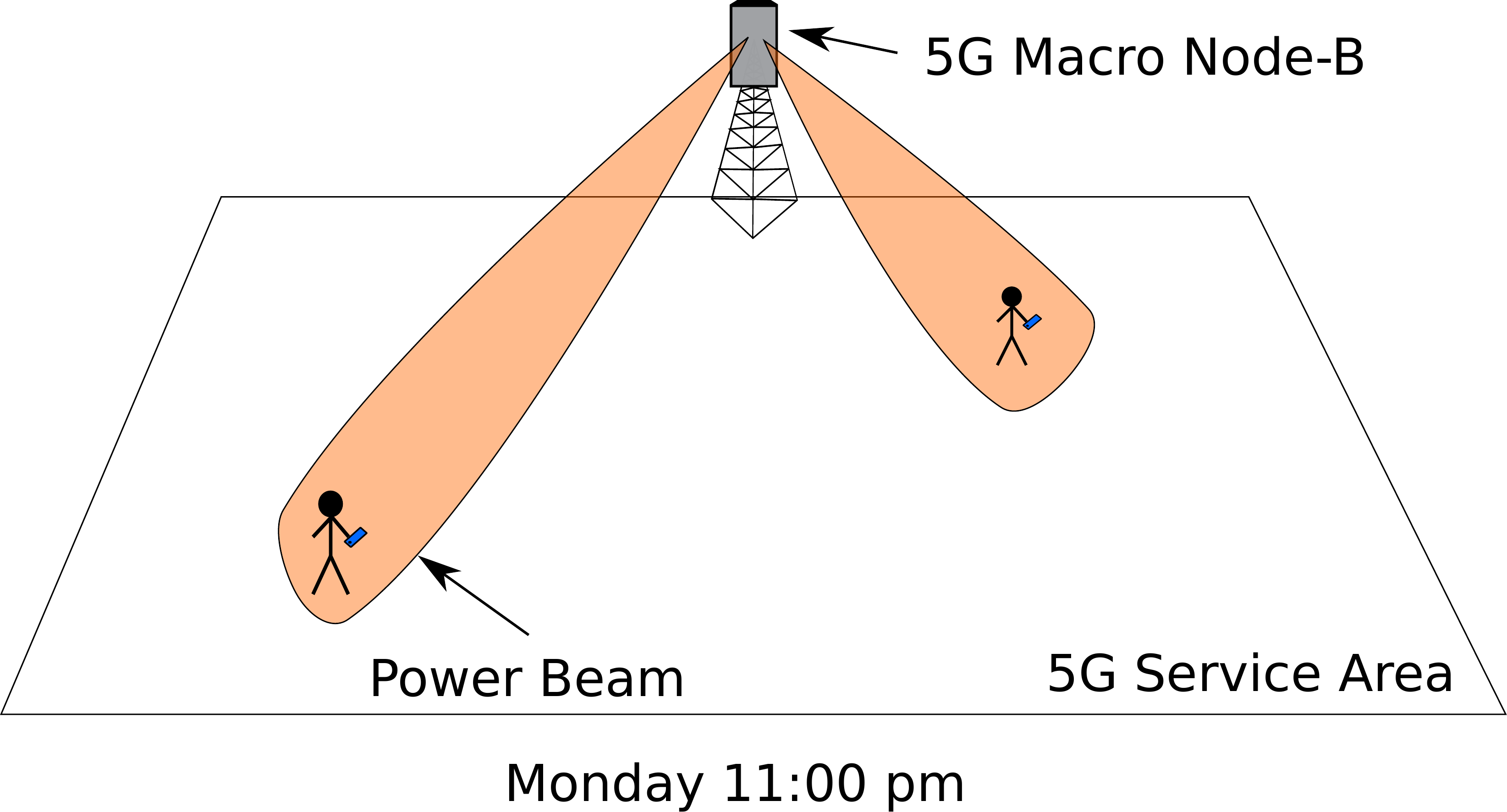}
    \label{fig:power_low_traffic}
    }
\caption{Dynamic management of the radiated power for a 5G macro \ac{gNB}. The power radiated over the territory varies in space and in time. For example, the same number of served users results into spatially different radiation patterns, as shown in (a),(b). The variation in the number of served users (e.g., between day and night) also impacts the radiation pattern, as shown in (a),(c).}
\label{fig:power}
\end{figure}

\textbf{Power Management.} In this part, we shed light about the power management adopted by the studies \cite{NTP:18a,NTP:18b,FalBua:18} w.r.t. realistic 5G \acp{gNB}. In general, the power management of a \ac{RF} source can be characterized according to two important aspects: \textit{i}) how the power is spatially radiated over the service area, and \textit{ii}) how the power is varied across time. We denote \textit{i}) as spatial power management, while \textit{ii}) is referred to as temporal power management. Focusing on the spatial power management, the goal of \cite{NTP:18a,NTP:18b,FalBua:18} is to keep a uniform exposure for all the rats inside the room. This is achieved by adopting radiation patterns of the \ac{RF} sources that tend to generate a uniform \ac{EMF} in the chamber. In addition, the adoption of a torus structure in \cite{FalBua:18} and of the stirrers in \cite{NTP:18a,NTP:18b} allows to achieve this design goal. When comparing these features against the spatial power management performed by a 5G macro \ac{gNB}, several notable differences emerge. As sketched in Fig.~\ref{fig:power_high_traffic} a 5G macro  \ac{gNB} does not uniformly radiate the power over the service area. On the contrary, the radiated power tends to be focused into beams, which are directed to the 5G users. Therefore, the different zones of the service area do not receive the same amount of radiated power. Also, another important feature implemented in 5G \acp{gNB} is the ability to dynamically vary the power beams in accordance to the locations of the served users \cite{NadKamAlouini:19}. To this aim, Fig.~\ref{fig:power_high_traffic_another_day} reports a scenario where the number of served users is the same as in Fig.~\ref{fig:power_high_traffic}, but with different positioning of the power beams, which then results in a different radiation pattern over the service area compared to Fig.~\ref{fig:power_high_traffic}. Consequently, we can claim that the approaches implemented in in \cite{NTP:18a,NTP:18b,FalBua:18} for the spatial power management are completely different w.r.t. the one pursued by a real 5G macro \ac{gNB}.

Focusing then on the spatial power variations for a \ac{UE}, the actual pattern radiated by the \ac{RF} sources installed on the terminal depends on their physical positioning in the terminal, as well as the placement of other nearby elements such as screen, battery, photo-camera, and \ac{RF} elements of other technologies (e.g., WiFi, Bluetooth, 2G/3G/4G). We refer the interested reader to \cite{huo20175g} for a detailed overview of these aspects. In addition, the actual exposure depends on how the device is held (e.g., horizontally or vertically, with one hand or with two hands) \cite{huo20175g}, and thus can not be precisely known a priori. In any case, however, the design choices tend to avoid a radiation pattern directed towards the user \cite{huo20175g}, which is instead assumed in \cite{NTP:18a,NTP:18b,FalBua:18}.

\begin{table}[t]
\centering
\caption{24h average ratio of radiated power $\delta^{\text{ON}}$ for the different devices.}
\label{tab:temporal_power_comp}
\footnotesize
\begin{tabular}{|c|c|c|}
\hline
\rowcolor{Lightred} \textbf{5G Device} & \textbf{Value} & \textbf{Reference} \\
\hline
\ac{RF} source - NTP & 0.38 & \cite{NTP:18a,NTP:18b} \\
\ac{RF} source - Ramazzini Institute & 0.79 & \cite{FalBua:18} \\
5G macro \ac{gNB} & 0.17 & \cite{ecoscienza} \\
5G \ac{UE} & 0.17 & 4~[h] of usage per day \cite{numberofhours}\\
\hline
\end{tabular}
\end{table}

In the following, we concentrate on the temporal power management aspect, by first considering the comparison of \cite{NTP:18a,NTP:18b,FalBua:18} against a 5G macro \ac{gNB}. As reported by \cite{NTP:18a,NTP:18b}, the \ac{RF} source is activated for 18 hours and 10 minutes per day, by imposing a repetition of an on period always followed by an off period, each of them lasting for 10 minutes. Since the goal of the studies \cite{NTP:18a,NTP:18b} is to keep a uniform exposure, the values of radiated power during the on period are almost constant.\footnote{As reported by \cite{NTP:18a,NTP:18b}, minor oscillations are possible in order to guarantee a uniform and stable \ac{SAR}.} Let us denote with $\delta^{\text{ON}}$ the ratio of time over 24~[h] during which the radiated power is on. Consequently, we can claim that the average ratio of radiated power computed over the 24h is equal to 38\% of the power radiated during the on periods, i.e., $\delta^{\text{ON}}=0.38$. Focusing then on the Ramazzini Institute study \cite{FalBua:18}, the \ac{RF} source is continuously activated for 19~[h] over the 24 hours. Therefore, the 24h average ratio of radiated power is 79\% of the power radiated during the on period, i.e., $\delta^{\text{ON}}=0.79$. A natural question is then: Are these values meaningful when compared to the temporal power variation of a real 5G \acp{gNB}? To answer this question, we consider the realistic values of 24h average radiated power available for 4G networks, which we assume to be meaningful also for 5G equipment. As reported by \cite{ecoscienza}, the 24h average ratio of radiated power from a 4G Node-B is at maximum equal to 17\% when considering the whole set of Node-Bs deployed in the city of Milan (Italy). Although this percentage may appear pretty low at first glance, we remind that different previous works (see, e.g., \cite{furno2016tale,grauwin2015towards}) have demonstrated that 4G networks are subject to strong temporal and spatial traffic variations. For instance, the traffic varies across the hours of the same day (daytime vs. nighttime),  the day of the week (e.g., weekday vs. weekend), and the location of the 4G Node-Bs (residential vs. business districts). Since the amount of traffic managed by a 4G Node-B heavily impacts the radiated power, it is natural that the 24h average radiated power (expressed as a fraction of maximum power) is clearly lower than unity. In line with this trend, 5G is expected to adapt the radiated power w.r.t the time-varying traffic conditions wisely. For example, the number of power beams can match the number of users that need to be served, as graphically shown in Fig.~\ref{fig:power_high_traffic}-\ref{fig:power_low_traffic}. Therefore, when the number of users is low (Fig.~\ref{fig:power_low_traffic}), the 5G macro \ac{gNB} can reduce radiated power. As a result, we can claim that the studies \cite{NTP:18a,NTP:18b,FalBua:18} adopt a temporal power management very conservative w.r.t. the one implemented by a 5G macro \ac{gNB}.

\renewcommand{\arraystretch}{1.2}
\begin{table}[t]
\caption{24h average \ac{EMF} $E^{\text{24h}}_{(s)}$ measured in the NTP study \cite{NTP:18a,NTP:18b} - GSM tests. The minimum \ac{EMF} value is highlighted in boldface.}
\label{tab:24_h_emf}
\centering
\footnotesize
\begin{tabular}{|c|c|c|}
\hline
\rowcolor{Lightred} \textbf{Target \ac{SAR} $s$} & \textbf{Frequency $f$} & \textbf{24h \ac{EMF} $E^{\text{24h}}_{(s)}$} \\
\hline
1.5 [W/kg] & 900~[MHz] & 56~[V/m]\\
\hline
\rowcolor{Linen} 3 [W/kg] & 900~[MHz] & 78~[V/m]\\
\hline
6 [W/kg] & 900~[MHz] &  111~[V/m]\\
\hline
\rowcolor{Linen} 1.5 [W/kg] & 1900~[MHz] & \textbf{48~[V/m]}\\
\hline
3 [W/kg] & 1900~[MHz] &  68~[V/m]\\ 
\hline
 \rowcolor{Linen} 6 [W/kg] & 1900~[MHz] &  98~[V/m]\\
\hline
\end{tabular}
\end{table}
\renewcommand{\arraystretch}{1.0}

\begin{table*}[t]
\centering
\caption{Computation of the \ac{EMF} level at a given distance for a 5G \ac{gNB}.}
\label{tab:parameter_EMF_computation}
\footnotesize
\begin{tabular}{|c|c|c|c|}
\hline
\rowcolor{Lightred} \textbf{Parameter} & \textbf{Notation} & \textbf{Value(s)/Formula} & \textbf{Comment and Reference} \\
\hline
Operating Frequency & $f$ & 3.7~[GHz] & In use in Italy for the mid-band\\
\hline
\rowcolor{Linen} Maximum Transmission Power & $P_{\text{MAX}}$ & 200~[W] & Value from real 5G equipment \cite{air5121} \\
\hline
\multirow{2}{*}{Statistical Reduction Factor} & \multirow{2}{*}{$\alpha_{\text{STAT}}$} & \multirow{2}{*}{\{0.25,1\}} & Value of 0.25 reported by \cite{ecoscienza}. \\
& & & Worst case value equal to 1.\\
\hline
\rowcolor{Linen}   &  &  & Value of 0.17 for the city of Milan reported by \cite{ecoscienza}.\\
\rowcolor{Linen} \multirow{-2}{*}{Time-average Reduction Factor} & \multirow{-2}{*}{$\alpha_{\text{24}}$} & \multirow{-2}{*}{\{0.17,1\}} & Worst case value equal to 1.\\
\hline
Average Transmission Power & $P_{\text{AVG}}$ & $P_{\text{MAX}} \cdot \alpha_{\text{STAT}} \cdot \alpha_{\text{24}}$ & Computation done in \cite{ecoscienza} \\
\hline
\rowcolor{Linen}  Transmission Gain & $G_{\text{TX}}$ & 15~[dB] & Gain of a transmitting antenna based on \cite{itutk70} \\
\hline
Transmission Loss & $L_{\text{TX}}$ & 2.32~[dB] & Loss reported in \cite{itutk70} \\
\hline
\rowcolor{Linen}  & & & \\[-0.8em]
\rowcolor{Linen}  Equivalent Isotropically Radiated Power &{\color{black} $\textrm{EIRP}$} & $\frac{P_{\text{AVG}} \text{[W]} \cdot G_{\text{TX}} \text{[linear]}}{L_{\text{TX}} \text{[linear]}}$ & Formula based on \cite{itutk70}\\[0.6em]
\hline
Normalized Antenna Numeric Gain & $G_{\text{N}}$ & 1 & Worst case based on \cite{itutk70}\\
\hline
\rowcolor{Linen}  Free Space Wave {Impedance} & $Z$ & 377~[$\Omega$] & Fixed parameter based on \cite{itutk70}\\
\hline
Distance from 5G \ac{gNB} & $d$ & 2-100 [m] & Varying parameter \\
\hline
\rowcolor{Linen}  Sight Condition & - & Line of Sight (LoS) & Worst case assumption \\
\hline
& & & \\[-0.8em]
\ac{EMF} level at distance $d$ & $E_{(d)}$ &{\color{black} $\sqrt{\frac{\textrm{EIRP} \cdot G_{\text{N}} \cdot Z}{4 \cdot \pi \cdot d^2}}$} & Point source model of \cite{itutk70}\\[0.6em]
\hline
\end{tabular}
\end{table*}

We now focus on the comparison between the temporal power management in \cite{NTP:18a,NTP:18b,FalBua:18} w.r.t. the one implemented in 5G \ac{UE}. First, we point out that the temporal variation of power depends on multiple factors, which include, e.g., the positioning of the \ac{UE} w.r.t. the serving \ac{gNB} as well as the channel conditions. For example, \ac{NLOS} conditions and distance from the serving \ac{gNB} in the order of hundreds of meters may result in a non-negligible amount of radiated power by the \ac{UE} \cite{ChiaraviglioGala:19}. In this context, we refer the interested reader to \cite{itutksupplement13} for a detailed overview of the main communications features affecting the temporal variation of the \ac{RF} output power. In addition, the temporal power management is heavily impacted by the type of applications (e.g., instant messaging vs. continuous downloading/uploading of photos/videos vs. continuous swapping of web pages enriched with multimedia content vs. notification-oriented applications), as well as the usage pattern of the user \cite{kim2019understanding}. According to recent trends (see, e.g., \cite{numberofhours}), the average usage of a \ac{UE} is currently equal to 3~[h] per day, with a projected increase to 4~[h] in 2021. Even by assuming a worst-case scenario, in which the \ac{UE} always transmits at full power during the whole usage time of 4~[h], the 24h temporal power variation is equal to 17\%, i.e., a value clearly lower than the one imposed in the laboratory studies  \cite{NTP:18a,NTP:18b,FalBua:18}.\footnote{We remind that when a \ac{UE} is not in use, the radiated power can be larger than zero due to, e.g., the App notifications and the pushing of multimedia content. However, the exposure zone tends to be different than the one during the active usage (e.g., a pocket vs. the front of the head and the chest).}

\textbf{ElectroMagnetic Field Exposure Levels.} During this step, we shed light on the values of \ac{EMF} imposed in \cite{NTP:18a,NTP:18b,FalBua:18}, and their comparison against the estimated exposure from a 5G macro \ac{gNB}. Focusing on the NTP studies 
\cite{NTP:18a,NTP:18b} the values of exposure are measured by \ac{EMF} meters that are placed in the chamber. More in depth, the average value of \ac{EMF} exposure during the on period in each chamber and in each experiment is reported as raw data in the Appendix of the studies \cite{NTP:18a,NTP:18b}. Let us denote this value as $E^{\text{ON}}_{(c,s)}$, where $c$ is the chamber index and $s$ is the level of target \ac{SAR}. Given $E^{\text{ON}}_{(c,s)}$, we initially compute the average 24 hours \ac{EMF} in each chamber and for each \ac{SAR} level. We denote this metric as $E^{\text{24h}}_{(c,s)}$. Intuitively, $E^{\text{24h}}_{(c,s)}$ allows to obtain a more conservative estimation of the level of exposure compared to $E^{\text{ON}}_{(c,s)}$. More formally, we have:
\begin{equation}
\label{eq:24havg}
E^{\text{24h}}_{(c,s)}=\delta^{\text{ON}} \cdot E^{\text{ON}}_{(c,s)}, \quad \text{[V/m]}
\end{equation}
Clearly, it holds that: $E^{\text{24h}}_{(c,s)}<E^{\text{ON}}_{(c,s)}$. Moreover, we point out that $E^{\text{24h}}_{(c,s)}$ is computed with a linear function of the \ac{EMF}, although time-averaged \acp{EMF} are commonly evaluated by applying a root mean square. However, we remark that a linear average results into lower values of \ac{EMF} compared to a root mean square, and thus allows to consider a very conservative scenario.
In the following, we compute the average 24 hours \ac{EMF} across all the chambers, denoted by $E^{\text{24h}}_{(s)}$, again computed for each \ac{SAR} level $s$.
$E^{\text{24h}}_{(s)}$ is formally expressed as:
\begin{equation}
\label{eq:24havgchamber}
E^{\text{24h}}_{(s)}=\frac{1}{|C|}\sum_c E^{\text{24h}}_{(c,s)}, \quad \text{[V/m]}
\end{equation}
where $|C|$ is the total number of chambers used in the study. Similarly to Eq.~(\ref{eq:24havg}), also $E^{\text{24h}}_{(s)}$ is conservatively computed as a linear average.
The final values of $E^{\text{24h}}_{(s)}$ are reported in Tab.~\ref{tab:24_h_emf}.\footnote{The table reports the value for the \ac{GSM} experiments. Similar values were obtained for the \ac{CDMA} experiments, not reported here for the sake of simplicity.} Interestingly, the 24 hours average \ac{EMF} ranges between 48~[V/m] and 111~[V/m], depending on the experiment.

In the following, we focus on the \ac{EMF} values of the Ramazzini Institute study \cite{FalBua:18}. Compared to \cite{NTP:18a,NTP:18b}, the goal of \cite{FalBua:18} is not to target a certain level of \ac{SAR}, but rather a given value of \ac{EMF}, which is selectively set to 5~[V/m], 25~[V/m] and 50~[V/m]. By considering the daily activation period of the experiments, which we recall is equal to 19~[h], we get the following values of 24 hours average \ac{EMF}: 4~[V/m], 20~[V/m] and 40~[V/m].\footnote{Similarly to Eq.~(\ref{eq:24havg}), a linear average is computed also here, in order to obtain a set of conservative values.}

We then focus our attention on the 24 hours average \ac{EMF} radiated by a 5G macro \ac{gNB}. Let us denote with $E_{(d)}$ the \ac{EMF} from a 5G macro \ac{gNB} placed a distance $d$ from the current position. Clearly, the value of $E_{(d)}$ is influenced by multiple factors (apart from $d$), including: the maximum transmission power of the 5G macro \ac{gNB}, the presence of transmission gains/losses in the \ac{RF} chain, the adopted power management schemes, the antenna gain and the sight conditions (e.g., \ac{LOS} or \ac{NLOS}). To this aim, Tab.~\ref{tab:parameter_EMF_computation} reports the main steps to compute $E_{(d)}$, by adopting a set of conservative (and worst case) assumptions and realistic parameters. In brief, the maximum radiated power $P_{\text{MAX}}$ is multiplied by the statistical reduction factor $\alpha_{\text{STAT}}$ and the time-average reduction factor $\alpha_{\text{24}}$. These two factors are introduced to take into account the spatial and temporal power management performed by 5G macro \ac{gNB}, and then obtain realistic values of the average radiated power $P_{\text{AVG}}$. Focusing on $\alpha_{\text{STAT}}$, we refer the interested reader to \cite{ThoFurTor:17} for a closed-form model to compute this parameter. In addition, recent studies in the literature (see e.g., \cite{ecoscienza}, which is based on the \ac{IEC} recommendations \cite{iec1,iec2}) suggest a value of $\alpha_{\text{STAT}}$ equal to 0.25. In this work, we consider two distinct values of $\alpha_{\text{STAT}}$, namely 0.25 and 1. In this way, we are able to assess the impact of adopting either realistic or worst case settings. Focusing then on $\alpha_{\text{24}}$, current works (see e.g., \cite{ecoscienza}) suggest that the 24h average variation of power is clearly lower than unity. However, also in this case we adopt two different values, namely $\alpha_{\text{24}}=0.17$ and $\alpha_{\text{24}}=1.0$, to consider both realistic and worst case assumptions. As a result, $P_{\text{AVG}}$ is computed as $P_{\text{MAX}}\cdot \alpha_{\text{STAT}} \cdot \alpha_{\text{24}}$. In the following step, we compute the \ac{EIRP}, by scaling $P_{\text{AVG}}$ with the transmission gain and losses reported in Tab.~\ref{tab:parameter_EMF_computation}. Given the \ac{EIRP}, we apply the point source model detailed by the \ac{ITU} in \cite{itutk70} to finally compute $E_{(d)}$. It is important to remark that, compared to other models (reported in \cite{itutk70}), the point source represents a worst case, being the measured level of \ac{EMF} exposure always lower than the one computed through this model in the far-field zone.

\begin{figure}[t]
\centering
\includegraphics[width=9cm]{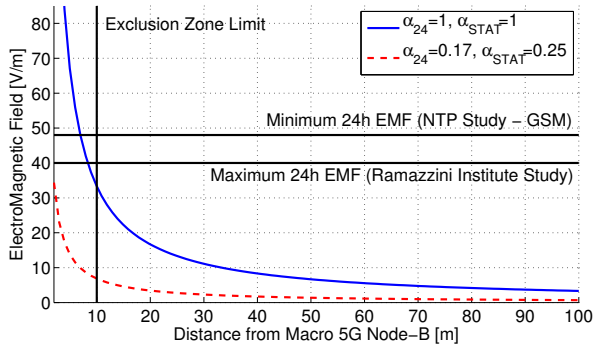}
\caption{\ac{EMF} (electric field strength) vs. distance for different settings of a 5G macro \ac{gNB}, based on the models and parameters detailed in Tab.~\ref{tab:parameter_EMF_computation}. The setting with $\alpha_{\text{STAT}}=1$ and $\alpha_{24}=1$ represents the worst-case. The setting with $\alpha_{\text{STAT}}=0.25$ and $\alpha_{24}=0.17$ is based on realistic considerations. The figure highlights also the positioning of the tests performed by NTP and Ramazzini Institute and the typical size of an exclusion zone for a 5G macro \ac{gNB} (figure best viewed in colors). }
\label{fig:tests_vs_realistc_exposure}
\end{figure}

Fig.~\ref{fig:tests_vs_realistc_exposure} reports the values of $E_{(d)}$ vs. the variation of $d$ and the two values imposed for $\alpha_{\text{STAT}}$ and $\alpha_{\text{24}}$. The figure also highlights the typical size of the exclusion zone with a vertical line, which we remind is the minimum distance between a user  and a 5G macro \ac{gNB} in \ac{LOS}. In addition, the horizontal lines mark the maximum 24h average \ac{EMF} imposed in the Ramazzini Institute study \cite{FalBua:18} and the minimum 24h average \ac{EMF} measured in the \ac{NTP} study \cite{NTP:18a,NTP:18b}. We select the maximum value for \cite{FalBua:18} because this is the only setting showing a statistically significant increase of critical diseases in the rats. On the other hand, we select the minimum \ac{EMF} for \cite{NTP:18a,NTP:18b} since some adverse health effects were found even with this level of exposure.

Several considerations hold by observing Fig.~\ref{fig:tests_vs_realistc_exposure}. First, $E_{(d)}$ is rapidly decreasing with $d$, with values lower than 10~[V/m] when $d>35$~[m]. Second, the introduction of realistic values for $\alpha_{\text{STAT}}$ and $\alpha_{\text{24}}$ results into an abrupt decrease of $E_{(d)}$, with an \ac{EMF} lower than 10~[V/m] already inside the exclusion zone, and values lower than 5~[V/m] when $d>20$~[m]. Third, the critical values of 24h \ac{EMF} used in the studies \cite{NTP:18a,NTP:18b,FalBua:18} are clearly larger than the $E_{(d)}$ values outside the exclusion zone, even for the worst case $\alpha_{\text{STAT}}=1$ and $\alpha_{\text{24}}=1$. Fourth, when adopting realistic settings for $\alpha_{\text{STAT}}$ and $\alpha_{\text{24}}$, $E_{(d)}$ is clearly lower than the minimum 24h average \ac{EMF} of \cite{NTP:18a,NTP:18b} and the maximum 24h average of \cite{FalBua:18}. As a result, we can claim that the critical \ac{EMF} levels used in \cite{NTP:18a,NTP:18b,FalBua:18} to argue the health impact from \ac{RF} sources are never reached outside the exclusion zone of a 5G macro \ac{gNB}. Therefore, the exposure levels for the general public are always far below the critical values of \cite{NTP:18a,NTP:18b,FalBua:18}, thus ensuring safety for the population.

Finally, the analysis on the \ac{EMF} exposure does not include the comparison against 5G \ac{UE}. The near-field conditions at which such devices operate impose to consider the \ac{SAR} metric, which is tackled in the next point.

\begin{table*}[t]
\centering
\caption{Comparison of \ac{SAR} values between the animal-based studies  \cite{NTP:18a,NTP:18b,FalBua:18} and the \ac{UE}.}
\label{tab:sar_value}
\footnotesize
\begin{tabular}{|c|p{6cm}|}
\hline
\rowcolor{Lightred} \textbf{Device} & \textbf{Value(s)} \\
\hline
\multirow{3}{*}{\ac{RF} Source - NTP Study \cite{NTP:18a,NTP:18b}} & 1.5 [W/kg] (whole body of the animal) \\
& 3~[W/kg] (whole body of the animal) \\
& 6~[W/kg] (whole body of the animal) \\ 
\hline
\rowcolor{Linen} & 0.001~[W/kg] (whole body of the animal)\\
\rowcolor{Linen}  & 0.03~[W/kg] (whole body of the animal) \\
\rowcolor{Linen}  \multirow{-3}{*}{\ac{RF} Source - Ramazzini Institute Study \cite{FalBua:18}}  & 0.1~[W/kg] (whole body of the animal) \\
\hline
 & 0.68~[W/kg] (local \ac{SAR} measured by placing the \ac{UE} close to the head during a call \cite{iec4})\\
\multirow{-2}{*}{\ac{UE} \cite{sardatabase}} &  0.98~[W/kg] (local \ac{SAR} measured by wearing the \ac{UE} on the body \cite{iec3})\\
\hline
\end{tabular}
\end{table*}

\textbf{Specific Absorption Rate Levels.} We consider here the comparison of \cite{NTP:18a,NTP:18b,FalBua:18} in terms of realistic \ac{SAR} values for 5G \ac{UE}. To this aim, Tab.~\ref{tab:sar_value} reports the \ac{SAR} values imposed by studies \cite{NTP:18a,NTP:18b,FalBua:18}, and their comparison against the \ac{SAR} of \ac{UE}. Due to the limited number of 5G mobile devices, we include in our analysis also pre-5G \ac{UE} with smartphone capabilities (whose data are retrieved from the publicly available database of \cite{sardatabase}). In particular, by adopting the standardized procedures of \cite{iec4,iec3}, two different \ac{SAR} values are provided by each manufacturer of \ac{UE}. The first one is referred to as a use case where the \ac{UE} is close to the head during a call. The second one is instead representative for a \ac{UE} worn on the body. The two average values, obtained over a wide set of \ac{UE} models, are reported in Tab.~\ref{tab:sar_value}. On the other hand, the \ac{NTP} study \cite{NTP:18a,NTP:18b} adopts three different values of \ac{SAR} over the whole animal body, corresponding to the different exposure levels imposed during the experiments. Similarly, three different whole-body \ac{SAR} levels are employed in the Ramazzini Institute study \cite{FalBua:18}.

Different considerations hold by analyzing the outcome of Tab.~\ref{tab:sar_value}. First, the \ac{SAR} values imposed by the NTP study are consistently higher than those of commercial \ac{UE}. As a result, the negative outcomes of \cite{NTP:18a,NTP:18b} can not be generalized to \ac{UE}. Second, the \ac{SAR} values estimated from the Ramazzini Institute study are consistently lower than those of commercial devices. Therefore, the outcomes of \cite{FalBua:18} may be relevant to the \ac{UE} in use (which we remind also include legacy technologies). However, we also point out that the measured \ac{SAR} of \ac{UE} is a local metric (i.e., not referred over the whole body), while the \ac{SAR} of the animal-based studies \cite{NTP:18a,NTP:18b} is measured over the whole mass of the rats/mice. Therefore, the local and whole-body \ac{SAR} values can not be directly compared, as they are referred to different absorption areas and volumes. We will shed light on this aspect when considering the \ac{SAR} regulations in Sec.~\ref{sec:metrics_regulations}. Intuitively, local \ac{SAR} may be higher than whole-body \ac{SAR}. However, it is also important to remark that the whole-body \ac{SAR} of the animal-based studies \cite{NTP:18a,NTP:18b} are referred to rats, whose absorption area and volume are much lower compared to a human body.  Eventually, the actual \ac{SAR} levels of \ac{UE} may differ from the values provided by manufacturers, since the \ac{SAR} metric is influenced by multiple factors, which may introduce strong variations, as pointed out by  \ac{ITU} \cite{itutksupplement13}.

\textbf{Transmission and Modulation Techniques.} In this part, we focus on the different transmission and modulation techniques implemented in \cite{NTP:18a,NTP:18b,FalBua:18}, and their comparison against the one adopted by real 5G equipment. 
Focusing on the NTP study \cite{NTP:18a,NTP:18b}, the authors evaluate two different technologies, namely \ac{GSM} and \ac{CDMA}. Focusing on \ac{GSM}, this technology leverages \ac{FDMA} and \ac{TDMA} techniques. More specifically, the \ac{GSM} band is divided in frequency with channels of 200~[kHz]-wide; then, each channel is temporally divided into eight different time slots that are used for voice communications. During a voice call of a \ac{UE}, a single time slot of a given channel is assigned to the terminal. The resulting signal shape is therefore clearly pulsed, as shown, e.g., in Fig.~2 of \cite{NTP:18a,NTP:18b}. Consequently, a large variation between average and peak power is observed. It is important to remark, however, that the useful metric for the evaluation of exposure and/or \ac{SAR} is the average power over the sequence of frames and not the instantaneous one. Eventually, \cite{NTP:18a,NTP:18b} adopted a \ac{GMSK} modulation scheme, which exploits a Gaussian filter to shape the digital data. Focusing then on the experiments based on \ac{CDMA}, we remind that this technology employs the \ac{DSSS} transmission scheme, i.e., the information to be transmitted is firstly multiplied by a random code and then modulated on the carrier. Differently from \ac{GSM}, each transmission employs the whole frequency band to transfer the information. In this case, a fundamental feature is the control of the emitted power, e.g., a \ac{UE} should always transmit at minimum power to reduce the interference to the other terminals in the same cell. In addition, the adopted modulation scheme is \ac{QPSK}, which employs a phase change solution. The resulting implemented \ac{CDMA} standard is \ac{IS-95}. 

Focusing then on the test conditions of \cite{NTP:18a,NTP:18b}, the signals imposed in the experiments are generated by a signal generator, with different uplink configurations, namely: one slot per frame active for \ac{GSM} chambers, and the \ac{IS-95} standard uplink signal generator settings for \ac{CDMA} chambers.\footnote{This information is available in \cite{capstick2017radio}.}

In the following, we move our attention to the Ramazzini Institute study \cite{FalBua:18}. In line with \cite{NTP:18a,NTP:18b}, also this work adopts \ac{FDMA} and \ac{TDMA} techniques, in combination with the \ac{GMSK} modulation scheme. More in-depth, the authors of \cite{FalBua:18} state that a complete-time slot assignment and the call operating mode are exploited. Although the number of used slots is not explicitly reported, it is natural to assume that one slot per frame is active also in the study.

Lastly, we analyze the main features in terms of transmission and modulation techniques implemented in the \ac{5G-NR}. Unless otherwise specified, we adopt the 3GPP release 16 specifications, whose working documents are publicly available in \cite{3gpprel16}. In order to support the variegate services offered by 5G, the features implemented in the physical layer are very flexible and highly customizable to the working conditions. More in-depth, the multiple access is realized with \ac{OFDM} with \ac{CP} in the downlink, and \ac{DFT-s-OFDM} or \ac{OFDM} with \ac{CP} in the uplink. These techniques are the evolution of \ac{OFDM}, which employs orthogonal subcarrier signals to transmit data information in parallel. Also, another great difference between 5G and legacy generations (like the one used in \cite{NTP:18a,NTP:18b,FalBua:18}) is the ability of employing a flexible (i.e., not fixed) subcarrier spacing. Eventually, 5G integrates the possibility of adopting different modulation techniques, which include \ac{BPSK}, \ac{QPSK}, and \ac{QAM}.

In conclusion, the transmission and modulation techniques adopted in \cite{NTP:18a,NTP:18b,FalBua:18} are representative for legacy devices, which assume voice as the only service provided by the mobile network. On the other hand, the transmission and modulation techniques adopted in 5G devices are radically different, to cope with the great level of flexibility that this technology guarantees w.r.t. \ac{GSM} or \ac{CDMA}. This level of flexibility is clearly neglected by \cite{NTP:18a,NTP:18b,FalBua:18}, thus posing limits on the applicability of their outcomes in the 5G context.

%{ \color{black} To Luca: What do you think the difference in health impact of OFDM as multiple access technique when compared to FDM/TDM}
%\textbf{LC: I do not have an answer. Intuitively, OFDM is proven to be more efficient than legacy TDM and FDM techniques. }
%{\color{red}
 %OFDM signaling can  provide a coding gain over single carrier modulation, which enables reducing the transmitted power from  the \ac{UE} and decreasing the \ac{SAR} \cite{Brown:19}. }

%{\color{black} . Also, the fact that NR access technique allows higher bandwidth up to 100 MHz. }
%\textbf{LC: Intuitively, the larger is the BW, the better the radio resources can be exploited. The increase of the BW is beneficial for the performance. I do not have an answer regarding the health effects at the moment.}

\textbf{Summary.} We have reviewed the works \cite{NTP:18a,NTP:18b,FalBua:18}  under the perspective of 5G communications engineering. Many settings and/or assumptions imposed in \cite{NTP:18a,NTP:18b,FalBua:18} appear to be completely different and/or far from reality when compared to those ones adopted in 5G equipment. Such differences include:
\begin{itemize}
\item very short distances compared to the real ones from a 5G macro \ac{gNB};
\item large amount of radiated power and almost absence of power management techniques;
\item very long exposure times;
\item very high \ac{EMF} levels - much larger the ones radiated by a 5G macro \ac{gNB};
\item whole-body \ac{SAR} levels not directly comparable to local \ac{SAR} in real smartphones;
\item basic transmission and modulation schemes. 
\end{itemize}
Therefore, it is not possible to claim that the health effects observed in \cite{NTP:18a,NTP:18b,FalBua:18} may appear in a real 5G deployment. To this aim, \ac{ICNIRP} pointed out in a specific note \cite{icnirpnote} that the studies \cite{NTP:18a,NTP:18b,FalBua:18} do not provide a consistent, reliable and generalizable body of evidence for revising the exposure guidelines. Further studies, tailored to address the limitations of \cite{NTP:18a,NTP:18b,FalBua:18}, are therefore needed.

\begin{table*}[t]
\centering
\caption{Comparison of the main communications features that are collected by the population-based studies \cite{Interphone:10,interphone2011acoustic,johansen2001cellular,frei2011use,benson2013mobile,aydin2011mobile} w.r.t the ones that are meaningful in the context of 5G.}
\label{tab:pop_5G_feature_comparison}
\footnotesize
\begin{tabular}{|c|m{7cm}|m{7cm}|}
\hline
\rowcolor{Lightred} \textbf{Feature} & \textbf{Population-Based Study} & \textbf{5G Communications} \\
\hline
Evaluation & Questionnaire, Personal Interviews, Remote Interview, Mobile-operator Log & Cloud-based App, Mobile-operator Log \\
\hline
\rowcolor{Linen} Evaluation Frequency & One shot, periodic & Continuous \\
%\hline
%Device Information & Model & Model\\
\hline
Activity Type & Call & Call, Streaming Video, Social Media, Instant Messaging \\
\hline
\rowcolor{Linen} Activity Intensity & Number of calls & Number of minutes spent for each App, Amount of content uploaded/Downloaded \\
\hline
Connectivity & Phone Number, Operator & Phone Number, Operator, Used Interfaces, Used Frequencies, Handover Information \\
\hline
\rowcolor{Linen} \ac{UE} position & Distance from head, Use of hands free devices & User Proximity, \ac{UE} handling grasp \\
\hline
User location & Country, Residence & Country, Residence, Mobility Patterns \\
%\hline
\hline
\rowcolor{Linen} \ac{UE} Information & Device Model & Device Model\\
\hline
\end{tabular}
\end{table*}

\subsubsection{Population-based Studies}
\label{subsubsec:population_based_review}

We then review the population-based studies \cite{Interphone:10,interphone2011acoustic,johansen2001cellular,frei2011use,benson2013mobile,aydin2011mobile} from the perspective of the 5G communications engineering. To this aim,  Tab.~\ref{tab:pop_5G_feature_comparison} compares the main communications features adopted in previous studies and how such metrics have to be (eventually) changed or enriched when considering 5G equipment.

First of all, the evaluation in \cite{Interphone:10,interphone2011acoustic,johansen2001cellular,frei2011use,benson2013mobile,aydin2011mobile} is done by applying traditional ways, e.g., questionnaires, personal/remote interviews, and (in few cases) analysis of the log files made available by network operators. Due to the variegate set of 5G services, which include the exchange of data and voice communications, it is not possible to rely upon questionnaires and/or interviews to measure the \ac{UE} activity. On the contrary, this information can be easily retrieved by running custom applications on the \ac{UE}, which automatically transfer the measured data on a controlled cloud. Eventually, when this approach can not be pursued (e.g., due to privacy issues), log files made available by the mobile operators should be used.

Focusing on the evaluation frequency, the population-based studies \cite{Interphone:10,interphone2011acoustic,johansen2001cellular,frei2011use,benson2013mobile,aydin2011mobile} assume that the information about \ac{UE} activity is retrieved with a small pace, i.e., either at the end of the considered period or on a periodic base. In contrast to them, 5G imposes continuous monitoring of \ac{UE} activities, due to the highly temporal variation of the amount of data exchanged by the applications installed on the smartphone with the 5G  services. 

As a third aspect, the primary goal of the population-based studies is to monitor the duration of the calls. Although 5G still provides voice services, for which the call duration should be monitored, it is also important to report the time spent over each service type, which may include, e.g., streaming video, social media, and instant messaging. This step is fundamental to build a precise user profile, with exposure information for each service type. Besides, previous studies adopted the number of calls as an indicator of the intensity of \ac{UE} activity. In the context of 5G, it is fundamental to track the amount of time spent in each application, as well as the amount of data uploaded/downloaded, to derive specific information tailored to the user and the adopted application(s). 

Focusing on the connectivity, the population-based studies mainly measure basic features such as the subscriber mobile phone number and the mobile operator. In the context of 5G, this information has to be enriched by including the temporal usage of each interface(s) (e.g., 5G, 4G, WiFi). In addition, another important information includes the adopted frequencies (e.g., sub GHz, mid-band, mm-Waves), as well as the indication about the performed handovers (which can affect the exposure patterns). 

Focusing then on the \ac{UE} position, the population-based studies \cite{Interphone:10,interphone2011acoustic,johansen2001cellular,frei2011use,benson2013mobile,aydin2011mobile} adopt simple metrics, like the distance from the head and the use of hands free devices. When considering 5G, it is essential to retrieve the proximity of the \ac{UE} w.r.t the user, which can be from head/chest or other parts of the body. Besides, since the \ac{UE} is used in different ways (e.g., talking, watching a video, texting, self-recording, environment recording), it is also fundamental to measure the \ac{UE} handling grasp (e.g., one hand, two hands, vertical handle, horizontal handle).
Eventually, the user location (in terms of country and residence) is used by population-based studies, e.g., to classify the users w.r.t. the living areas (e.g., urban, rural). In the context of 5G, user mobility is key information that should be also recorded. 

\renewcommand{\arraystretch}{1.2}
\begin{table*}[t]
    \caption{\ac{PD} Limits for frequencies $f$ up to 300~[GHz].}
    \label{Tab:PDUE}
    \centering
    \footnotesize
    \begin{tabular}{|m{0.2cm}|m{0.8cm}|m{2.8cm}|m{2.8cm}|m{2.8cm}|m{2.8cm}|m{2.8cm}|}
    \hline
    \rowcolor{Lightred} & & \multicolumn{2}{c|}{\textbf{ICNIRP}} & \multicolumn{2}{c|}{\textbf{IEEE C95.1}} & \\
    \cline{3-6}
    \rowcolor{Lightred} \multirow{-2}{*}{\textbf{PD}}  & \multicolumn{1}{c|}{\multirow{-2}{*}{\textbf{Metric}}} & \multicolumn{1}{c|}{\textbf{1998} \cite{ICNIRPGuidelines:98}} & \multicolumn{1}{c|}{\textbf{2020} \cite{ICNIRPGuidelines:20}} & \multicolumn{1}{c|}{\textbf{2005} \cite{IEEEC95:05}} & \multicolumn{1}{c|}{\textbf{2019}\cite{IEEEC95:19}}  & \multicolumn{1}{c|}{\multirow{-2}{*}{\textbf{FCC-1997} \cite{FCC:97}}}\\
\rowcolor{Linen} \cellcolor{White} &  & \multicolumn{4}{c|}{2~[W/m$^2$], $30 < f \leq 400$~[MHz]} & 2~[W/m$^2$], $30 < f \leq 300$~[MHz] \\
    \cline{3-7}
\rowcolor{Linen} \cellcolor{White}    &  & \multicolumn{4}{c|}{$f$/200~[W/m$^2$], $400 < f \leq 2000$~[MHz]}  & $f$/150~[Wm/$^2$], $300 < f \leq 1500$~[MHz] \\
         \cline{3-7}
\rowcolor{Linen} \cellcolor{White}    & & \multicolumn{2}{c|}{} & 10~[W/m$^2$], $2 < f < 100$~[GHz]  & 10~[W/m$^2$], & 10~[W/m$^2$],  \\
    \cline{5-5}
\rowcolor{Linen} \cellcolor{White}    & \multirow{-6}{*}{\begin{sideways}Gen. Public\end{sideways}} & \multicolumn{2}{c|}{\multirow{-3}{*}{10~[W/m$^2$], $2 < f < 300$~[GHz]}} & $(90 \cdot f -7000)$~[W/m$^2$], $100 < f < 300$~[GHz] & $2 < f < 300$~[GHz] & $1.5 < f < 100$~[GHz] \\
    \cline{2-7}
        &  & \multicolumn{2}{c|}{10~[W/m$^2$], $30 < f \leq 400$~[MHz]} & 10~[W/m$^2$], $30 < f \leq 300$~[MHz] & 10~[W/m$^2$], $30 < f \leq 400$~[MHz] & 10~[W/m$^2$], $30 < f \leq 300$~[MHz] \\
    \cline{3-7}
    &  & \multicolumn{2}{c|}{$f$/40~[W/m$^2$], $400 < f \leq 2000$~[MHz]}  & $f$/30~[W/m$^2$], $300 < f \leq 3000$~[MHz] & $f$/40~[W/m$^2$], $400 < f \leq 2000$~[MHz] & $f$/30~[Wm/$^2$], $300 < f \leq 1500$~[MHz] \\
         \cline{3-7}
    %& & \multicolumn{2}{c|}{} &  & 50~[W/m$^2$], & 50~[W/m$^2$],  \\
\multirow{-12}{*}{\begin{sideways}Whole Body\end{sideways}}    &  \multirow{-5}{*}{\begin{sideways}Occupational\end{sideways}} & \multicolumn{2}{c|}{50~[W/m$^2$], $2 < f < 300$~[GHz]} & 100~[W/m$^2$], $3 < f < 300$~[GHz] & 50~[W/m$^2$], $2 < f < 300$~[GHz] & 50~[W/m$^2$], $1.5 < f < 100$~[GHz] \\
\hline
 \rowcolor{Linen} \cellcolor{White} &  & \ac{SAR} Limits, $f < 10$~[GHz]  & $0.058 \cdot f^{0.86}$~[W/m$^2$], $400 < f \leq 2000$~[MHz] & 40~[W/m$^2$], $400 < f \leq 3000$~[MHz] & $1.19 \cdot f^{0.463}$~[W/m$^2$], $100 < f < 2000$~[MHz] &  \ac{SAR} Limits, $f < 6$~[GHz] \\
 \cline{3-3}
 \cline{4-7}
\rowcolor{Linen} \cellcolor{White}& & 10~[W/m$^2$], $10 < f < 300$~[GHz] & 40~[W/m$^2$], $2 < f \leq 6$~[GHz] & $18.56 \cdot f^{0.699}$, $3 < f \leq 30$~[GHz] & 40~[W/m$^2$], $2 < f < 6$~[GHz] &  10~[W/m$^2$], $6 < f < 100$~[GHz]  \\
\cline{4-6}
\rowcolor{Linen} \cellcolor{White} & \multirow{-6}{*}{\begin{sideways}Gen. Public\end{sideways}} &  & $55/f^{0.177}$~[W/m$^2$], $6 < f \leq 300$~[GHz] & 200~[W/m$^2$, $30 < f \leq 300$~[GHz] & $55 / f^{0.177}$~[W/m$^2$], $6 < f < 300$~[GHz] & \\
\cline{2-7}
 & & \ac{SAR} Limits, $f < 10$~[GHz]  & $0.29 \cdot f^{0.86}$~[W/m$^2$], $400 < f \leq 2000$~[MHz] & 200~[W/m$^2$], $300 < f \leq 3000$~[MHz]  & $5.93 \cdot f^{0.463}$~[W/m$^2$], $100 < f < 2000$~[MHz] &  \ac{SAR} Limits, $f < 6$~[GHz] \\
  \cline{3-3}
\cline{4-7}
&  & 50~[W/m$^2$], $10 < f < 300$~[GHz]  & 100~[W/m$^2$], $2 < f \leq 6$~[GHz] & $200 \cdot (f/3)^{1/5}$, $3 <  f \leq 96$~[GHz]  & 200~[W/m$^2$], $2 < f < 6$~[GHz] &  50~[W/m$^2$], $6 < f < 100$~[GHz]  \\
\cline{4-6}
\multirow{-12}{*}{\begin{sideways}Localized Exposure\end{sideways}} & \multirow{-6}{*}{\begin{sideways}Occupational\end{sideways}}& & $275/f^{0.177}$~[W/m$^2$], $6 < f \leq 300$~[GHz] & 400~[W/m$^2$], $96 < f \leq 300$~[GHz] & $274.8 / f^{0.177}$~[W/m$^2$], $6 < f < 300$~[GHz] & \\
\hline
    \end{tabular}
\end{table*}
\renewcommand{\arraystretch}{1.0}

Finally, population-based studies store the device model as \ac{UE} information. Since the \ac{UE} exposure varies across the different models, this information should be also recorded when considering 5G equipment.

Summarizing, although large efforts have been done by previous population-based studies \cite{Interphone:10,interphone2011acoustic,johansen2001cellular,frei2011use,benson2013mobile,aydin2011mobile} to assess the exposure from \ac{UE} in legacy generation networks, their findings can not be entirely generalized also to 5G \ac{UE}. Therefore, a new set of population-based studies, explicitly focused on 5G, should be put into place. This step would require to radically change the measurement techniques, the parameters that need to be measured, and the methodology to share the data. However, we  point out all these steps are completely feasible from a technological point of view, even when considering currently available devices. {Clearly, security and privacy issues should be carefully taken into account when considering the exchange of exposure information from \ac{UE}, e.g., to avoid that malicious users inject misleading exposure information, thus  making the health risk evaluation inefficient. We refer the interested reader to} \cite{nguyen2020blockchain} {for an overview of Blockchain-based solutions that may be put into place to secure 5G communications from/to \ac{UE}}.
%%%%%%%%%%%%%%%%%%%%%%%%%%%%%%%%%%%%%%%%%%%%%%%%%%%%%%%%%%%%%%%%%%%%%%%%%%%%%%%%%%%%%%%%%%%%%%%%%%%%%%%%%%%%%%%%%%%%%%%%%%%%%%%%%%%%%%%%%%%%%%%%%%%%%%%%%%%%%%%%%%%%%%%%%%%%%%%%%%%%%%%%%%%%%%%%%%%%%%%%%%%%%%%%%%%%%%%%%%%%%%%%%%%%%%%%%%%%%%%%%%%%%%%%%%%%%%%%%%%%%%%%%%%%%%%%%%%%%%%%%%%%%%%%%%%%%%%%%%%%%%%%%%%%%%%%%%%%%%%%%%%%%%%%%%%%%%%%%%%%%%%%%%%%%%%%%%%%%%%%%%%%%%%%%%%%%%%%%%%%%%%%%%%%%%%%%%%%%%%%%%%%%%%%%%%%%%%%%%%%%%%%%%%%%%%%%%%%%%%%%%%%%%%%%%%%%%%%%%%%%%%%%%%%%%%%%%%%%%%%%%%%%%%%%%%%%%
\section{5G Exposure: Regulations and Compliance Assessment Procedures}
\label{sec:metrics_regulations}

A key aspect to {minimize} the health risks is {the verification of compliance with regulatory limits}. To face this point, we focus on the following aspects: \textit{i}) analysis of the international {guidelines} defining limits on 5G exposure, \textit{ii}) analysis of the impact of national regulations on the health risks, \textit{iii}) overview of the policies to assess the exposure compliance of 5G equipment w.r.t the limits.

%%%%%%%%%%%%%%%
\renewcommand{\arraystretch}{1.3}
\begin{table*}[t]
	\caption{\ac{EMF} strength (incident E field) limits for frequencies $f$ up to 300~[GHz]. }
	\label{Tab:EMFBS}
	\centering
		\footnotesize
	\begin{tabular}{|m{0.75cm}|m{2.95cm}|m{2.95cm}|m{2.95cm}|m{2.95cm}|m{2.95cm}|}
		\hline
	\rowcolor{Lightred}  & \multicolumn{2}{c|}{\textbf{ICNIRP}} & \multicolumn{2}{c|}{\textbf{IEEE C95.1}} & \\
	\cline{2-5}
	\rowcolor{Lightred}   \multirow{-2}{*}{\textbf{Metric}} & \multicolumn{1}{c|}{\textbf{1998} \cite{ICNIRPGuidelines:98}} & \multicolumn{1}{c|}{\textbf{2020} \cite{ICNIRPGuidelines:20}} & \multicolumn{1}{c|}{\textbf{2005} \cite{IEEEC95:05}} & \multicolumn{1}{c|}{\textbf{2019} \cite{IEEEC95:19}}  & \multicolumn{1}{c|}{\multirow{-2}{*}{\textbf{FCC-1997} \cite{FCC:97}}}\\
	\hline
	 & 28~[V/m], $10 < f \leq 400$~[MHz] & 27.7~[V/m], $30 < f \leq 400$~[MHz]  & \multicolumn{2}{c|}{27.5~[V/m], $30 < f \leq 400$~[MHz]} & 27.5~[V/m], $30 < f \leq 300$~[MHz] \\
	 \cline{2-6}
	 & \multicolumn{2}{c|}{$1.375 \cdot f^{0.5}$~[V/m], $400 < f \leq 2000$~[MHz]} & \multicolumn{2}{c|}{Power density limits,} & \multicolumn{1}{c|}{Power density limits,}\\
	\cline{2-3} 
	 \multirow{-4}{*}{\begin{sideways}General Public\end{sideways}} & 61~[V/m], $2 < f < 300$~[GHz] & Power density limits, $2 < f < 300$~[GHz]  & \multicolumn{2}{c|}{$f > 400$~[MHz]}  &  \multicolumn{1}{c|}{$f > 300$~[MHz]} \\
	 \hline
\rowcolor{Linen} \cellcolor{White}	 & 61~[V/m], $10 < f \leq 400$~[MHz] &  61~[V/m], $30 < f \leq 400$~[MHz]  & \multicolumn{2}{c|}{61.4~[V/m], $30 < f \leq 400$~[MHz]} &  61.4~[V/m], $30 < f \leq 300$~[MHz] \\
	 \cline{2-6}
\rowcolor{Linen} \cellcolor{White}	 & \multicolumn{2}{c|}{$3 \cdot f^{0.5}$~[V/m], $400 < f \leq 2000$~[MHz]} &  \multicolumn{2}{c|}{Power density limits,} & \multicolumn{1}{c|}{Power density limits,}\\
	\cline{2-3} 
\rowcolor{Linen} \cellcolor{White}	 \multirow{-4}{*}{\begin{sideways}Occupational\end{sideways}} & 137~[V/m], $2 < f < 300$~[GHz] & Power density limits, $2 < f < 300$~[GHz] & \multicolumn{2}{c|}{$f > 400$~[MHz]}   & \multicolumn{1}{c|}{$f > 300$~[MHz]}\\
	\hline
	  & 6~[min], $f \leq 10$~[GHz] &  & 30~[min] (General Public) & & 30~[min] (General Public)\\
	  \cline{2-2} \cline{4-4} \cline{6-6}
	 \multirow{-2}{*}{\begin{sideways}Time\end{sideways}} & $68/f^{1.05}$~[min], $f > 10$~[GHz] ($f$ in GHz)& \multicolumn{1}{c|}{\multirow{-2}{*}{30~[min]}} & 6~[min] (Occupational) & \multicolumn{1}{c|}{\multirow{-2}{*}{30~[min]}} & 6~[min] (Occupational)\\
	
	\hline
	\end{tabular}
\end{table*}
\renewcommand{\arraystretch}{1.0}

%%%%%%%%%%%%%%%%%%%%%
\subsection{International {guidelines} on 5G \ac{EMF} Exposure}

The main international organizations defining {guidelines} on \ac{RF} exposure are \ac{ICNIRP}, \ac{IEEE} and \ac{FCC}. Both \ac{ICNIRP} and \ac{IEEE} revised the {guidelines} throughout the years. More in-depth, \ac{ICNIRP} defined the \ac{EMF} guidelines in 1998 \cite{ICNIRPGuidelines:98}, and then revised them in 2020 \cite{ICNIRPGuidelines:20}. In a similar way, \ac{IEEE} defined \ac{RF} safety guidelines in the C95.1 standard, which was updated in 2005 \cite{IEEEC95:05}, and then updated again in 2019 \cite{IEEEC95:19}. Finally, the \ac{FCC} released the \ac{RF} guidelines in \cite{FCC:97}, which, to the best of our knowledge, are still in force since their release, dated back to 1997. The reason for reporting various regulations of each organization is twofold. On one side, it is possible to track changes over the different {guidelines} and check whether the different {guidelines} are converging into a common set of limit values. On the other hand, different countries in the world {implement} different {guidelines in their} regulations \cite{gsmamap}. For example, the \ac{ICNIRP} 1998 guidelines \cite{ICNIRPGuidelines:98} are still in force in many countries, with plans to gradually switch to the \ac{ICNIRP} 2020 guidelines \cite{ICNIRPGuidelines:20} in the forthcoming years.

In general, the \ac{EMF} {guidelines} consider two distinct sets of limits for human exposure, namely general public limits and occupational limits.\footnote{{A further differentiation exhists between basic restrictions and reference levels. In this work, unless otherwise specificed, we consider the reference levels as the representative exposure limits.}} The first set is tailored to the general public, who may be not aware of being exposed to radiation (e.g., \ac{EMF} from \ac{gNB}). On the other hand, occupational limits are defined for workers subject to \ac{RF} exposure in a controlled environment, and therefore may take some precautionary procedures to reduce the exposure. A typical example of this second set is a technician performing a maintenance operation on a cellular tower under operation. The general public limits are, in general, more stringent than the occupational ones. In the following, we discuss the international limits in terms of \ac{PD}, \ac{EMF} strength, and \ac{SAR}, under the 5G communications perspective. 

{As a side comment, the intrinsic temporal variability of 5G exposure requires to integrate in the national regulations also short-term exposure limits, which are e.g., defined in the \ac{ICNIRP} 2020 guidelines} \cite{ICNIRPGuidelines:20}. {We leave the discussion of this aspect as a future work, while in the rest of the section we concentrate on long-term limits (i.e., typically on time scales of several minutes). } 

\subsubsection{\ac{PD} Limits} 

We initially analyze the \ac{PD} limits, shown in Tab.~\ref{Tab:PDUE}. {For the sake of clarity, the table reports the limits without the averaging times.} Several considerations hold by exploring the table values. First, a huge variability in terms of limits emerges, even by considering different versions of the same organization (e.g., \ac{ICNIRP} or \ac{IEEE}). Second, \ac{PD} limits notably change across the frequencies, being some limits fixed for a given range of frequencies, and other ones varying with the adopted frequency. Third, the values of occupational limits are, in general, higher than the general public ones (as expected). Fourth, when going towards mm-Wave frequencies, most of the limits employ fixed values (i.e., not varying with frequency). Fifth, the latest versions of \ac{ICNIRP} and \ac{IEEE} adopt a common set of limits when the \ac{PD} over the whole body is considered. Sixth, both the \ac{ICNIRP} 1998 \cite{ICNIRPGuidelines:98} and the \ac{FCC} guidelines \cite{FCC:97} enforce \ac{SAR} limits (which are going to be detailed later on) for frequencies below the threshold when considering local exposure. Finally, \ac{PD} limits for the local exposure are extensively defined for all 5G frequencies in the \ac{ICNIRP} 2020 \cite{ICNIRPGuidelines:20} and \ac{IEEE} C95.1 guidelines \cite{IEEEC95:19}.  

\subsubsection{\ac{EMF} Strength Limits} 

We then move our attention to the limits on the \ac{EMF} strength, which are reported in Tab.~\ref{Tab:EMFBS}. {For the sake of clarity, the table reports the limits for the electric field, while the limits for the magnetic field are intentionally omitted.} In general, these limits are enforced when considering the \ac{EMF} from \ac{gNB}. Interestingly, the latest versions of the limits define two working regions. In the first one, which is typically below 300~[MHz] of frequency, the limits are expressed in terms of maximum incident $E$ field, with values very close among the different regulations. On the other hand, for very high frequencies (i.e., in the order of dozens of GHz), the limits are defined in terms of \ac{PD}. For intermediate frequencies, \ac{ICNIRP} 2020 \cite{ICNIRPGuidelines:20} considers the maximum \ac{EMF} strength up to 2~[GHz], then \ac{PD} is taken into account for higher frequencies. Note that many countries in the world (see, e.g., the network limit map of \cite{gsmamap}) still adopt the  \ac{ICNIRP} 1998 limits \cite{ICNIRPGuidelines:98}, which are instead defined in terms of the maximum incident electric strength for all the frequencies up to 300~[GHz]. Similarly to the \ac{PD} case, general public limits are much more conservative than occupational ones (as expected). Finally, the table reports the minimum amount of time needed to measure the incident electric field. Interestingly, the last versions of \ac{ICNIRP} \cite{ICNIRPGuidelines:20} and \ac{IEEE} C95.1 \cite{IEEEC95:19} guidelines converge to a 30~[min] of time duration for both general public and occupational. On the other hand, the previous version of \ac{IEEE} C95.1 \cite{IEEEC95:05}, as well as the \ac{FCC} guidelines \cite{FCC:97}, adopt a 6~[min] time duration when considering occupational exposure.

\renewcommand{\arraystretch}{1.3}
\begin{table*}[t]
	\caption{\ac{SAR} Limits (including dose metrics) for frequencies $f$ up to 300~[GHz].}
	\label{Tab:SARUE}
	\centering
		\footnotesize
	\begin{tabular}{|c|c|m{2.0cm}|m{2.5cm}|m{2.0cm}|m{2.0cm}|m{2.0cm}|}
	\hline
	\rowcolor{Lightred} \multicolumn{2}{c|}{} & \multicolumn{2}{c|}{\textbf{ICNIRP}} & \multicolumn{2}{c|}{\textbf{IEEE C95.1}} & \\
	\cline{3-6}
	\rowcolor{Lightred} \multicolumn{2}{c|}{\multirow{-2}{*}{\textbf{Metric}}} & \multicolumn{1}{c|}{\textbf{1998}  \cite{ICNIRPGuidelines:98}} & \multicolumn{1}{c|}{\textbf{2020} \cite{ICNIRPGuidelines:20}} & \multicolumn{1}{c|}{\textbf{2005} \cite{IEEEC95:05}} & \multicolumn{1}{c|}{\textbf{2019} \cite{IEEEC95:19}}  & \multicolumn{1}{c|}{\multirow{-2}{*}{\textbf{FCC-1997}  \cite{FCC:97}}}\\
	\hline
	  \multicolumn{2}{|c|}{} & & \multicolumn{1}{c|}{None (Whole Body)} & & \multicolumn{2}{c|}{} \\
	   \multicolumn{2}{|c|}{\multirow{-2}{*}{\ac{SAR} to \ac{PD} Switching Frequency $\ft$}} & \multicolumn{1}{c|}{\multirow{-2}{*}{10~[GHz]}} & \multicolumn{1}{c|}{6~[GHz] (Local)} & \multicolumn{1}{c|}{\multirow{-2}{*}{3~[GHz]}} & \multicolumn{2}{c|}{\multirow{-2}{*}{6~[GHz]}}\\
	\hline
  & General Public  & \multicolumn{5}{c|}{0.08~[W/kg]}\\
  \cline{2-7}
\rowcolor{Linen} \cellcolor{White}  & Occupational  & \multicolumn{5}{c|}{0.4~[W/kg]} \\
  \cline{2-7}
%  &  & \multicolumn{1}{c|}{6 [min], $f\leq10$~[GHz]}  & & \multicolumn{1}{c|}{6 [min], $f\leq3$~[GHz]} & & \\
   & Averaging Time & \multicolumn{1}{c|}{$6$~[min]} & \multicolumn{1}{c|}{30 [min]} &  \multicolumn{1}{c|}{$6$ [min]} & \multicolumn{1}{c|}{30 [min]} & \multicolumn{1}{c|}{Defined in \cite{IEEEC95:91}} \\
   \cline{2-7}
 \rowcolor{Linen} \cellcolor{White}    &   Dose Metric for  $f\leq\ft$ & \multicolumn{5}{c|}{\ac{SAR}}  \\
 \cline{2-7}
\multirow{-5}{*}{\begin{sideways}Whole Body \end{sideways}}    & Dose Metric for  $f>\ft$ & incident \ac{PD} $S_{\text{inc}}$ & \multicolumn{1}{c|}{\ac{SAR}} & incident \ac{PD} $S_{\text{inc}}$ & plane-wave equivalent \ac{PD} $S_{\text{eq}}$ & plane-wave equivalent \ac{PD} $S_{\text{eq}}$ \\
   \hline
\rowcolor{Linen} \cellcolor{White}   & General Public & \multicolumn{4}{c|}{2~[W/kg]} & \multicolumn{1}{c|}{1.6~[W/kg]}\\
      \cline{2-7}
 & Occupational  & \multicolumn{4}{c|}{10~[W/kg]} & \multicolumn{1}{c|}{8~[W/kg]} \\
    \cline{2-7}
% \rowcolor{Linen} \cellcolor{White} &  & \multicolumn{1}{c|}{6 [min], $f\leq10$~[GHz] } & & \multicolumn{1}{c|}{6 [min], $f\leq3$~[GHz]} & & \\
\rowcolor{Linen} \cellcolor{White}  & Averaging Time & \multicolumn{4}{c|}{6~[min]}  & \multicolumn{1}{c|}{Defined in \cite{IEEEC95:91}} \\
     \cline{2-7}
  & Averaging Mass & \multicolumn{4}{c|}{10~[g]} & \multicolumn{1}{c|}{1~[g]}\\
     \cline{2-7}
\rowcolor{Linen} \cellcolor{White}  & Averaging Shape & \multicolumn{1}{c|}{Not defined} & \multicolumn{4}{c|}{cubic}\\
  \cline{2-7}    
  & Dose Metric for  $f\leq\ft$ & \multicolumn{5}{c|}{\ac{SAR}}  \\
   \cline{2-7}    
\rowcolor{Linen} \cellcolor{White} \multirow{-7}{*}{\begin{sideways}Localized Exposure\end{sideways}}   &  Dose Metric for  $f>\ft$ & incident $S_{\text{inc}}$ &  absorbed \ac{PD} $S_{\text{ab}}$ & incident \ac{PD} $S_{\text{inc}}$ & epithelial \ac{PD} & plane-wave equivalent \ac{PD} $S_{\text{eq}}$    \\
   \hline
	\end{tabular}
\end{table*}
\renewcommand{\arraystretch}{1.0}

\subsubsection{\ac{SAR} Limits}

In the final part of this step, we consider the \ac{SAR} limits, which are reported in Tab.~\ref{Tab:SARUE}. We initially focus on the \ac{SAR} to \ac{PD} switching frequency $\ft$. In the context of 5G, $\ft$ will differentiate between limits (and metrics) applied to mm-Waves w.r.t. the mid-band and the sub-GHz frequencies used by this technology. Interestingly, the values of $\ft$ are not the same across the regulations. For example, the \ac{ICNIRP} 2020 guidelines \cite{ICNIRPGuidelines:20} do not impose any frequency threshold on the whole body exposure, and thus \ac{SAR}-based limits are assumed over the whole range of 5G frequencies. However, a threshold $\ft=6$~[GHz] is imposed for the local exposure, and this setting is in common with the FCC 1997 \cite{FCC:97} and the \ac{IEEE} C95.1 2019 \cite{IEEEC95:19} guidelines.  Moreover, many countries in the world currently adopt the \ac{ICNIRP} 1998 \cite{ICNIRPGuidelines:98} and FCC 1997 \cite{FCC:97}  regulations, which enforce $\ft=10$~[GHz] and $\ft=6$~[GHz], respectively. In this case, \ac{PD}-based limits will be enforced for mm-Wave frequencies. 

 Focusing then on the whole body \ac{SAR} limits, we can note that the same values are used across the different regulations. In addition, both \ac{ICNIRP} 2020 \cite{ICNIRPGuidelines:20}  and \ac{IEEE} C95.1 2019 \cite{IEEEC95:19} adopt the same value of averaging time (i.e., 30~[min] {for whole body exposure limits}). Eventually, the averaging time for \ac{SAR} in the \ac{FCC} guidelines {is set equal to the one defined in} the \ac{IEEE} C95.1 1991 standard \cite{IEEEC95:91}. Specifically, when considering occupational exposure, an averaging time equal to 6~[min] is assumed. When considering instead general public exposure, multiple times (reported in Table 2 of \cite{IEEEC95:91}) are defined, ranging however between 6~[min] and 30~[min] for the adopted \ac{SAR} frequencies. 
 
We then move our attention on the dose metrics for the whole body exposure. Clearly, \ac{SAR} is always used for frequencies $f\leq\ft$. When considering instead $f> \ft$, different metrics are used (e.g., incident \ac{PD}, \ac{SAR}, plane-wave equivalent \ac{PD}). However, it is important to remark that the \ac{ICNIRP} 2020 guidelines \cite{ICNIRPGuidelines:20} conservatively enforce \ac{SAR} limits even for $f> \ft$ {for whole body exposure limits}. 

In the following, we consider the \ac{SAR} limits for local exposure, reported on bottom of Tab.~\ref{Tab:SARUE}. In this case, \ac{ICNIRP} and \ac{IEEE} differentiate from \ac{FCC} in terms of: limits, averaging time, and averaging mass. However, the latest versions of the regulations agree on a cubic mass, thus adopting a uniform metric. Focusing then on the dose metrics, the same consideration of the whole body exposure hold for $f\leq\ft$. When considering $f> \ft$, all the regulations adopt \ac{PD}-based metrics. However, it is important to remark that the adopted \ac{PD} metrics are not the same across the regulations. For example, the \ac{ICNIRP} 1998 guidelines \cite{ICNIRPGuidelines:98} adopt the incident \ac{PD}, while the \ac{IEEE} C95.1 2019 regulations \cite{IEEEC95:19} employ the ephitelial \ac{PD} (i.e., the power flow through the epithelium per unit area directly under the body surface). In this case, it is important to remark that custom \ac{PD} limits (different from the ones reported in Tab.~\ref{Tab:PDUE}) are defined for the ephitelial \ac{PD}, i.e., 20~[W/m$^2$] for $6 < f < 300$~[GHz] (general public) and 100~[W/m$^2$] for $6 < f < 300$~[GHz] (occupational).

%%%%%%%%%%%%%%
\begin{comment}
\renewcommand{\arraystretch}{1.3}
\begin{table}
	\caption{ICNIRP/FCC maximum \ac{EMF} limits over three representative 5G frequencies.}
	\label{Tap:Referencelevels}
	% \usepackage{multirow}
	% \usepackage{graphicx}
	%\scriptsize
	\centering
	\small
	%\resizebox{\textwidth}{!}{%
	\begin{tabular}{|c|c|c|c|}
		\hline \rowcolor{Lightred} & \multicolumn{3}{c|}{\textbf{Frequency}} \\ 
		\cline{2-4} 
	    \rowcolor{Lightred} \multirow{-2}{*}{\textbf{Standard}} & \SI{700}{[\mega\hertz]}      & \SI{3.5}{[\giga\hertz]}      & \SI{26}{[\giga\hertz]}       \\ \hline
		
		ICNIRP               & 36.4~[V/m] & 61~[V/m]   & 61~[V/m]   \\ 
		\hline
		FCC                  & 41.9~[V/m] & 61.4~[V/m]  & 61.4~[V/m]  \\ \hline
	\end{tabular}
\end{table}
\renewcommand{\arraystretch}{1.0}
\end{comment}
%}
%%%%%%%%%%%%%%%%%%

\subsubsection{Summary}
We have considered international {guidelines} that define exposure limits in terms of \ac{PD}, \ac{EMF} strength, and \ac{SAR}. Although some efforts in making uniform rules across the different organizations obviously emerge, we can claim that 5G devices will be subject to different limits, due to the different frequencies at which they operate, as well as to the different thresholds and metrics used in the compliance assessment. This {fragmentation} {may} increase the health risks {of 5G} that are perceived by the population, since a lack of common limits and/or metrics may be (improperly) associated to a lack of a universal view among the different guidelines. Moreover, several countries in the world adopt more stringent exposure limits than the international ones, on the basis of precautionary principles. This issue, which notably complicates the health risks perception and the 5G deployment, is analyzed in detail in the following subsection. 

\subsection{Impact of National Regulations}
\label{sec:national_regulations}

We provide a comprehensive review of the national \ac{EMF} exposure regulations and their impact on 5G deployment. We divide our research under the following avenues: \textit{i}) overview of national exposure regulations stricter than the \ac{ICNIRP} 1998 \cite{ICNIRPGuidelines:98} and/or \ac{FCC} 1997 \cite{FCC:97} guidelines\footnote{We adopt the \ac{EMF} limits defined in \ac{ICNIRP} 1998 \cite{ICNIRPGuidelines:98} and \ac{FCC} 1997 \cite{FCC:97} guidelines, as these regulations are currently still in force in many countries in the world. To the best of our knowledge, the adoption of the \ac{ICNIRP} 2020 guidelines \cite{ICNIRPGuidelines:98} by the national governments is an ongoing process, not yet completed at the time of preparing this work.} (henceforth simply referred as ICNIRP and FCC, respectively), \textit{ii}) impact of national regulations on 5G \ac{gNB} deployment, \textit{iii}) impact of national regulations on 5G \ac{UE} adoption, \textit{iv}) population-based analysis, and \textit{v}) geographical analysis. 

\subsubsection{Overview of National Exposure Regulations Stricter than \ac{ICNIRP}/\ac{FCC} Guidelines}

We preliminary perform an in-depth search of the exposure regulations in each country in the world. Our primary sources are the data made available by \ac{GSMA} in \cite{gsmamap,gsmadata} and by \ac{WHO} in \cite{whodata}, the work of Madjar \cite{madjar2016human}, the report of Stam \cite{stam2011comparison}, the information retrieved from national \ac{EMF} regulation authorities \cite{serbiaemf,croatiaemf,algeriaemf,monacoemf}, and other relevant documents \cite{montenegroemf5g,chiang2009rationale,serbiaemf5g,sarvarious,hkemf,safetyissues,liechtensteinemf,sanmarinoemf,thailandemf}. 

We initially focus on the national \ac{EMF} regulations for \ac{gNB} exposure. As a consequence, we focus on \ac{EMF} strength with far field conditions. When a national regulation expresses the limit in terms of \ac{PD}, we employ Eq.~(\ref{eq:sincff}) to compute the \ac{EMF} strength. In this way, we obtain a set of homogeneous limits.

\begin{figure}[t]
    \centering
    \includegraphics[width=0.9\linewidth]{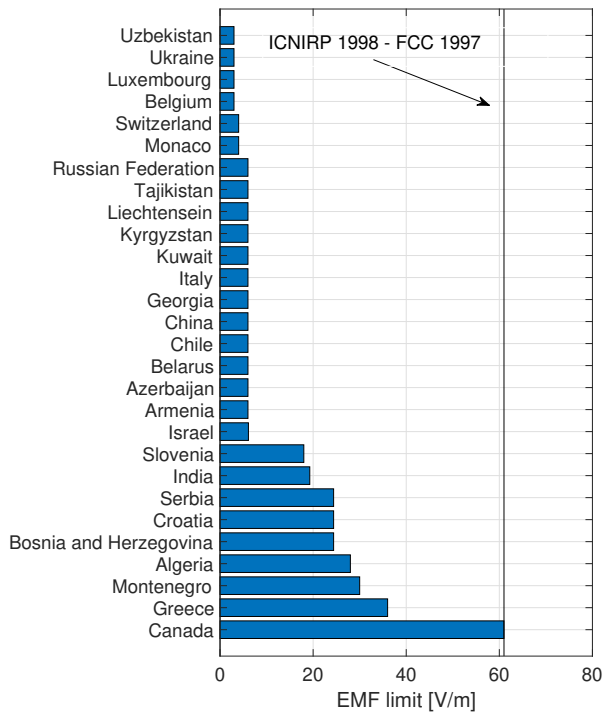}
    \caption{\color{black}Maximum 5G \ac{gNB} limits on the field strength in residential areas for the countries adopting \ac{EMF} regulations stricter than \ac{ICNIRP} 1998 \cite{ICNIRPGuidelines:98} and \ac{FCC} 1997 \cite{FCC:97} guidelines (frequency under consideration: 26~[GHz]). }
    \label{fig:strict_limits_mm_wave}
\end{figure}

\begin{figure*}
    \centering
    \subfigure[5G \ac{gNB} (\ac{EMF} limit in terms of field strength)]
    {
        \includegraphics[width=7cm]{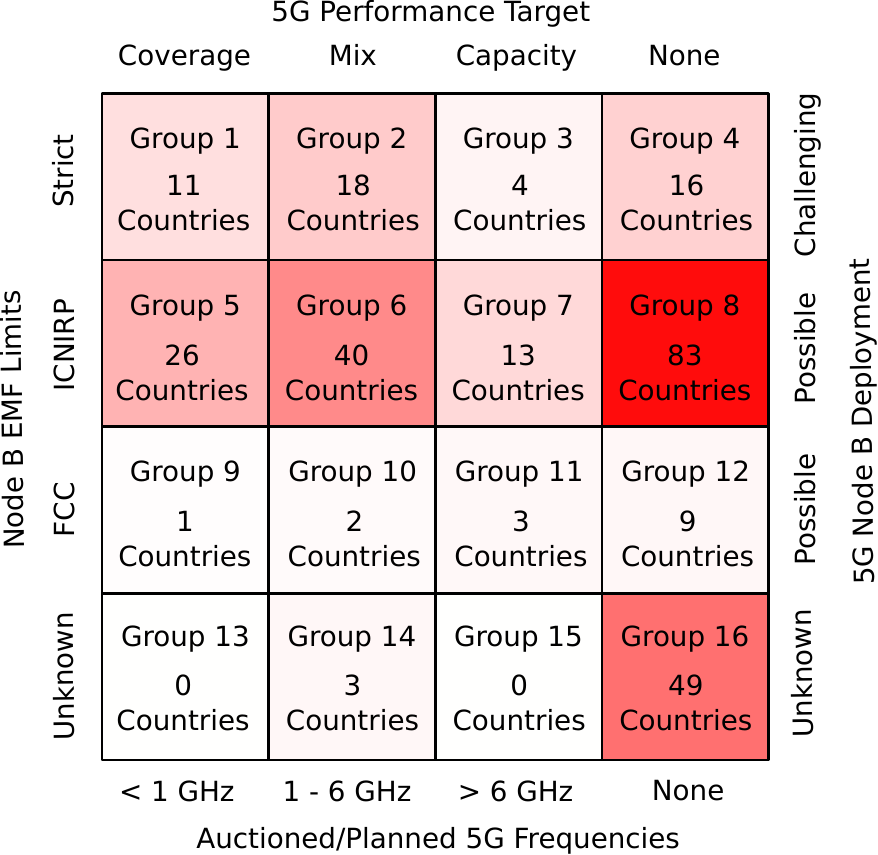}
        \label{fig:limitscapfreqmatrix}
        %    \vspace{2mm}
    }
    \subfigure[5G \ac{UE} (\ac{EMF} limit in terms of whole body \ac{SAR} or \ac{PD})]
    {
        \includegraphics[width=7cm]{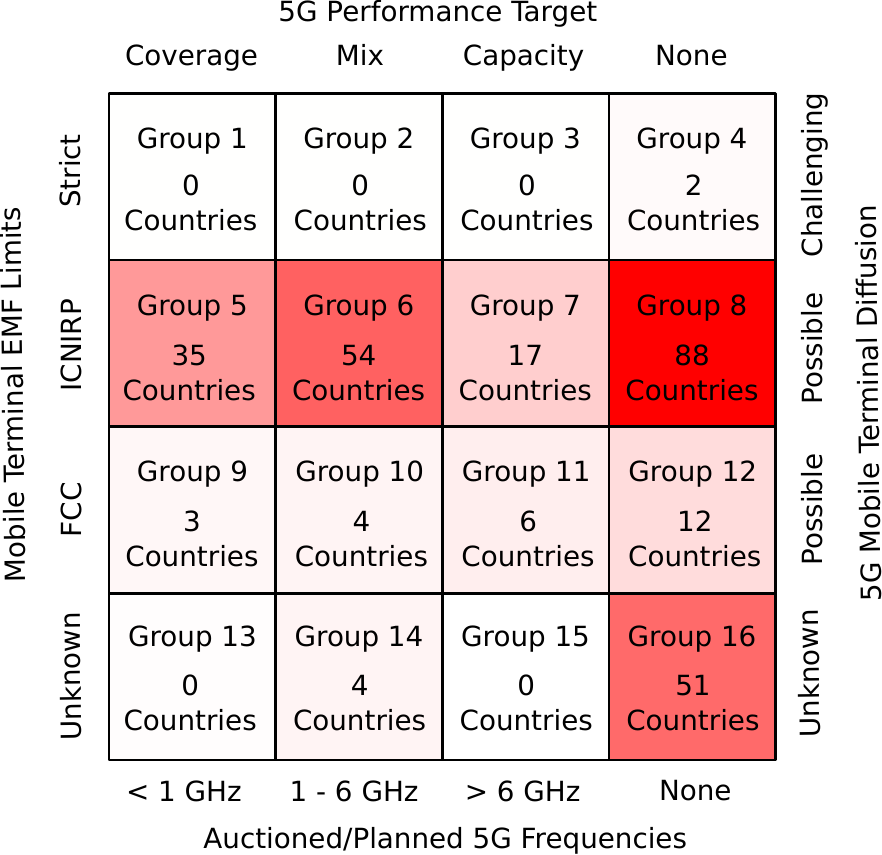}
        \label{fig:limitscapfreqmatrix_device}
        %    \vspace{2mm}
    }
    \caption{\ac{EMF}-limits / 5G frequencies matrices. Each cell in the matrix includes the group ID and the number of countries belonging to the group. The cell color is proportional to the number of countries belonging to the group, from white (zero weight) to red (weight equal to 88).}
    \label{fig:limitscapfreqmatrix_all}
\end{figure*}

Fig.~\ref{fig:strict_limits_mm_wave} reports a graphical overview of the maximum \ac{EMF} limits {for different} countries imposing strict regulations for \ac{gNB} exposure, and their positioning w.r.t. the \ac{EMF} strength limits of the \ac{ICNIRP}/\ac{FCC} guidelines.\footnote{For the sake of simplicity, \ac{ICNIRP} and \ac{FCC} limits are collapsed in a single line. {Moreover, a single limit is taken for those countries (e.g., Belgium) imposing different limits over different regions.}} In this case, we report the limits by considering a 5G frequency equal 26~[GHz]\footnote{The \ac{EMF} limits in Canada are stricter than \ac{ICNIRP} 1998  \cite{ICNIRPGuidelines:98} and \ac{FCC} 1997 \cite{FCC:97} guidelines only for 5G frequencies lower than 6~[GHz] \cite{madjar2016human}. When considering mm-Waves, like in this case, the \ac{EMF} limits enforced in Canada correspond to the \ac{ICNIRP} 1998 \cite{ICNIRPGuidelines:98} and \ac{FCC} 1997 \cite{FCC:97}  ones. However, we report Canada in the figure for completeness.} for multiple reasons, namely: \textit{i}) this frequency is representative of the mm-Wave band in 5G, which triggers the highest exposure concerns from the population, \textit{ii}) frequencies in the mid-band and sub-GHz bands are in general varying across the different countries (especially in the sub-GHz band); as a result, the \ac{EMF} limit considerably changes w.r.t. the adopted frequency, making the comparison of the limits across the different countries a challenging task, \textit{iii}) although not all countries in the world adopt the 26~[GHz] frequency in the mm-Wave band, the enforced limit does not generally vary across other frequencies in the mm-Wave band.

Several considerations hold by analyzing Fig.~\ref{fig:strict_limits_mm_wave} in detail. First, a huge variability across the national limits apparently emerges. Second, when enforcing a limit stricter than \ac{ICNIRP}/\ac{FCC}, the reduction factor is considerably large, i.e., more than 10 times for the majority of the countries adopting strict limits. This reduction factor heavily impacts the installation of 5G \acp{gNB} in residential areas, as a strict \ac{EMF} limit may be easily saturated in the presence of multiple operators and/or multiple technologies (e.g., 2G/3G/4G/5G) operating over the territory \cite{itutksupplement14,chiaraviglio2018planning}. Third, the perception of health risks connected to 5G \ac{gNB} in countries with strict \ac{EMF} limits may be higher compared to the ones enforcing \ac{ICNIRP}/\ac{FCC} limits, due to the fact that the measured \ac{EMF} levels may be close to the limits.

We then move our attention on the national \ac{UE} exposure regulations that are stricter than \ac{ICNIRP} 1998 \cite{ICNIRPGuidelines:98} and \ac{FCC} 1997 \cite{FCC:97} guidelines. Interestingly, the only countries in the world falling in this category are Belarus and Armenia, which still adopt regulations based on legacy Soviet Union limits, expressed in terms of maximum \ac{PD} of 100~[QW/cm$^2$] at an unknown distance. On the other hand, most of the other countries adopt \ac{ICNIRP}/\ac{FCC} limits, typically expressed in terms of \ac{SAR}. Therefore, we can conclude that the perception of health risks connected to the adoption of 5G \ac{UE} is typically lower than the \ac{gNB} case.

\subsubsection{Impact of National Regulations on 5G \ac{gNB} Deployment}
\begin{figure*}
    \centering
    \subfigure[5G \ac{gNB} (\ac{EMF} limit in terms of field strength)]
    {
        \includegraphics[width=7cm]{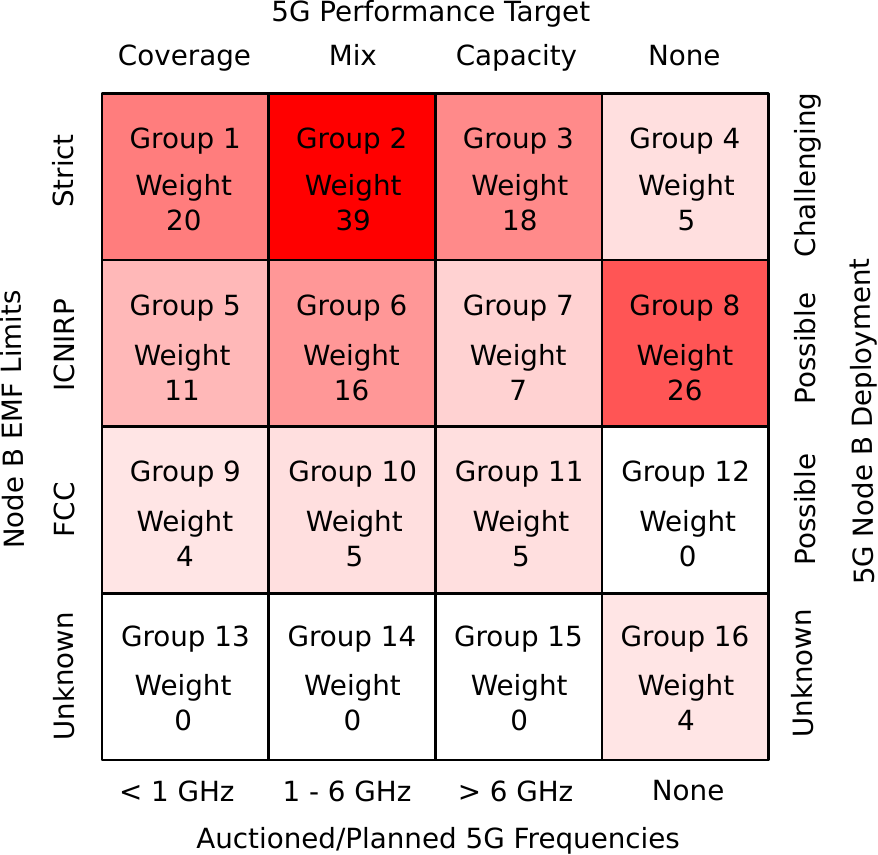}
        \label{fig:limitscapfreqmatri_pop_weight}
        %    \vspace{2mm}
    }
    \subfigure[5G \ac{UE} (\ac{EMF} limit in terms of whole body \ac{SAR} or \ac{PD})]
    {
        \includegraphics[width=7cm]{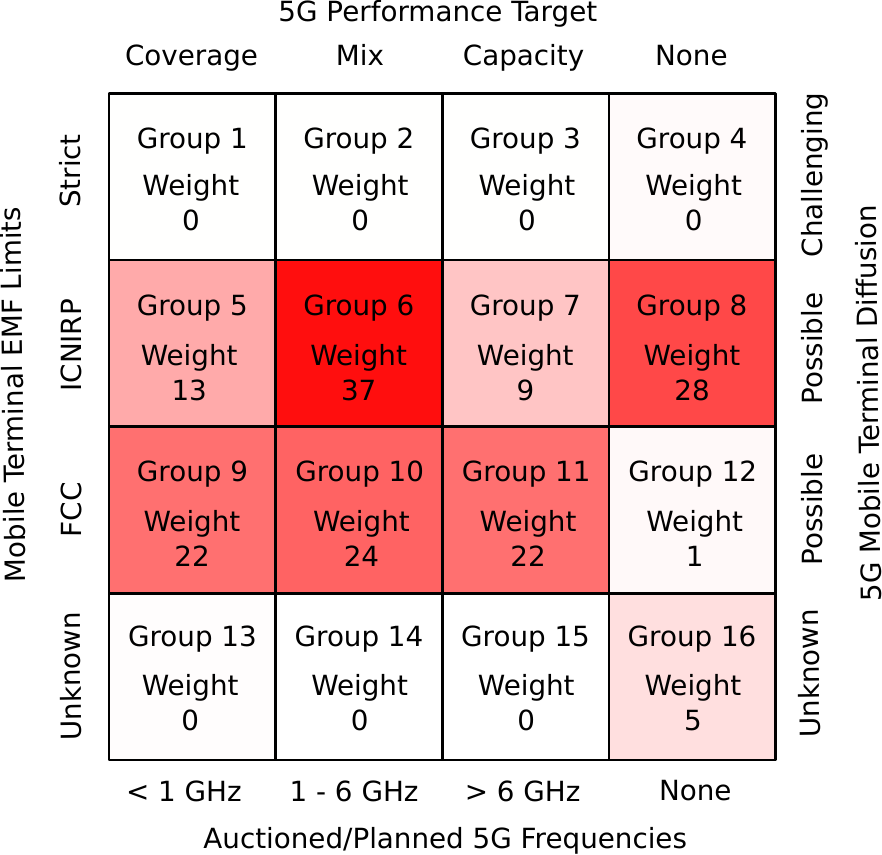}
        \label{fig:limitscapfreqmatrix_device_pop_weight}
        %    \vspace{2mm}
    }
    \caption{\ac{EMF}-limits / 5G frequencies matrices. Each cell in the matrix includes the group ID and the weight based on the population of each country belonging to the group. The cell color is proportional to the cell weight, from white (zero weight) to red (weight equal to 39).}
    \label{fig:limitscapfreqmatrix_all_pop_weight}
\end{figure*}

In the following, we jointly consider the impact of different \ac{EMF} strength limits with the deployment of  5G \acp{gNB} in each country of the world. Specifically, we initially retrieve the information about 5G deployment in each country in the world, by considering nations that have already auctioned the 5G frequencies or have clear plans of forthcoming 5G auctions. Our primary sources are the \ac{GSMA} documents about 5G spectrum management  \cite{gsmaspectrum,gsmaasean}, the European 5G Observatory \cite{5gspectrumobservatory} and other relevant (and up-to-date) national references  \cite{asiaspectrum,thailandspectrum,malaysiaspectrum,indonesiaspectrum,philippinesspectrum,ukrainespectrum,belaruslatviaspectrum,montenegroemf5g,serbiaemf5g,latamspectrum,brazilspectrum, brazilspectrum2,uruguayspectrum,chilespectrum,lesotho5g,nigeriaspectrum,indiaspectrum,liechtensteinspectrum,sanmarinospectrum}.

We then consider the following taxonomy for the \ac{EMF} limits: \textit{L1}) stricter than \ac{ICNIRP}/\ac{FCC}, \textit{L2}) \ac{ICNIRP}-based, \textit{L3}) \ac{FCC}-based, or \textit{L4}) unknown. Clearly, the deployment of the 5G \acp{gNB} will be a challenging step in countries with strict \ac{EMF} limits (\textit{L1}), a possible step in countries adopting \ac{ICNIRP}/\ac{FCC} limits (\textit{L2},\textit{L3}), and an unknown step in countries without a regulation on the limits (\textit{L4}). 

Focusing then on the frequencies used by 5G \acp{gNB}, we consider the taxonomy in terms of 5G frequency bands that have been auctioned/planned in the country, and namely: \textit{F1}) below 1~[GHz], \textit{F2}) between 1~[Ghz] and 6~[Ghz], \textit{F3}) above 6~[Ghz], \textit{F4}) none. Each frequency in \textit{F1})-\textit{F3}) has a different 5G performance target. The sub-GHz frequencies in \textit{F1}) will be used to provide coverage, the mm-Wave frequencies in \textit{F3}) will be exploited to guarantee capacity, while the mid band frequencies in \textit{F2}) will provide a mixture of coverage and capacity. It is important to note that \textit{F1})-\textit{F3}) are not exclusive, i.e., a country may exploit 5G frequencies of any combination of \textit{F1}), \textit{F2}), \textit{F3}). For example, Italy will deploy 5G networks over frequencies in \textit{F1})-\textit{F3}), while 5G frequencies in \textit{F1}) and \textit{F2}) are exploited in Saudi Arabia. Finally, a country is listed in \textit{F4}) if the frequency plans for 5G have not been (yet) defined.

\begin{figure*}[t]
    \centering
    \includegraphics[width=18cm]{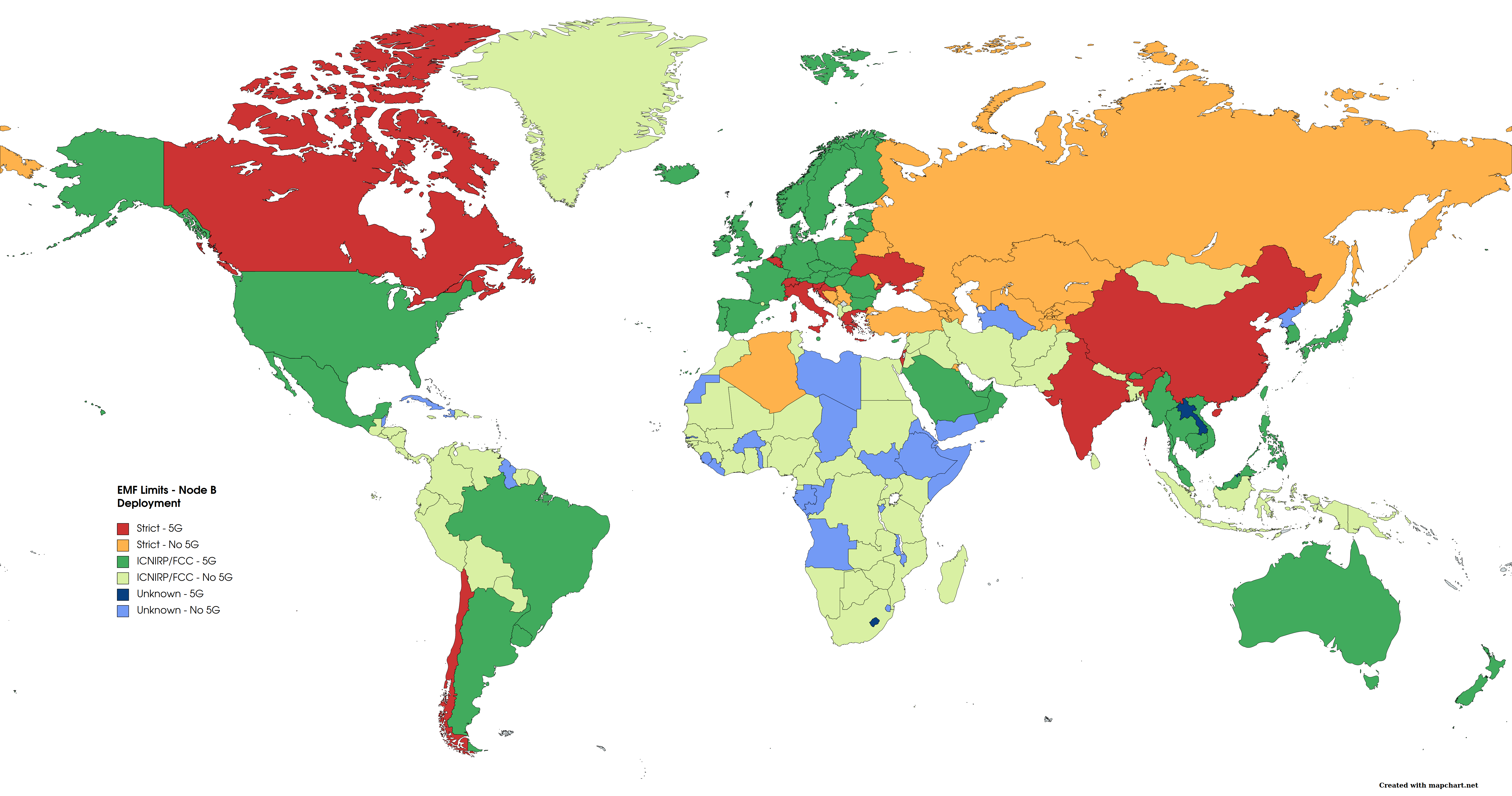}
    \caption{\color{black}World map with countries colored according to the \ac{EMF} limits on field strength (strict, \ac{ICNIRP}/\ac{FCC}, unknown) and \ac{gNB} deployment (5G, no 5G) - Most of micro states are omitted (Figure best viewed in colors) {- data retrieved in June 2020.}}
    \label{fig:worlmap5gnetwork}
\end{figure*}

Fig.~\ref{fig:limitscapfreqmatrix} reports the matrix of the possible combinations between adopted \ac{EMF} strength limits (strict, \ac{ICNIRP}, \ac{FCC}, none) and planned/auctioned 5G frequencies (below 1~[GHz], between 1~[Ghz] and 6~[Ghz], above 6~[Ghz], none). Each cell in the matrix is identified by a group ID (between 1 and 16). Each cell's color is proportional to the number of countries belonging to the group (from white to red), whose value is also reported in the cell. Moreover, we stress the fact that each country may be repeated across the following frequency options: below 1~[GHz], between 1~[Ghz] and 6~[Ghz], and above 6~[Ghz] (depending on the auctioned/planned 5G frequencies).

Several considerations hold by analyzing in more detail Fig.~\ref{fig:limitscapfreqmatrix}. First, as clearly shown by the intense red color of group 8, the majority of the countries adopt \ac{ICNIRP} limits without any plan (so far) to deploy the 5G technology. Second, when observing the countries deploying 5G \acp{gNB} with \ac{ICNIRP} limits (group 6), a typical setting is to target a mixture of coverage and capacity. Third, the highest health risks will be perceived in groups 1-3, i.e., the countries where 5G \acp{gNB} will be deployed under strict \ac{EMF} constraints. Interestingly, the cardinality of groups 1-3 is not negligible. Fourth, the number of countries with unknown limits and without any 5G \acs{gNB} implementation is also relevant (i.e., group 16). The population of these countries will perceive high health risks in case of future deployment of 5G networks. However, we also stress the fact that operators generally apply \ac{ICNIRP}/\ac{FCC} limits on a volunteer basis in countries without \ac{EMF} limits. 

\begin{figure*}[t]
    \centering
    \includegraphics[width=18cm]{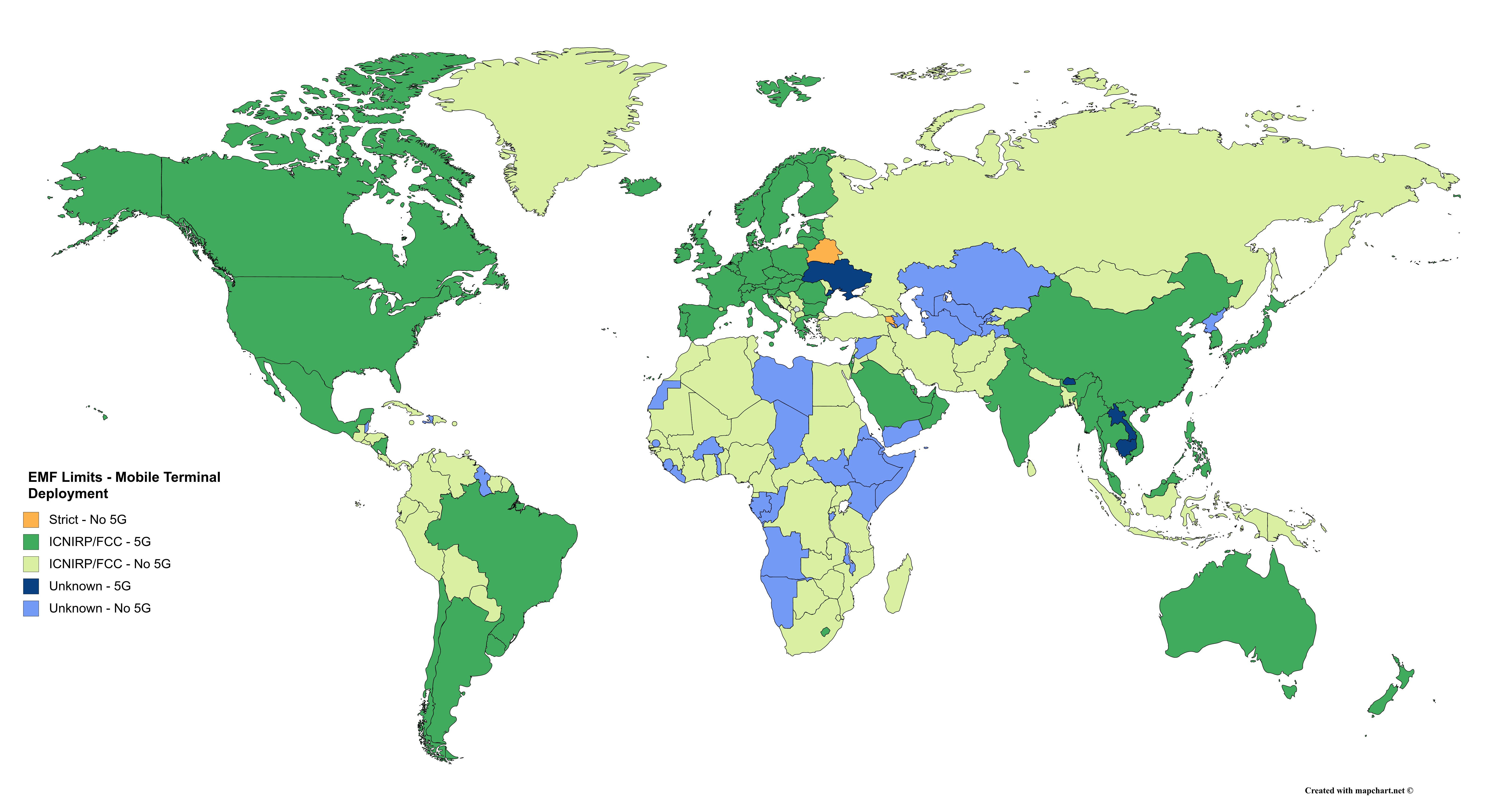}
    \caption{World map with countries colored according to the \ac{EMF} limits on \ac{SAR} and/or \ac{PD} (strict, ICNIRP/FCC, unknown) and \ac{UE} deployment (5G, no 5G) - Most of micro states are omitted (Figure best viewed in colors) {- data retrieved in June 2020.}}
    \label{fig:worlmap5device}
\end{figure*}

In a nutshell, a considerable {fragmentation} emerges when the different \ac{EMF} limits and the deployment of 5G \acp{gNB} are jointly considered. This {fragmentation} will also affect the perceived health risks associated with 5G in different countries.

\subsubsection{Impact of National Regulations on 5G \ac{UE} Adoption}

We then move our attention to the impact of national regulations on the adoption of 5G \ac{UE}. Similarly to the 5G \acs{gNB} case, we consider the following taxonomy for the \ac{EMF} limits: \textit{L1}) stricter than \ac{ICNIRP}, \textit{L2}) \ac{ICNIRP}-based, \textit{L3}) \ac{FCC}-based, or \textit{L4}) unknown.  Differently from 5G \acp{gNB}, in this case, we focus on the limits for \ac{UE} expressed in terms of \ac{SAR} and/or \ac{PD}. In line with the 5G \ac{gNB} analysis, we consider the same taxonomy (and same references) for the planned/auctioned 5G frequencies already used before, and namely: \textit{F1}) below 1~[GHz], \textit{F2}) between 1~[Ghz] and 6~[Ghz], \textit{F3}) above 6~[Ghz], or \textit{F4}) none. Fig.~\ref{fig:limitscapfreqmatrix_device} details the obtained matrix, with colors from white to red highlighting the number of countries falling in each group. By observing in more detail the figure, we can note that the group with the highest cardinality is composed of countries enforcing \ac{ICNIRP} limits for the \ac{UE} and no plans to adopt 5G devices (group 8).
Moreover, we remind that the number of countries implementing strict EMF limits for \ac{UE} is extremely limited in contrast to the 5G \ac{gNB} limits reported of Fig.~\ref{fig:limitscapfreqmatrix} (groups 1-4). On the other hand, the number of countries adopting \ac{ICNIRP}/\ac{FCC} limits with plans to exploit 5G \ac{UE} is consistently higher compared to the 5G \ac{gNB} case (groups 5-7, 9-11). Eventually, the number of countries with unknown limits and no plan to exploit 5G devices is not negligible (group 16), and similar to the 5G \acp{gNB} case.

In summary, the analysis conducted so far on \ac{UE} reveals that there is a lower {fragmentation} of limits across the countries compared to the 5G \acp{gNB} case. While this condition will decrease the health risks perceptions associated to 5G \ac{UE}, we need to remind that the number of countries without any plan to adopt 5G devices (with \ac{ICNIRP}/\ac{FCC} limits or with unknown limits) is very large.

\subsubsection{Population-based Analysis}

So far, we have conducted our analysis by counting the number of countries that fall within each group in the matrices of Fig.~\ref{fig:limitscapfreqmatrix_all}. However, a natural question emerges here: What is the impact of each group when we consider the population in each country? To answer this interesting question, we have weighed each country by its population (in percentage w.r.t. the global population), and we summed up the weighed countries falling in each group. Fig.~\ref{fig:limitscapfreqmatri_pop_weight}-\ref{fig:limitscapfreqmatrix_device_pop_weight} report the obtained matrices for the 5G \ac{gNB} and \ac{UE} cases, respectively. When the population weight is introduced for the 5G \ac{gNB} case (Fig.~\ref{fig:limitscapfreqmatri_pop_weight}), we can note that almost 40\% of the world population is living in countries where 5G is implemented as a mixture of coverage and capacity, under strict \ac{EMF} limits (group 2). As a result, the perception of health risks associated to 5G \acp{gNB} will be extremely high in those countries. However, we can note that the percentage of people living in countries with 5G \acp{gNB} implementations under \ac{ICNIRP}/\ac{FCC} limits is not negligible (groups 5-7, 9-11). Eventually, 26\% of the world population will be subject to \ac{ICNIRP} limits, without any implementation of 5G \acp{gNB} (group 8). Finally, the percentage of people living in countries with unknown limits is overall pretty limited, i.e., lower than 5\% (groups 13-16).  

We then repeat the population-based analysis by considering the \ac{UE}, as shown in Fig.~\ref{fig:limitscapfreqmatrix_device_pop_weight}. Interestingly, the outcome appears to be more homogeneous compared to the 5G \acp{gNB} population-based case (Fig.~\ref{fig:limitscapfreqmatri_pop_weight}) as well as the country-based analysis (Fig.~\ref{fig:limitscapfreqmatrix_all}). In particular, the percentage of people living in countries imposing \ac{ICNIRP}/\ac{FCC} \ac{UE} limits and having plans to deploy 5G networks (groups 5-7, 9-11 in Fig.~\ref{fig:limitscapfreqmatrix_device_pop_weight}) is predominant w.r.t. the unknown (groups 13-16) and strict cases (groups 1-4). In addition, we can note that, although the number of countries imposing FCC limits and exploiting 5G terminals appears to be limited (groups 5-7 in Fig.~\ref{fig:limitscapfreqmatrix_device}), their population weight is very large (groups 5-7 in Fig.~\ref{fig:limitscapfreqmatrix_device_pop_weight}), i.e., always higher than 20\% w.r.t the world population.
As a result, we believe that the risk conditions (either perceived or potential) will be avoided for most of the population when considering 5G \ac{UE}.

\subsubsection{Geographical Analysis}

In the following, we move our attention to the geographical {fragmentation} of \ac{EMF} limits and 5G implementation at a country level. We initially focus on the 5G \acp{gNB}. To this aim, Fig.~\ref{fig:worlmap5gnetwork} reports the world map, by differentiating with different colors: \textit{i}) countries enforcing strict \ac{EMF} limits with 5G \acp{gNB} implementation (coral color), \textit{ii}) countries enforcing strict \ac{EMF} limits without 5G \acp{gNB} implementation (dark yellow color), \textit{iii}) countries enforcing \ac{ICNIRP}/\ac{FCC} limits with 5G \ac{gNB} implementation (dark green color), \textit{iv}) countries enforcing \ac{ICNIRP}/\ac{FCC} limits without 5G \ac{gNB} implementation (light green color), \textit{v}) countries enforcing unknown \ac{EMF} limits with 5G \acp{gNB} (dark blue color), and \textit{vi}) countries enforcing unknown \ac{EMF} limits without any plan of 5G \acp{gNB} deployment (light blue). 

Several considerations hold by analyzing in more detail Fig.~\ref{fig:worlmap5gnetwork}. First, most of the countries in Europe are deploying/have plans to install 5G \ac{gNB}. However, the \ac{EMF} limits notably vary across Europe, with different countries imposing strict limits and other countries enforcing \ac{ICNIRP}/\ac{FCC} limits. Second, a large number of countries previously included in the Soviet Union are still implementing strict \ac{EMF} limits, without any plan to deploy 5G \acp{gNB}. Third, countries in the Mid East typically employ \ac{FCC}/\ac{ICNIRP} limits. However, the deployment of 5G \acp{gNB} is planned only in a limited subset of countries (e.g., Saudi Arabia, Oman, United Arab Emirates, Qatar, and Djibouti). However, in this region, there are also countries enforcing strict \ac{EMF} limits, e.g., Israel and Kuwait. Fourth, many countries in the Far East are planning to deploy 5G \acp{gNB}. However, a considerable variability in terms of \ac{EMF} limits is experienced in these countries, being China and India enforcing strict \ac{EMF} limits. Fourth, countries in Oceania generally enforce \ac{ICNIRP}/\ac{FCC} limits. Despite this fact, the implementation of 5G \acp{gNB} is limited to a subset of countries (e.g., Australia and New Zealand). Besides, we remark that the majority of the micro-states in Oceania (not reported in the map due to their limited land size) are applying \ac{ICNIRP}/\ac{FCC} limits without any 5G implementation. Fifth, many countries in Africa are enforcing \ac{ICNIRP}/\ac{FCC} limits. However, the number of states with unknown limits is far to be negligible. In any case, the implementation of 5G networks in this continent will be extremely limited. Sixth, Chile is the only country in South America with strict \ac{EMF} limits. On the other hand, \ac{ICNIRP}/\ac{FCC} limits are widely adopted in this continent. Moreover, the deployment of 5G \acp{gNB} will be realized in different countries of the continent. Seventh, North America will implement 5G networks by applying \ac{ICNIRP}/\ac{FCC} limits. The only country imposing (slightly) stricter limits than \ac{ICNIRP}/\ac{FCC} is Canada.

Summarizing, the geographical {fragmentation} of EMF limits and 5G implementation clearly emerges when considering the deployment of 5G \acp{gNB}. Consequently, the fear of the associated health effects will be very different across different countries. The lack of 5G implementations, coupled with the fact that in many countries the \ac{EMF} limits are still unset, is an obvious barrier for the deployment of 5G \acp{gNB} in the African continent. On the other hand, the differences in terms of \ac{EMF} limits in Europe, as well as in the Far East, will inevitably impact the deployment of 5G networks in these regions.

In the final part of our analysis, we consider the geographical {fragmentation} of \ac{EMF} limits and 5G implementations when considering the \ac{UE}. Fig.~\ref{fig:worlmap5device} reports the obtained outcome by adopting the same set of colors used in Fig.~\ref{fig:worlmap5gnetwork}. By mutually comparing  Fig.~\ref{fig:worlmap5device} against Fig.~\ref{fig:worlmap5gnetwork}, we can see that the \ac{EMF} limits for 5G \ac{UE} are much more homogeneous across the world compared to the ones adopted by \acp{gNB}. Specifically, the \ac{ICNIRP}/\ac{FCC} limits on \ac{SAR} and/or \ac{PD} of the \ac{UE} are adopted by most of the countries in the world. As previously pointed out, one exception is represented by Belarus and Armenia, which adopt \ac{UE} limits stricter than \ac{ICNIRP}/\ac{FCC}. In addition, different countries in Africa and Asia do not adopt any limit, and they are not planning to exploit 5G \ac{UE}. Therefore, the perceived health risks will be higher in such countries. In any case, we can conclude that the level of geographical {fragmentation} appears to be much more limited compared to the 5G \acp{gNB} case.

\subsection{Compliance Assessment of 5G Exposure}
\label{sec:compliace_assessment}

We now review the main methodologies to perform the compliance assessment of 5G exposure w.r.t. the \ac{RF} limits. We focus on the policies defined in the \ac{IEEE} \cite{ieee1,ieee2} and \ac{IEC} standards \cite{iec1,iec2,iec3,iec4}, as well as in the \ac{ITU} recommendations \cite{itutk52,itutk61,itutk70,itutk83,itutk91,itutk100,itutk121,itutk122} (complemented by supplements \cite{itutksupplement13,itutksupplement14,itutksupplement16}), which assume \ac{ICNIRP} 1998 \cite{ICNIRPGuidelines:98} or \ac{IEEE} C95.1 \cite{IEEEC95:05,IEEEC95:19} as underlying limits. However, we also point out that national regulations may impose specific rules for the compliance assessment of \ac{RF} exposure. For example, in Italy, local municipalities often impose a minimum distance constraint between a site hosting \ac{RF} equipment and a sensitive place (e.g., school, hospital, church). This constraint is additive w.r.t.  national {regulations} and international {guidelines}. In this work, however, we concentrate on international guidelines for the compliance assessment, due to multiple reasons. First, since 5G is a relatively new technology, the local regulations may not include revisions of the assessment of compliance tailored to 5G equipment. Second, it is expected that the compliance assessment policies defined by \ac{ITU}, \ac{IEC}, and \ac{IEEE} will be implemented in the national regulations in the forthcoming years.

In this context, a natural question arises: If the current regulations do not integrate the compliance assessment of 5G exposure, is it safe to install 5G equipment and to adopt 5G \ac{UE} at present? The answer is affirmative: current \ac{RF} limits are already defined for all the frequencies (including the ones used by 5G). Besides, current regulations for the compliance assessment can also be applied to 5G devices by introducing very conservative (and worst-case) assumptions, which always guarantee the population's safety. For example, the installation of 5G \acp{gNB} used for experimental trials in Italy is authorized by assuming an ideal maximum power that is radiated when all the beams are simultaneously activated in all the directions \cite{ecoscienza}. However, we  stress the importance of revising the current regulations by considering the realistic modeling and measurement of 5G features, to better assess the 5G exposure.\footnote{We refer the interested reader \cite{ecoscienza} for an overview of the modifications planned for the Italian country.}

Tab.~\ref{tab:compliance_assessment} overviews the main methodologies for the compliance assessment of 5G exposure. We divide the standards based on the following indicators: \textit{i}) simulation-based procedures for 5G \ac{gNB}, \textit{ii}) simulation-based procedures for 5G \ac{UE}, \textit{iii}) measurement-based procedures for 5G \ac{gNB}, and \textit{iv}) measurement-based procedures for 5G \ac{UE}. Clearly, the methodologies in \textit{i})-\textit{ii}) can be applied during the planning phase of the 5G network and during the design of \ac{UE}. On the other hand, the procedures in \textit{iii})-\textit{iv}) are useful in order to assess the compliance of 5G networks under operation and/or already designed \ac{UE} that have to be tested/monitored. In addition, the table reports a brief summary for each document, by detailing the features that are relevant to 5G exposure.

\renewcommand{\arraystretch}{1.2}
\begin{table*}
    \caption{Compliance assessment methodologies for 5G exposure.}
    \label{tab:compliance_assessment}
    \centering
    \footnotesize
    \begin{tabular}{|c|c|c|c|c|c|p{8cm}|}
    \hline
       \rowcolor{Lightred} \multicolumn{2}{|c|}{} & \multicolumn{4}{|c|}{\textbf{Assessment Methodology}} & \multicolumn{1}{|c|}{} \\    
       \rowcolor{Lightred} \multicolumn{2}{|c|}{} & \multicolumn{2}{|c|}{\textbf{Simulation-Based}} & \multicolumn{2}{|c|}{\textbf{Measurement-Based}} & \multicolumn{1}{|c|}{} \\
       \rowcolor{Lightred} \multicolumn{2}{|c|}{\multirow{-3}{*}{\textbf{Document}}} & \multicolumn{1}{|c|}{\textbf{\ac{gNB}}} & \multicolumn{1}{|c|}{\textbf{\ac{UE}}} & \multicolumn{1}{|c|}{\textbf{\ac{gNB}}} & \multicolumn{1}{|c|}{\textbf{\ac{UE}}} & \multicolumn{1}{|c|}{\multirow{-3}{*}{\textbf{Relevance for 5G Exposure}}} \\ 
       \hline
       & & & & & & - Reproducible and conservative measurement procedures of \ac{PD};\\
       & & & & & & - Multiple transmitters or antennas for \ac{UE};\\
       & & & & & & - Different \ac{UE} positions (including in front of the face);\\
       & \multirow{-4}{*}{P63195-1 \cite{ieee1}}& & & & \multirow{-4}{*}{Yes} & - Frequencies from 6~[GHz] to 300~[GHz];\\
        \cline{2-7}
\rowcolor{Linen} \cellcolor{White}  & & & & &  & - Conservative, repeatable and reproducible computation procedures of \ac{PD};\\
\rowcolor{Linen} \cellcolor{White}  & & & & & & - Multiple transmitters or antennas for \ac{UE};\\
\rowcolor{Linen} \cellcolor{White} & & & & & & - Different \ac{UE} positions (including in front of the face);\\
\rowcolor{Linen} \cellcolor{White}  \multirow{-8}{*}{\begin{sideways}\acs{IEEE}\end{sideways}}  & \multirow{-5}{*}{P63195-2 \cite{ieee2}}& & \multirow{-5}{*}{Yes}  & & & - Frequencies from 6~[GHz] to 300~[GHz];\\
       \hline
        &   & & & & &  - Procedures for determining the field strength and \ac{SAR} in the vicinity of \ac{gNB};\\
        & & & & & & - \ac{RF} source may be a single antenna or a set of antennas; \\
        &  \multirow{-4}{*}{62232 \cite{iec1}} & \multirow{-4}{*}{Yes} & & \multirow{-4}{*}{Yes} & & - \ac{gNB} frequencies up to 100~[GHz] are considered;\\ 
           \cline{2-7}
\rowcolor{Linen} \cellcolor{White}&  & & & & & - Procedures for measuring the \ac{UE} \ac{SAR}; \\
\rowcolor{Linen} \cellcolor{White}& & & & & & - Frequencies up to 6~[GHz] are considered;\\
 \rowcolor{Linen} \cellcolor{White}  & \multirow{-3}{*}{62209 \cite{iec3,iec4}} & & & & \multirow{-3}{*}{Yes} &  - Different \ac{UE} positions (including in front of the face); \\
         \cline{2-7}
           & & & & & & - Case studies implementing the procedures detailed in \ac{IEC} 62232 \cite{iec1}; \\
           & & & & & & - Considered metrics include incident field, \ac{SAR} and \ac{PD};\\
           & & & & & & - Different categories of \ac{gNB};\\
 \multirow{-12}{*}{\begin{sideways}\acs{IEC}\end{sideways}}          & \multirow{-5}{*}{62669 \cite{iec2}} & \multirow{-5}{*}{Yes} & &\multirow{-5}{*}{Yes} &  & - One case study includes a massive \ac{MIMO} compliance assessment;\\
%           \cline{2-7}
%\rowcolor{Linen} \cellcolor{White}        & 62311 \cite{iec5} & & & & & TBD \\
        \hline
\rowcolor{Linen} \cellcolor{White}       & & & & & & - {Reference} \ac{IEC} 62232 \cite{iec1} and \ac{IEC} 62209 \cite{iec3,iec4}; \\
\rowcolor{Linen} \cellcolor{White}       & \multirow{-2}{*}{K.52 \cite{itutk52}} & \multirow{-2}{*}{Yes} &  & & \multirow{-2}{*}{Yes} &  - \ac{SAR} methodologies up to 6~[GHz] \\
       \cline{2-7}
 &  & & & & &  - {Reference} \ac{IEC} 62232 \cite{iec1}; \\
 & \multirow{-2}{*}{K.61 \cite{itutk61}} & \multirow{-2}{*}{Yes} & & \multirow{-2}{*}{Yes} & & - Frequencies up to 300~[GHz] are considered.\\
       \cline{2-7}
\rowcolor{Linen} \cellcolor{White}       & & & & & & - {Reference} \ac{IEC} 62232 \cite{iec1}; \\
\rowcolor{Linen} \cellcolor{White}       & \multirow{1}{*}{K.70 \cite{itutk70}} & \multirow{1}{*}{Yes} & & & & - A simplified method for the calculation of the compliance distances (which takes into account also 5G) and an \ac{EMF} software (which needs to be updated for 5G sites) are included.\\
       \cline{2-7}
 & & & & & &    - Site evaluation procedures already include frequencies used by 5G \\   
       & \multirow{-2}{*}{K.83 \cite{itutk83}} & & & \multirow{-2}{*}{Yes} & &  equipment. \\
       \cline{2-7}
\rowcolor{Linen} \cellcolor{White}       & & & & & & - Based on \ac{IEC} 62232 \cite{iec1}, \ac{IEC} 62209 \cite{iec3,iec4};\\
\rowcolor{Linen} \cellcolor{White}       & & & & & & - Frequencies up to 300~[GHz] are considered; \\
\rowcolor{Linen} \cellcolor{White}       & & & & & & - Typical sources of radiation do not include 5G; \\
\rowcolor{Linen} \cellcolor{White}        & \multirow{-4}{*}{K.91 \cite{itutk91}} & \multirow{-4}{*}{Yes} &  & \multirow{-4}{*}{Yes} & \multirow{-4}{*}{Yes} & - Examples of real measurements do not include 5G equipment.\\
          \cline{2-7}
     & & & & & &  - Conservative \ac{EIRP} computation (applicable also to 5G \ac{gNB}); \\    
    & & & & & &  - Introduced the idea of continuous monitoring of maximum transmitted power or \ac{EIRP}; \\ 
    & & & & & &  - Simplified assessment procedures covering frequencies up to 40~[GHz]; \\ 
 \multirow{-7}{*}{\begin{sideways}\ac{ITU}-T\end{sideways}}       & \multirow{-5}{*}{K.100 \cite{itutk100}} & \multirow{-5}{*}{Yes} & & \multirow{-5}{*}{Yes} & &  - Guidance to compute the power density for different technologies given selective measurements does not include 5G equipment.\\
        \cline{2-7}
\rowcolor{Linen} \cellcolor{White}        & & & & & & - General document integrating previous ITU-T K recommendations;\\
\rowcolor{Linen} \cellcolor{White}       & \multirow{-2}{*}{K.121 \cite{itutk121}} & \multirow{-2}{*}{Yes} & & \multirow{-2}{*}{Yes} & & - Useful as a starting reference for 5G operators and governments.\\
        \cline{2-7}
             & & & & & & - Useful for operators performing maintenance on the \ac{gNB}; \\
             & & & & & & - Two sub-6 GHz frequencies are considered;\\
       & \multirow{-2}{*}{K.122 \cite{itutk122}} & \multirow{-2}{*}{Yes} & & & & - Two case studies based on point-to-point links exploiting frequencies comparable to mm-Waves are also reported.\\
          \cline{2-7}
\rowcolor{Linen} \cellcolor{White}       & K Supplement 13 \cite{itutksupplement13} & & & & Yes & - Description of \ac{IEC} 62209 \cite{iec3,iec4}. \\
          \cline{2-7}
    & & & & & & - General discussion about the impact of \ac{EMF} limits stricter than\\
       & \multirow{-2}{*}{K Supplement 14 \cite{itutksupplement14}} & \multirow{-2}{*}{Yes} & & & &  \ac{ICNIRP} on the deployment of 5G networks. \\
          \cline{2-7}
\rowcolor{Linen} \cellcolor{White}          & & & & & & - Based on \ac{IEC} 62669 \cite{iec2}, \ac{IEC} 62232 \cite{iec1} and \ac{IEC} 62209 \cite{iec3,iec4};\\
  \rowcolor{Linen} \cellcolor{White}             & & & & & & - Indications of future releases of relevant \ac{IEC} documents are provided;\\
\rowcolor{Linen} \cellcolor{White}       & \multirow{-4}{*}{K Supplement 16 \cite{itutksupplement16}} & \multirow{-4}{*}{Yes} &  &  \multirow{-4}{*}{Yes} &  & - Compliance assessment tailored to 5G equipment. \\
    \hline
       \rowcolor{Lightred} \multicolumn{2}{|c|}{}  & \multicolumn{1}{|c|}{\textbf{\ac{gNB}}} & \multicolumn{1}{|c|}{\textbf{\ac{UE}}} & \multicolumn{1}{|c|}{\textbf{\ac{gNB}}} & \multicolumn{1}{|c|}{\textbf{\ac{UE}}} &  \multicolumn{1}{|c|}{} \\ 
       \rowcolor{Lightred} \multicolumn{2}{|c|}{} & \multicolumn{2}{|c|}{\textbf{Simulation-Based}} & \multicolumn{2}{|c|}{\textbf{Measurement-Based}} & \multicolumn{1}{|c|}{} \\
       \rowcolor{Lightred} \multicolumn{2}{|c|}{\multirow{-3}{*}{\textbf{Document}}}  & \multicolumn{4}{|c|}{\textbf{Assessment Methodology}} & \multicolumn{1}{|c|}{\multirow{-3}{*}{\textbf{Relevance for 5G Exposure}}}  \\
       \hline
    \end{tabular}
\end{table*}
\renewcommand{\arraystretch}{1.0}

Several considerations hold by analyzing the methodologies in Tab.~\ref{tab:compliance_assessment} from the perspective of 5G communications engineering. First, \ac{IEEE} define in  \cite{ieee1,ieee2} the methodologies to assess the compliance of \ac{UE} at both simulation and measurement levels. Specifically, the \ac{PD} is taken as a reference metric, with a range of frequencies between 6~[GHz] and 300~[GHz]. Therefore, the procedures in  \cite{ieee1,ieee2} are of high interest in the context of 5G communications, and in particular, for \ac{UE} exploiting mm-Waves.  Also, different \ac{UE} positions (including the one in front of the head) are taken into account, thus matching the actual usage of \ac{UE} during gaming, social networking, and video streaming. Second, the \ac{IEC} standards \cite{iec1,iec3,iec4} are focused on the assessment of compliance of 5G \ac{gNB} and 5G \ac{UE}. More in-depth, \ac{IEC} 62232 \cite{iec1} targets the assessment of compliance of {\ac{SAR}/\ac{PD} and electric field strength} from \ac{gNB}, by considering approaches based on simulation and/or measurement of the exposure. In addition, the frequencies taken into account include mm-Waves (up to 100~[GHz]).  Eventually, \ac{IEC} 62232 \cite{iec1} is complemented by \ac{IEC} 62669 \cite{iec2}, which includes a set of representative case studies that implement the procedures of \cite{iec1}. The document \cite{iec2} is particularly relevant for the compliance assessment of 5G \ac{gNB} exposure. For example, one case study is tailored to the compliance assessment of a 5G \ac{MIMO} \ac{gNB}. Moreover, different types of \acp{gNB} are taken into account (e.g., macro cells and small cells). Moreover, the procedures in \cite{iec2} include the evaluation of \ac{PD} in addition to \ac{SAR}.  Hence, they can be directly mapped to the corresponding \ac{PD}-based limits defined by the international organizations for 5G \ac{gNB} exposure. Focusing on the \ac{UE}, the \ac{IEC} 62209 documents \cite{iec3,iec4} detail the procedures for the compliance assessment of \ac{UE} \ac{SAR}. In this case, frequencies up to 6~[GHz] are considered (thus excluding mm-Waves). In line with the \ac{IEEE} procedures \cite{ieee1,ieee2} different \ac{UE} positions are taken into account by \cite{iec3,iec4}. However, in contrast to \cite{ieee1,ieee2}, the considered metric is \ac{SAR} (and not \ac{PD}).

In the following, we move our attention to the \ac{ITU}-T recommendations \cite{itutk52,itutk61,itutk70,itutk83,itutk91,itutk100,itutk121,itutk122}, which are reported in the bottom of Tab.~\ref{tab:compliance_assessment}. In general, \ac{ITU}-T provides brief documents, which can be used by the governments in order to build national specific regulations for the compliance assessment of 5G exposure. For this reason, most of \ac{ITU} documents refer to the \ac{IEC} standards for the details about the compliance assessment procedures. Moreover, the \ac{ITU} documents integrate the previous standards by: \textit{i}) defining simplified installation procedures for \ac{gNB}, based on different installation types \cite{itutk52,itutk100}, \textit{ii}) defining clear rules to differentiate between far field and near field exposure assessment \cite{itutk61}, \textit{iii}) proposing mitigation techniques to reduce the exposure in case the limits are not met \cite{itutk52,itutk70}, \textit{iv}) providing software and simplified models for the computation of the exposure in the different field regions \cite{itutk70}, \textit{v}) defining solutions to monitor the \ac{EMF} levels \cite{itutk83,itutk100}, with both broadband and frequency-selective measurements, \textit{vi}) providing procedures to compute the actual maximum \ac{EIRP} \cite{itutk100}, which is then used in the compliance assessment procedures (e.g., the \ac{IEC} ones), \textit{vii}) providing high level views of the compliance assessment that may be useful for decision-makers \cite{itutk121} and \textit{viii}) providing information for the assessment procedures in the vicinity of base stations \cite{itutk122} (which can be applied to workers operating on the site for maintenance operations).

Finally, we review the \ac{ITU}-T supplements \cite{itutksupplement13,itutksupplement14,itutksupplement16}, which also include relevant information for the compliance assessment of 5G exposure. Specifically, K Supplement 13 \cite{itutksupplement13} is tailored to the identification of the factors to determine the \ac{SAR} from \ac{UE}{, based on \ac{IEC} 62209-1 and -2}. K Supplement 14 \cite{itutksupplement14} is instead tailored to the evaluation of the impact of national limits stricter than \ac{ICNIRP} and/or \ac{IEEE} on the planning of 4G and 5G networks. In particular, strict regulations introduce several negative aspects, such as difficulty in using the full available spectrum, a limitation in the network densification, and a significant barrier to the technology innovation. Eventually, K Supplement 16 \cite{itutksupplement16} is devoted to the compliance assessment of exposure from 5G \ac{gNB}, by providing indications to the relevant \ac{IEC} standards, as well as by including different case studies based on 5G (e.g., a massive \ac{MIMO} \ac{gNB} and a small cell).

\renewcommand{\arraystretch}{1.2}
\begin{table*}
    \caption{Taxonomy of health risks associated with 5G features and relevant aspects in the context of communications engineering.}
    \label{tab:health_risks_taxonomy}
    \centering
    \footnotesize
    \begin{tabular}{|c|c|c|}
    \hline
       \rowcolor{Lightred} \textbf{5G Feature} & \textbf{Relevant Aspects} & \textbf{References} \\
       & Increase of power and number of radiating elements & \cite{WerKniIsk:19,KelHel:19,ShiThiVer:19,FraColGriMig:20,XuZhaHe:19} \\
       \cline{2-3}
       & Introduction of statistical exposure models & \cite{BarWebWil:18,ThoFurTor:17,wiart2018statistical}\\
       \cline{2-3}
       \multirow{-3}{*}{Extensive adoption of massive \ac{MIMO} and beamforming} & Measurement of exposure levels & \cite{PawKraZur:19,Migliore:19,colombi2019assessment,WerKniIsk:19,AerVerVanTor:19,FraCotGriPavMig:20,FraColGriMig:20}\\
        \hline
\rowcolor{Linen}        & Computation of \ac{RF} pollution at selected locations & \cite{ChiaraviglioBiaBleFiore:19}\\
         \cline{2-3}
\rowcolor{Linen}       & Computation of average received power & \cite{5gwhichrisks}\\
        \cline{2-3}
\rowcolor{Linen}       \multirow{-3}{*}{Densification of 5G sites over the territory} & Impact of strict \ac{EMF} limits on densification & \cite{itutksupplement14,ChiaraviglioGala:19} \\
       \hline
        & Deployment status of mm-Waves & \cite{5gwhichrisks,XuZhaHe:19}\\
       \cline{2-3}
      \multirow{-2}{*}{Adoption of frequencies in the mm-Wave bands} & mm-Waves device exposure evaluation & \cite{he2020fast,he2020incident,XuZhaHe:19}\\
     \hline
\rowcolor{Linen}        & Maximum output power levels & \cite{etsitr3,etsitr2} \\
  \cline{2-3}
\rowcolor{Linen}       \multirow{-2}{*}{Connection of millions of IoT devices} & Data-rate and delay requirements & \cite{iotstandards,iotfuture} \\
\hline
        & Saturation levels of legacy pre-5G networks & \cite{PerCarBar:18,PerCarEliSum:18,chiaraviglio2020safe} \\
  \cline{2-3}
       & Impact of radio and TV broadcasting & \cite{tell1980population,tell2014survey,chiaraviglio2020safe} \\
         \cline{2-3}
       \multirow{-3}{*}{Coexistence of 5G with legacy technologies} & Interaction with weather satellites & \cite{marcus20195g,nasanoaa,nasanoaa2} \\
       \hline
     \end{tabular}
\end{table*}  
\renewcommand{\arraystretch}{1.0}        

Summarizing, different organizations (\ac{IEEE}, \ac{IEC}, \ac{ITU}) provide guidelines (or draft of guidelines) for the compliance assessment of 5G exposure. Moreover, a great effort is currently devoted to the compliance assessment of exposure from \ac{gNB} (with both simulation-based and measurement-based approach). When focusing on the \ac{UE}, most of the approaches are based on \ac{SAR} and \ac{PD} measurement. Although revisions of different procedures (e.g., \cite{itutk91,itutk100}) are still needed to integrate case studies tailored to 5G, we can conclude that the compliance assessment of 5G exposure is overall already defined.
%%%%%%%%%%%%%%%%%%%%%%%%%%%%%%%%%%%%%%%%%%%%%%%%%%%%%%%%%%%%%%%%%%%%%%%%%%%%%%%%%%%%%%%%%%%%%%%%%%%%%%%%%%%%%%%%%%%%%%%%%%%%%%%%%%%%%%%%%%%%%%%%%%%%%%%%%%
%%%%%%%%%%%%%%%%%%%%%%%%%%%%%%%%%%%%%%%%%%%%%%%%%%%%%%%%%%%%%%%%%%%%%%%%%%%%%%%%%%%%%%%%%%%%%%%%%%%%%%%%%%%%%%%%%%%%%%%%%%%%%%%%%%%%%%%%%%%%%%%%%%%%%%%%%%%%%%%%%%%%
\section{Health Risks Associated with 5G Features}
\label{sec:health_risks_5G_features}
%{5G communication system and its implication on Human health }

In this section, we analyze the health risks associated with key 5G features from the communications engineering perspective. Our goal is, in fact, not to survey the entire set of 5G features, but to concentrate on the ones that trigger health concerns among the population. More in-depth, we focus on the following controversial aspects:
\begin{itemize}
\item extensive adoption of massive \ac{MIMO} and beamforming;
\item densification of 5G sites over the territory;
\item adoption of frequencies in the mm-Wave bands;
\item connection of millions of \ac{IoT} devices;
\item coexistence of 5G with legacy technologies.
\end{itemize}
Since our goal is tailored to the communications engineering perspective, we consider health risks in terms of exposure generated by 5G \ac{gNB} and by 5G \ac{UE}. To this aim, Tab.~\ref{tab:health_risks_taxonomy} reports the considered 5G features, the corresponding aspects in the context of 5G communications, together with the relevant references. In the following, we provide more details about each feature and each work reported in Tab.~\ref{tab:health_risks_taxonomy} .

\subsection{Extensive Adoption of Massive \ac{MIMO} and Beamforming}

We initially analyze the impact of massive \ac{MIMO} and beamforming on the exposure from 5G devices.
We focus on the following features: \textit{i}) increase of power and number of radiating elements, \textit{ii}) introduction of statistical exposure models, \textit{iii}) measurement of exposure levels.

\subsubsection{Increase of Power and Number of Radiating Elements}

%5G will rely on two key functionalities, namely massive \ac{MIMO} and beamforming \cite{BocHeathPop:14}. 

When considering 5G devices implementing \ac{MIMO} and beamforming, two essential differences emerge w.r.t. legacy ones, and namely: \textit{i}) a general increase in the maximum output power,\footnote{{Clearly, the maximum output power depends on the equipment class (e.g, macro cell vs. small cells), and therefore the maximum output power may be subject to strong variations.}} and \textit{ii}) an increase in the number of radiating elements.  Focusing on the total power radiated by 5G \ac{gNB}, data sheets of macro equipment available in the market report a maximum output power equal to 200~[W] \cite{air5121}. On the other hand, 4G base stations {typically} radiate a consistent lower amount of power, e.g., in the order of 10-100~[W] \cite{bsoutputpower}.\footnote{{Clearly, deviations from these numbers are also possible for 4G base stations.}} Therefore, a natural question arises: Is this increase of maximum power directly translated into an increase of exposure, and consequently, in an increase of the health risks?
To answer this question, we need to recall how 5G \acp{gNB} will exploit \ac{MIMO}. In fact, the \ac{MIMO} technology is not new, and it has been in use for several years \cite{WerKniIsk:19}. The main idea of \ac{MIMO} is to exploit multiple antennas taking advantage of independent propagation paths to improve the transmission. With massive \ac{MIMO} the number of antenna elements is radically increased (with a typical size of more than 64 elements) to further improve the system capacity. 

In general, spatial multiplexing and beamforming are two key features implemented in 5G systems exploiting massive \ac{MIMO}. As clearly detailed by \cite{WerKniIsk:19}, spatial multiplexing allows transmitting independent data over multiple uncorrelated paths. In contrast, beamforming allows concentrating the power of each antenna element on a specific user who needs to be served. Thanks to such features, the radiation pattern of 5G \ac{gNB} is radically different compared to those of legacy technologies. In particular, the radiation pattern implemented by 4G base stations with \ac{MIMO} is mostly static, i.e., with fixed beams over the territory. On the other hand, 5G \acp{gNB} exploiting massive \ac{MIMO} adapt radiation patterns that are dynamically varied in space and time, i.e., to match the traffic conditions and/or the positioning of the users over the territory. Therefore, although the total power consumption of a 5G \ac{gNB} is consistently higher than the one of a 4G \ac{gNB}, the exposure exhibits a different pattern in time and space. As a consequence, the total power that is radiated by a 5G \ac{MIMO} \ac{gNB} is not spread over the entire coverage area, but it tends to be concentrated on specific portions of the territory \textit{and} wisely modulated based on the network and traffic conditions. For example, according to \cite{KelHel:19} (and references therein), the current exposure from 5G \ac{gNB} is four times lower than the maximum exposure in 95\% of all cases. In any case, however, it is very unlikely that the whole power radiated by a 5G \ac{gNB} will concentrate on a single beam with the maximum antenna gain for a time period sufficiently long, i.e., in the order of minutes. Therefore, despite the increase in the maximum radiated power of 5G \ac{gNB}, the expected exposure from 5G \ac{gNB} will be in line (and in general lower) compared to legacy technologies.

Eventually, the authors of \cite{ShiThiVer:19} present a numerical approach for assessing the exposure of massive \ac{MIMO} \ac{gNB} in indoor environments, by combining a ray-tracing technique and the time-domain method to estimate the \ac{SAR} on a phantom. The authors then compute the maximum power admissible for a 5G \ac{gNB} to ensure that the estimated \ac{SAR} is below the \ac{ICNIRP} limit of 2~[W/kg] at a distance of 8~[m]. Interestingly, the maximum power per antenna is at most equal to 110~[W] in the worst scenario. However, since the considered environment is an indoor scenario, the 5G \ac{gNB} can be implemented with a small cell (and not with a macro one), thus being able to employ an output power consistently lower than the maximum values extracted by the authors. Therefore, the perceived health risks are minimized in this case.

In parallel with the increase of power, another aspect that characterizes 5G devices is the increase in the
the number of radiating elements. In general, the size of 5G \ac{gNB} tends to be larger than that of legacy technologies, due to the need to host the circuitry to power the antenna elements\cite{FraColGriMig:20}. Although this aspect is not a problem in cellular deployments (especially for roof-mounted and poles installations), the increase of size may be associated to a higher exposure. Focusing on 5G \ac{UE}, it is expected that multiple antenna elements (up to 8) will be exploited by terminals implementing full 5G functionalities \cite{XuZhaHe:19}. However, no change in the size of the \ac{UE} is planned. Therefore, the expected impact on the user side in terms of perceived health risks will be marginal.

\subsubsection{Introduction of Statistical Exposure Models}

Traditional methods to estimate the exposure from base stations are based on very conservative assumptions, including maximum transmission power and static beams in all the covered area directions. Although such assumptions are, in general, valid for legacy technologies, they tend to be overly conservative when considering 5G \ac{gNB} \cite{BarWebWil:18}. In general, the application of conservative assumptions to estimate the exposure from 5G \ac{gNB} is detrimental for the health risks due to two main reasons. On one side, the exclusion zone of each 5G \ac{gNB} tends to be very large, i.e., in the order of several dozen meters \cite{5gimpact}. On the other hand, the predicted exposure levels tend to be pretty high \cite{5gimpact}, thus triggering health concerns by the population. Therefore, the exposure estimation of \ac{5G} \ac{gNB} is based on the introduction of statistical models \cite{ThoFurTor:17,BarWebWil:18,wiart2018statistical}, which allow on one side to better assess the size of the exclusion zone of the \ac{gNB}, and on the other one to estimate the predicted exposure levels over the territory in a more realistic way. 

In this context, \cite{ThoFurTor:17} introduces a statistical model to take into account multiple factors, such as the \ac{gNB} utilization, the time-division duplex, the scheduling time, as well as the spatial distribution of the users in the covered area. Results show that, by applying the presented model, the largest maximum power is less than 15\% w.r.t. the corresponding theoretical one. Consequently, the exclusion zone can be reduced by a factor of 2.6 compared to a traditional methodology. Similarly,  \cite{BarWebWil:18} presents a statistical approach by leveraging on the three-dimensional spatial channel model standardized by \ac{3GPP}. Results show that the exclusion zone of a massive \ac{MIMO} \ac{gNB}, computed through the statistical model, is reduced by half compared to the ones obtained by a traditional approach (i.e., not based on statistical parameters). Eventually, the authors of \cite{wiart2018statistical} compute the probability that multiple antenna elements of 5G massive \ac{MIMO} \ac{gNB} are radiating with the actual maximum power over the same point of the territory and at the same time. Results show that the probability of this event is clearly lower than the case with a single antenna element.

In summary, the high dynamicity introduced in power radiated by 5G \ac{gNB} implementing \ac{MIMO} and beamforming imposes to consider statistical models to more realistically compute both the exposure levels and the size of exclusion zones compared to traditional approaches. This step could be beneficial to reduce the health risks perceived by the population.

\subsubsection{Measurement of Exposure Levels}

A third aspect that has to be considered is the measurement of exposure levels due to the large adoption of \ac{MIMO} and beamforming features. Focusing on \ac{gNB}, the authors of \cite{PawKraZur:19} point out that the methodologies used to measure the exposure in legacy networks are not always suitable for assessing the exposure of 5G \acp{gNB} exploiting massive \ac{MIMO} and beamforming. In general, such features may cause uncertainties in the estimation of the field strength, according to \cite{Migliore:19}. This aspect may be an issue for the health risks that are perceived by the population. However, as also suggested by \cite{Migliore:19}, a possible solution could be to force the system to generate a maximum toward the direction of the measurement position. Obviously, this step requires either to position one or more \ac{UE} in the vicinity of the measurement probe and/or to perform the measurement in cooperation with the operator {owning} the 5G \ac{gNB}.

In general, the measurement procedure of 5G \ac{gNB} involves either wide-band probes operating on a given range of frequencies, or narrow-band probes that are able to retrieve information on the field strength on a set of selected frequencies. Focusing on the former methodology, the authors of \cite{colombi2019assessment} measure the output power levels of a 5G \ac{gNB}, by exploiting massive \ac{MIMO} in an operational network. Interestingly, the time-averaged power transmitted on a given beam direction is lower than the maximum theoretical output power. In addition, the maximum field strength measured in the proximity of the 5G \ac{gNB} represents a tiny fraction (lower than 6\%) compared to the one that is estimated by assuming a maximum power transmission. In line with \cite{colombi2019assessment}, the authors of \cite{WerKniIsk:19} perform a measurement campaign of the field strength of a 4G base station implementing massive \ac{MIMO}. Although the considered base station belongs to legacy technologies, the measured exposure levels are meaningful in the context of 5G, thanks to the adoption of massive \ac{MIMO} in the considered scenario. The obtained results demonstrate that, even when the base station is fully loaded, the measured field strength is a small ratio (lower than 17\%) compared to the maximum \ac{ICNIRP} limit for occupational exposure. Therefore, both the works \cite{colombi2019assessment,WerKniIsk:19} indicate exposure levels lower than the theoretical ones, and in general lower than the limits. Although further assessments are required (e.g., by extending the measurement to other operational networks and to different traffic conditions), current results indicate that the exposure from \ac{gNB} implementing massive \ac{MIMO} will be overall limited, thus minimizing the overall risks for the population.

In the following step, we focus on the assessment of exposure through narrow-band measurements. More in detail, \cite{AerVerVanTor:19} aims at identifying the Synchronization Signal Block, in order to assess the power density carried by its resources and to finally extrapolate the theoretical maximum exposure level. The authors consider a location at around 60~[m] from the 5G \ac{gNB} (in \ac{LOS} conditions), at a close distance (around 7~[m]) from the \ac{UE}. Also, a constant fixed beam, oriented towards the position of the \ac{UE}, is enforced at the 5G \ac{gNB} site. Interestingly, the measured exposure is at most equal to 3.716~[V/m] in the worst case (achieved by imposing a 100\% of downlink traffic load). By applying the methodology detailed in \ac{IEC} 62232 \cite{iec1}, a theoretical maximum field strength of 5.537~[V/m] is obtained. It is important to remark that this value is lower than the maximum limit for countries adopting \ac{ICNIRP}/\ac{FCC}-based regulations (see Tab.~\ref{Tab:EMFBS}), and thus being able to limit the health risks perceived by the population. On the other hand, this value is very close or above the maximum limit for different countries imposing regulations stricter than \ac{ICNIRP}/\ac{FCC} (reported in Fig.~\ref{fig:strict_limits_mm_wave}). Clearly, in such countries, the associated health risks of 5G \ac{gNB} deployments similar to \cite{AerVerVanTor:19} may be highly perceived by the population.

Moreover, the authors of \cite{FraCotGriPavMig:20} point out that the maximum \ac{EMF} level in a given location is a combination of three factors, namely: \textit{i}) the total number of subcarriers of the carrier, which depends on the signal bandwidth and the numerology,  \textit{i}) the fraction of the signal frame reserved for downlink transmission, \textit{iii}) the maximum EMF level measured for a single resource element, which in turns depends on different other metrics (including a factor depending on the serving beam). Interestingly, the importance of adopting \ac{UE} forcing full load traffic in the vicinity of the measurement point is stressed by the authors. Eventually, the authors of \cite{FraColGriMig:20} define the experimental procedures for estimating the relevant factors associated with time division duplexing and beam sweeping, which are then used to extrapolate the maximum field strength from the exposure measurements.

Summarizing, current works tailored to the measurement of the incident \ac{EMF} field strength from 5G \ac{gNB} exploiting \ac{MIMO} and beamforming reveal that the overall exposure is limited and in general lower than the maximum theoretical values. Although further efforts are needed, e.g., to extend the outcomes by measuring the \ac{EMF} over different operational networks and different traffic conditions, the current literature indicates that the health risks from exposure can be minimized. However, we stress the fact that the measured \ac{EMF} is highly influenced by the traffic and the user activity  in the proximity of the measurement probe. Therefore, it is of fundamental importance to setup a proper (and meaningful) measurement scenario.

\subsection{Densification of 5G sites over the Territory}
\label{sec:densfication_risks}

A second controversial aspect among the population is that the pervasive installation of 5G \acp{gNB} over the territory results in an exponential increase of exposure, thus leading to an unacceptable increase of the health risks. The closest works investigating this issue from a scientific point of view are \cite{ChiaraviglioBiaBleFiore:19,5gwhichrisks}. More in detail, the authors of \cite{ChiaraviglioBiaBleFiore:19} develop a very simple model to evaluate the \ac{RF} pollution {(in terms of total received power)} at selected locations of the territory (i.e., at an average or a minimum distance from the serving \ac{gNB}). A set of closed-form expressions are then derived from the model, to evaluate the increase/decrease of \ac{RF} pollution among a pair of candidate \ac{gNB} deployments that are characterized by different \ac{gNB} densities over the same service area. By leveraging on a set of worst-case and common assumptions (which include, e.g., a homogeneous set of \ac{gNB} of regular size, maximum radiated power, and simultaneous activation of all the beams by each \ac{gNB}), the authors demonstrate that, when a given performance level has to be ensured (e.g., in terms of minimum received power), the densification of the 5G network allows to promptly reduce the \ac{RF} pollution. This result can also be explained in a very intuitive way: in a network with a high density of \ac{gNB}, each site has to cover a small service area, and hence the required output power can be limited.
On the other hand, a network composed of few \ac{gNB} is characterized by a huge coverage area for each site, and hence higher radiated power. Therefore, in contrast to the common opinion of the population, the increase in the number of 5G \acp{gNB} allows to reduce the exposure at the selected locations (i.e., at an average distance or a minimum one from the serving \ac{gNB}). Eventually, the authors of \cite{ChiaraviglioBiaBleFiore:19} considers a scenario where the minimum received power \textit{and} the number of 5G \acp{gNB} are jointly increased, showing that, even in this case, the \ac{RF} pollution estimated at the selected locations is limited.

The outcomes of \cite{ChiaraviglioBiaBleFiore:19} are further corroborated by \cite{5gwhichrisks}, in which the authors evaluate the average received power over a whole territory and a set of candidate deployments. Results demonstrate that the average received power is dramatically reduced when the number of 5G \ac{gNB} is increased. Consequently, the associated health risks are minimized. Moreover, another aspect that can be observed from the network densification considered in \cite{ChiaraviglioBiaBleFiore:19} is the harmonization of exposure. When a network is composed of few \acp{gNB}, the users in close proximity to the sites tend to be exposed to higher levels of exposure compared to the ones that are far from the \ac{gNB}. On the other hand, when the number of \ac{gNB} is increased, the exposure tends to be more uniform over the territory. This issue is usually neglected by the population and may have a significant impact on the perceived health risks. 

In any case, however, it is essential to remark that the densification of the network is impacted by the \ac{EMF} regulations, which tend to limit the installation of 5G \acp{gNB} over the territory. This is especially true in countries adopting exposure limits stricter than \ac{ICNIRP}/\ac{FCC} \cite{itutksupplement14}, for which the installation of 5G \ac{gNB} is prevented, e.g., in proximity to sensitivity places and/or in the presence of other \ac{RF} installations. Although this aspect may appear beneficial for the health risks at first glance, the actual exposure levels are negatively impacted by strict regulations. To this aim, the authors of \cite{ChiaraviglioGala:19} perform a broad set of exposure measurements in a 4G operational network that is deployed under very strict regulations. Results show that strict regulations limiting the installation of 4G base stations have a negative impact on the exposure levels generated by \ac{UE} and on the performance perceived by users. In particular, the lack of 4G base stations in the neighborhood under consideration forces the \ac{UE} to be served by base stations that are typically far (i.e., more than 1000~[m]) and in \ac{NLOS} conditions. This issue results in a large electric field activity generated by the \ac{UE} and poor performance levels in terms of low throughput and large amount of time to transfer data in the uplink direction.

\subsection{Adoption of Frequencies in the mm-Wave bands}

The third controversial aspect triggering concerns by the population is the adoption of mm-Waves in 5G. To this aim, we remind that the biological impact of mm-Waves have been studied in the past years, although not in the context of cellular communications (see e.g., \cite{erwin1981assessment,gandhi1986absorption,ryan2000radio,riu1997thermal,walters2000heating,kues1999absence}). However, previous works investigating the health impact of mm-Waves did not find any adverse effect for exposure below the limits enforced by law. A similar observation is also reported by \ac{WHO} \cite{whostatement}. Moreover, the same organization is currently conducting a health risk assessment of exposure over the entire range of \ac{RF} range (including mm-Waves), which will be completed by 2022 (i.e., in parallel with the deployment of the 5G networks). This step would be beneficial to reduce the health risks of 5G that are perceived by the population. {However, we point out the current lack of well-done biomedical studies focused on the assessment of (potential) health effects from 5G devices operating on mm-Waves.}

In the following, we move our attention to radio communications from 5G \acp{gNB} exploiting mm-Waves. In general, such devices will be installed in scenarios where very high capacity is required \cite{timmmwave}. However, it is important to remark that mm-Waves are subject to very large path losses compared to micro-waves \cite{rappaport2013millimeter}. Also, other effects, including, e.g., low penetration capabilities inside the buildings, severely impact the maximum distance between a \ac{gNB} and a \ac{UE} operating at these frequencies. As a consequence, 5G deployments exploiting mm-Waves will be mainly realized through micro and small \ac{gNB}, which will be placed in close proximity to the service area \cite{timmmwave}. This, in turn, naturally limits the scope of application of mm-Waves, which will be not deployed on the whole territory, but rather at traffic-demanding hotspots (e.g., airports, stadiums, shopping malls). However, it is also important to remind that 5G will be mainly realized with sub-Ghz and sub-6~[GHz] in many countries in the world, and thus already limiting the exploitation of mm-Waves in the near future. For example, in Italy, the operators are not subject to any coverage constraint over the mm-Wave frequencies, while strict coverage constraints for lower 5G frequencies are required \cite{5gwhichrisks}. As a result, the auction on 5G frequencies in Italy resulted in a large competition among the operators  over sub-6~[GHz] frequencies, while a very limited competition was observed for mm-Waves.
Moreover, current international {guidelines} and current compliance assessment procedures already hold for mm-Waves, thus ensuring health risk minimization. However, we need to point out that measurement studies, tailored to the assessment of exposure of 5G \ac{gNB}, are needed, in order to limit the perceived health risks by the population. %Clearly, these studies can be made available as soon as the deployment of 5G \ac{gNB} will be ongoing in different countries in the world.

In the following part of our work, we focus on the exposure from \ac{UE} with mm-Waves. According to \cite{XuZhaHe:19}, 5G devices exploiting mm-Waves will be not realized with a very large number of antenna elements, being 4-element or 8-element antenna arrays the most promising solutions. However,  previous works (e.g., \cite{hong2017millimeter}) consider the design of \ac{UE} with antenna arrays composed of a larger number of radiating elements. Interestingly, the authors of \cite{XuZhaHe:19} demonstrate that the minimum peak \ac{EIRP} of a 5G \ac{UE} with mm-Waves satisfies both \ac{ICNIRP}, \ac{FCC} and \ac{IEEE} exposure limits. In addition, the authors of \cite{he2020fast,he2020incident} point out the importance of evaluating the \ac{PD} in proximity to 5G \ac{UE} with mm-Waves, claiming that a traditional approach based on magnitude-based field combination may led to very conservative estimation of the peak spatial-average \ac{PD}. Moreover, a more accurate \ac{PD} assessment, based on the magnitude and phase of the \acp{EMF}, is advocated.

\subsection{Connection of Millions of IoT Devices}

A fourth controversial aspect among the population is the effect on exposure due to the huge number of 5G terminals that will be pervasively connected in the same area. In this context, a common opinion is that massive deployments of \ac{IoT} terminals connected through 5G networks will result in an unacceptable and continuous exposure for users. To this aim, we analyze the problem from the perspective of the communications engineering by reporting a set of evidences, summarized as follows. First, current specifications defined by 3GPP always impose very low values of maximum transmitted power for each terminal, even for 5G ones \cite{etsitr3} (i.e., generally at most equal to 23~[dBm] in the majority of the cases, and in any case no higher than 35~[dBm]). Second, when considering \ac{IoT} terminals, more stringent power requirements may be introduced \cite{iotstandards}, in order to reduce the consumption and to increase the battery lifetime, in line with goals of \aclp{LPWAN} architectures \cite{RazKulSoo:17,ChiElz:19}. For example, typical values range between 23~[dBm] and 14~[dBm] \cite{etsitr2}. Third, international {guidelines} always impose maximum \ac{SAR} and/or \ac{PD} values to control the exposure from the terminals, thus guaranteeing safety for the population. Fourth, even in the presence of millions of terminals in the same area, the distance between the user and the terminal(s) will play a major role in determining the exposure. For example, when the distance is in the order of (few) meters, the exposure will be negligible, due to the aforementioned very limited maximum output power generated by the terminals. Also, the level of exposure may be further reduced due to the presence of obstacles, e.g., walls in the proximity of the terminals. 
Fifth, \ac{IoT} communications are in general very different compared to human communications \cite{iotstandards,iotfuture}. In most of the cases, \ac{IoT} devices will need to communicate with the rest of the world at a small pace, with a limited data rate, and with pretty large delays compared to human-centered communications. This will be translated into extremely low power levels in the uplink directions, and consequently in very low levels of exposure. 

\subsection{Coexistence of 5G with Legacy Technologies}

The last concern triggering health risks from the population is the coexistence of 5G with legacy technologies. We analyze the problem from the perspective of communications engineering, under the following avenues: \textit{i}) saturation levels in pre-5G networks (i.e., 2G/3G/4G), \textit{ii}) impact of radio and TV broadcasting, and \textit{iii}) interaction of 5G with weather satellites.

\subsubsection{Saturation Levels of Legacy pre-5G Networks}

As reported by \ac{ITU} \cite{itutksupplement14}, the installation of 5G sites is a challenging step in countries adopting \ac{EMF} regulations stricter than \ac{ICNIRP}/\ac{FCC} guidelines. The main effect that is observed is the saturation of \ac{EMF} levels to the maximum limits, especially in urban zones served by multiple operators and by multiple cellular technologies. In the presence of a saturation zone {(i.e., a portion of territory in which the total exposure is already close to the limit defined by law)}, the deployment of new 5G \acp{gNB} is not possible, since otherwise, the composite \ac{EMF} levels from the new \ac{gNB} and the already-deployed base stations would surpass the (strict) limits. On the other hand, the presence of these zones may also alarm the population living in their neighborhood, and thus increasing the perceived health risks associated to the installation of new 5G \ac{gNB}. 

In the literature, different works \cite{chiaraviglio2018planning,PerCarBar:18,CarAnaBar:19,chiaraviglio2020safe}  focus on the analysis of saturation zones in cellular networks subject to strict regulations. In this context, the \ac{EMF} levels are either estimated \cite{chiaraviglio2018planning,PerCarBar:18} or measured in proximity to the installations \cite{CarAnaBar:19,chiaraviglio2020safe}. Focusing on the former category, the authors of \cite{chiaraviglio2018planning} take into account the real base station deployment in an urban area in Naples (Italy). Results show that large saturation zones, in which the estimated \ac{EMF} exposure is close to the limits, already emerge. The problem is also studied in \cite{PerCarBar:18}, which is focused on the city of Bologna (Italy). By applying a set of conservative assumptions, which include, e.g., free space path loss and maximum radiated power from each deployed base station, the authors demonstrate the presence of high saturation levels for almost all the sites in the city center. Clearly, these sites can not be used to host any new 5G \ac{gNB}. 

In the following, we focus on the works tailored to the evaluation of saturation zones through measurements \cite{CarAnaBar:19,chiaraviglio2020safe}. Interestingly, the average exposure observed by \cite{CarAnaBar:19,chiaraviglio2020safe} is in general lower than the limits imposed by law. This outcome is expected, as the works based on \ac{EMF} estimations \cite{chiaraviglio2018planning,PerCarBar:18} typically introduce different assumptions, in terms of, e.g., path loss models or maximum output power, which may be very conservative in a real environment. Eventually, the authors of \cite{CarAnaBar:19} performed in an situ-measurement campaign that was conducted in the same city of  \cite{PerCarBar:18}, showing that only less than $1\%$ of the total base stations locations are actually saturated. In addition, the authors of \cite{chiaraviglio2020safe} corroborate the finding of \cite{CarAnaBar:19}, by extending the analysis over a whole region, and by taking into consideration the measurement logs which were collected over almost 20 years. Interestingly, the average \ac{EMF} levels present an increasing trend over the years, due to the installation of subsequent technologies and operators in the territory under consideration. In particular, if the \ac{EMF} levels will continue to grow with the current trend, a complete saturation will occur in the forthcoming years. Hence, there will be no possibility to install any further cellular equipment co-located or in the vicinity of the already deployed base stations. 

Summarizing, saturation zones are a consequence of strict regulations on \ac{EMF} limits. In such zones, the installation of 5G \acp{gNB} would be very limited or even prevented at all. Despite this fact may be (wrongly) perceived by the population as an advantage in terms of health risks, it is solely due to the application of the strict regulations, which are not based on any scientific evidence for both short term and long term health effects. 

\subsubsection{Impact of Radio and TV Broadcasting} 

In the following, we move our attention to the coexistence of 5G with non-cellular technologies, and in particular, on radio and/or TV broadcasting. In this context, the authors of \cite{tell1980population} performed a wide-scale measurement study to assess the \ac{EMF} levels from radio and TV broadcasting in the USA, showing that the exposure was higher than 1~[$\mu$ W/cm$^2$] for more than 440000 residents. The study was then updated 40 years later, showing that radio broadcasting radiates a consistently higher amount of power compared to cellular equipment \cite{tell2014survey}. The exposure from radio and TV repeaters w.r.t. cellular base stations is also analyzed by \cite{chiaraviglio2020safe}. Results prove that people living in proximity to repeaters used for radio and/or TV broadcasting are subject to exposure levels higher than those living in proximity to base stations.

\begin{figure}[t]
\centering
\includegraphics[width=6cm]{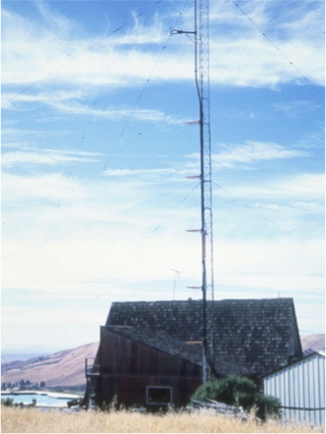}
\caption{House close to a radio broadcasting tower in San Jose - CA (photo by Richard A. Tell). The \ac{EMF} field measured in the proximity of the house was higher than 61.4~[V/m].}
\label{fig:fmproximity}
\end{figure}

Summarizing, the exposure levels in the vicinity of radio/TV broadcasting towers are far to be negligible. Therefore, these sources should be carefully taken into account when deploying 5G \acp{gNB}, in order to minimize the health risks over the population. However, we need to remind that the emissions from radio/TV broadcasting are often {\color{black}under rated} by the general public, which is apparently more concerned with the exposure from cellular networks. For example, Fig.~\ref{fig:fmproximity} shows a photo of a house almost co-located with a radio tower. The \ac{EMF} levels measured in the proximity of the house are higher than the maximum limit imposed by law (set to 61.4~[V/m]).  

\subsubsection{Interaction of 5G with Weather Satellites}

The 5G frequencies belonging to the 24.25-27.5~[GHz] and 37-40.5~[GHz] bands are close to the ones used by satellites for weather observation, i.e., 23.6-24~[GHz] and 36-37~[GHz]. Therefore, the power radiated by 5G \ac{gNB} and \ac{UE} may interfere with the sensing of water vapor and oxygen levels collected by the weather satellites, thus (possibly) impacting the weather information that is collected to monitor severe climate events, and consequently posing a health risk for the population \cite{marcus20195g}. {Not surprisingly, this aspect is frequently reported by the opponents of 5G to increase the negative perception of such technology.} {More scientifically, the} problem has been extensively studied by NASA and NOAA in \cite{nasanoaa,nasanoaa2}, which performed simulations with parameters set in accordance with ITU-R M.2101 recommendation\cite{itutrmodelling}.  Results show that a substantial noise limitation has to be imposed to 5G \acp{gNB} and 5G \ac{UE}, in order to avoid interference problems with weather satellites. The outcomes of \cite{nasanoaa,nasanoaa2} were also discussed during the ITU World Radiocommunication Conference 2019 \cite{wrc2019}, which instead defined limits on unwanted emissions for the total radiated power that are less conservative than \cite{nasanoaa,nasanoaa2}.

Several considerations hold when analyzing these outcomes from the perspective of communications engineering. First, the works \cite{nasanoaa,nasanoaa2} assumed a pervasive deployment of 5G \ac{gNB} and 5G \ac{UE} operating at mm-Waves in urban zones. However, current indications point out that the adoption of 5G \ac{gNB} will be rather limited, i.e., not deployed in whole urban areas like in \cite{nasanoaa,nasanoaa2}, but rather on specific locations (i.e., airport, stadiums, shopping malls). Moreover, the communications on mm-Waves will be one option among a set of possibilities, which will also encompass lower frequencies that do not interfere with the weather satellites. Fourth, as pointed out by \cite{lettercongress}, the input parameters used in the simulations of \cite{nasanoaa,nasanoaa2} are based on very conservative assumptions, i.e., no beamforming capabilities, simultaneous transmission of \ac{gNB} and \ac{UE} in the same time slot, power overestimation for \ac{UE} and \ac{gNB}, lack of the 250~[MHz] guard band for 5G, and over simplified propagation conditions (without buildings and foliage). Therefore, the outcomes of \cite{nasanoaa,nasanoaa2} may be not consistent with the ones achieved in a realistic setting.
Third, the set of limits defined by \cite{wrc2019} is incremental. More in-depth, looser limits will be initially applied to allow the installation of devices operating on mm-Waves. Then, after 1st September 2027, a new set of limits, more conservative than the current ones, will be applied. This choice appears to be meaningful, as the impact of interference may increase with the number of deployed devices. Fourth, as suggested by \cite{wrc2019}, solutions to avoid the antenna pointing in the direction of weather satellite sensors may be put into place in case interference problems are detected.  
%%%%%%%%%%%%%%%%%%%%%%%%%%%%%%%%%%%%%%%%%%%%%%%%%%%%%%%%%%%%%%%%%%%%%%%%%%%%%%%%%%%%%%%%%%%%%%%%%%%%%%%%%%%%%%%%%%%%%%%%%%%%%%%%%%%%%%%%%%
\begin{figure*}[t]
\centering
\begin{tikzpicture}[node distance = 0.5cm, thick, nodes = {align = center}, 
    >=latex] 
\small

%\definecolor{lightred}{rgb}{255,97,97};
%\definecolor{linen}{RGB}{250,240,230};
%\definecolor{coral}{RGB}{255,127,80};

 \node[rectangle, draw=black, rounded corners, fill=Lightred, text centered, text=black, text width=4cm] (riskmittech) {\textbf{Risk mitigation techniques}\\\textbf{of 5G exposure}}; 

\node[rectangle, text width=0.1em, draw=white, white,  minimum width=0.1em, minimum height=3em, opacity=0.0, below = of riskmittech] (dummy) {};
  
\node[rectangle, draw=black, rounded corners,fill=Grayblue, text centered, text=black, text width=3cm, right = of dummy] (netbased) {\textbf{Network based}};
  
  \node[rectangle, draw=black, rounded corners, fill=Linen, text centered, text=black, text width=3cm, below = of netbased] (emfawaredesign) {\ac{EMF}-aware cellular network planning};
    
\node[rectangle, draw=black, rounded corners, text centered, fill=Linen, text=black, text width=3cm, below = of emfawaredesign] (emfawareresall) {\ac{EMF}-aware resource allocation and communications protocols};
     
\node[rectangle, draw=black, rounded corners, text centered, fill=Grayblue, text=black, text width=3cm, left = of dummy] (archbased) {\textbf{Architectural based}};

 \node[rectangle, draw=black, rounded corners, text centered, fill=Grayblue, text=black, text width=3cm, left = of archbased] (devicebased) {\textbf{Device based}};
 
 \node[rectangle, draw=black, rounded corners, text centered, fill=Linen, text=black, text width=3cm, below = of devicebased] (sarawaredesign) {\ac{SAR}-aware \ac{UE} design};   
 
 \node[rectangle, draw=black, rounded corners, text centered, fill=Linen, text=black, text width=3cm, below = of sarawaredesign] (emfwaregnb) {\ac{EMF}-aware \ac{gNB} design};

\node[rectangle, draw=black, rounded corners, text centered, fill=Linen, text=black, text width=3cm, below = of archbased] (largesurfaces) {Large intelligent surfaces aided communications};   
  
\node[rectangle, draw=black, rounded corners, text centered, fill=Linen, text=black, text width=3cm, below = of largesurfaces] (verthozdens) {Vertical and horizontal densification};

\node[rectangle, draw=black, rounded corners, text centered, fill=Linen, text=black, text width=3cm, below = of verthozdens] (offloading) {Network offloading};
  
\node[rectangle, draw=black, rounded corners, text centered, fill=Grayblue, text=black, text width=3cm, right = of netbased] (regbased) {\textbf{Regulation based}};
   
\node[rectangle, draw=black, rounded corners, text centered, fill=Linen, text=black, text width=3cm, below = of regbased] (dismission) {Dismission of legacy 2G/3G/4G networks};

\node[rectangle, draw=black, rounded corners, text centered, fill=Linen, text=black, text width=3cm, below = of dismission] (harmonization) {Harmonization of exposure limits and assessment of compliance procedures};

\node[rectangle, draw=black, rounded corners, text centered, fill=Linen, text=black, text width=3cm, below = of harmonization] (reducingexposure) {Reduction of emissions from non-cellular \ac{RF} sources};

\node[rectangle, draw=black, rounded corners, text centered, fill=Linen, text=black, text width=3cm, below = of reducingexposure] (emfmeasureport) {Pervasive \ac{EMF} measurements and integrated \ac{EMF} reporting systems};

 \node[rectangle, draw=black, rounded corners, text centered, text=black, text width=4cm] (riskmittech2) {\textbf{Risk mitigation techniques}\\\textbf{of 5G exposure}};

\node[rectangle, draw=black, rounded corners, text centered, text=black, text width=3cm, right = of dummy] (netbased2) {\textbf{Network based}};
  
  \node[rectangle, draw=black, rounded corners, text centered, text=black, text width=3cm, below = of netbased] (emfawaredesign2) {\ac{EMF}-aware cellular network planning};
    
\node[rectangle, draw=black, rounded corners, text centered, text=black, text width=3cm, below = of emfawaredesign] (emfawareresall2) {\ac{EMF}-aware resource allocation and communications protocols};
     
\node[rectangle, draw=black, rounded corners, text centered, text=black, text width=3cm, left = of dummy] (archbased2) {\textbf{Architectural based}};

 \node[rectangle, draw=black, rounded corners, text centered, text=black, text width=3cm, left = of archbased] (devicebased2) {\textbf{Device based}};
 
 \node[rectangle, draw=black, rounded corners, text centered, text=black, text width=3cm, below = of devicebased] (sarawaredesign2) {\ac{SAR}-aware \ac{UE} design};   
 
 \node[rectangle, draw=black, rounded corners, text centered, text=black, text width=3cm, below = of sarawaredesign] (emfwaregnb2) {\ac{EMF}-aware \ac{gNB} design};

\node[rectangle, draw=black, rounded corners, text centered, text=black, text width=3cm, below = of archbased] (largesurfaces2) {Large intelligent surfaces aided communications};   
  
\node[rectangle, draw=black, rounded corners, text centered, text=black, text width=3cm, below = of largesurfaces] (verthozdens2) {Vertical and horizontal densification};

\node[rectangle, draw=black, rounded corners, text centered, text=black, text width=3cm, below = of verthozdens] (offloading2) {Network offloading};
  
\node[rectangle, draw=black, rounded corners, text centered, text=black, text width=3cm, right = of netbased] (regbased2) {\textbf{Regulation based}};
   
\node[rectangle, draw=black, rounded corners, text centered, text=black, text width=3cm, below = of regbased] (dismission2) {Dismission of legacy 2G/3G/4G networks};

\node[rectangle, draw=black, rounded corners, text centered, text=black, text width=3cm, below = of dismission] (harmonization2) {Harmonization of exposure limits and assessment of compliance procedures};

\node[rectangle, draw=black, rounded corners, text centered, text=black, text width=3cm, below = of harmonization] (reducingexposure2) {Reduction of emissions from non-cellular \ac{RF} sources};

\node[rectangle, draw=black, rounded corners, text centered, text=black, text width=3cm, below = of reducingexposure] (emfmeasureport2) {Pervasive \ac{EMF} measurements and integrated \ac{EMF} reporting systems};

\draw[-,-,->, thick,] (riskmittech.south) -- (devicebased.north);   
\draw[-,-,->, thick,] (riskmittech.south) -- (archbased.north);   
\draw[-,-,->, thick,] (riskmittech.south) -- (netbased.north);    
\draw[-,-,->, thick,] (riskmittech.south) -- (regbased.north);    

\draw[|-,-|,-, thick,] (largesurfaces.west) -- +(-0.5em,0) |- (archbased.west);
\draw[|-,-|,-, thick,] (verthozdens.west) -- +(-0.5em,0) |- (archbased.west);
\draw[|-,-|,-, thick,] (offloading.west) -- +(-0.5em,0) |- (archbased.west);
 
\draw[|-,-|,-, thick,] (emfawaredesign.west) -- +(-0.5em,0) |- (netbased.west);
\draw[|-,-|,-, thick,] (emfawareresall.west) -- +(-0.5em,0) |- (netbased.west);

\draw[|-,-|,-, thick,] (dismission.west) -- +(-0.5em,0) |- (regbased.west); 
\draw[|-,-|,-, thick,] (harmonization.west) -- +(-0.5em,0) |- (regbased.west); 
\draw[|-,-|,-, thick,] (reducingexposure.west) -- +(-0.5em,0) |- (regbased.west);  
\draw[|-,-|,-, thick,] (emfmeasureport.west) -- +(-0.5em,0) |- (regbased.west);  

\draw[|-,-|,-, thick,] (sarawaredesign.west) -- +(-0.5em,0) |- (devicebased.west); 
\draw[|-,-|,-, thick,] (emfwaregnb.west) -- +(-0.5em,0) |- (devicebased.west); 
%%%%%%%%%%%%%%%%%%%%%%%%%%%%%%%%%%%%%%%%%%%%%%%%%%%%%%%%%%%%%%%%%%%

\end{tikzpicture} 
\caption{Main techniques from the perspective of 5G communications to tackle the risk mitigation of 5G exposure.}
\label{fig:risk_mititigation_techniques}
\end{figure*}
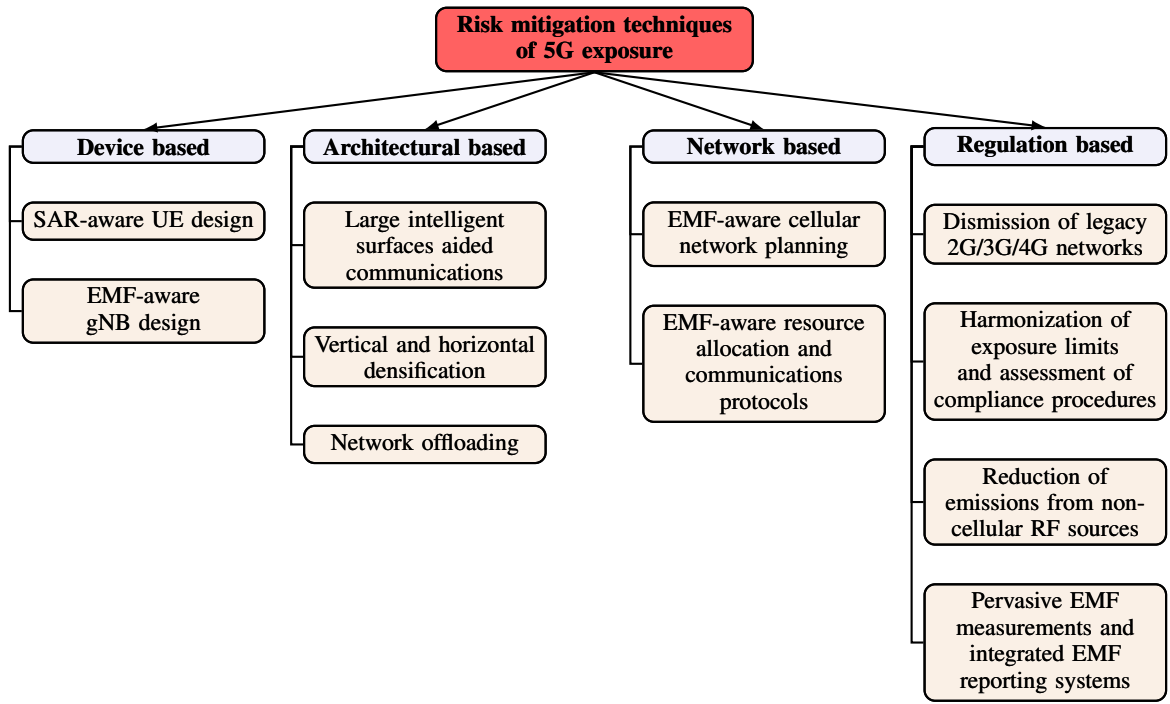

    \section{Risk Mitigation Techniques for 5G Exposure}
    \label{sec:risk_mitigation_techniques}
    
    We then move our attention to the possible techniques that can be put into place to reduce the health risks from the \ac{EMF} exposure from cellular networks. We refer the reader to \cite{TesConWia:14,SurveySamHelImr:15} for an overview of the risk mitigation techniques in cellular networks. In contrast to \cite{TesConWia:14,SurveySamHelImr:15}, this section is explicitly focused on the risk mitigation techniques tailored to 5G and beyond 5G networks.  

Fig.~\ref{fig:risk_mititigation_techniques} reports a graphical overview of the taxonomy that we employ to analyze the risk mitigation, which is observed through the lens of communications engineering. More in detail, we group the techniques into the following categories:
    \begin{enumerate}
    \item device-based solutions aimed at designing \ac{SAR}-aware 5G \ac{UE} or \ac{EMF}-aware  \ac{gNB};
    \item architectural-based approaches aimed at reducing the risks by introducing new architectural features in 5G and beyond 5G networks, i.e., large intelligent surfaces aided communications, vertical/horizontal densification and network offloading;
    \item network-based solutions aimed at developing \ac{EMF}-aware planning solutions for cellular networks or at managing the radio resources and communication protocols to reduce the \ac{EMF}; 
    \item regulation-based approaches are targeting risk reduction through the dismission of legacy 2G/3G/4G networks, harmonization of exposure limits, compliance assessment procedures across the countries, definition of constraints to limit the emissions from non-cellular \ac{RF} sources, and pervasively supporting \ac{EMF} measurement campaigns and the \ac{EMF} data integration across national and international databases.
    \end{enumerate}

    \subsection{Device-based Approaches}

We initially focus on solutions targeting the reduction of exposure at the level of individual devices.
Henceforth, we detail the design of \ac{SAR}-aware \ac{UE} and \ac{EMF}-aware \ac{gNB}.
    
    \subsubsection{\ac{SAR}-aware \ac{UE} Design}

Traditionally, the goal of \ac{SAR}-aware \ac{UE} design has been pursued since the advent of cellular communications \cite{Brown:19}. More in depth, previous techniques were focused on the reduction of the head exposure due to voice communications \cite{GanLazFur:96,VauSco:99}. To this aim, several techniques have been developed in the literature to shield the \ac{SAR} generated towards the \ac{UE} during voice calls \cite{WanOsa:97,HwaChe:06,IslamAli:11}). The main shielding methodologies can be classified into: \textit{i}) ferrite shields \cite{WanOsa:97}, \textit{ii}) metamaterials \cite{HwaChe:06} and \textit{iii}) parasitic radiators \cite{IslamAli:11}. In the following, we shed light on each of the aforementioned solutions. {The review of other shielding methods (such as multi-antenna \ac{SAR} codes, planar inverted-F antenna design, and angled helix antenna) are intentionally omitted here and left for future work.}

In approaches based on ferrite shields, a ferrite sheet is introduced between the \ac{UE} antenna and the external \ac{UE} cover, in order to reduce the exposure to the head. Although the target of lowering \ac{EMF} exposure is, in general, accomplished, the presence of the ferrite sheet may introduce negative impacts on the antenna properties. For example, according to \cite{WanOsa:97}, the antenna gain tends to be consistently reduced. In the context of 5G, this aspect may be a significant drawback, due to the fact that the antenna features have to be preserved, especially for mm-Wave frequencies.

A second approach to provide \ac{RF} shielding is based on the exploitation of metamaterials, which are able to absorb the \ac{EMF} from the \ac{UE} antenna and consequently to protect the head \cite{HwaChe:06}. Metamaterials are artificially fabricated materials with customized electromagnetic characteristics that do not exist in nature, e.g., a negative permittivity or permeability. The metamaterial shield acts as a band-stop filter that can be tuned to the operating frequency of the antenna by adjusting the metamaterial dimension. Interestingly, simulation results indicate a $30\%$ reduction in the \ac{SAR} at the expense of a loss of almost $5\%$ in the radiated power  \cite{HwaChe:06}. Eventually, the performance of ferrite sheets and metamaterials are compared using numerical simulation in \cite{IslMisLinFar:09}, showing that ferrite sheets are in general more effective than metamaterials in reducing the \ac{SAR} in the human head. However, it is important to remark that both \cite{HwaChe:06,IslMisLinFar:09} are not tailored to 5G communications, which require the deployment of multiple antenna elements on the \ac{UE}, and for which finding spare space for metamaterials may be a concrete issue.

Regarding parasitic radiators based approaches, the main idea is to employ a parasitic element that is embedded on the \ac{UE} ground plane \cite{IslamAli:11}. More in detail, the parasitic radiator is a passive element that is designed to control the current distributions on the ground plane, thus leading to a decreased \ac{SAR} and an enhanced radiation pattern. However, the passive element tends to occupy space on the ground plane, which is already crowded with other integrated circuits required for the \ac{UE} operation. This aspect may be an issue when considering 5G \ac{UE}, which have to include a large set of circuits, and, in particular, the ones realizing wireless interfaces for 2G/3G/4G/5G and IEEE 802.11 connectivity.

In the following, we concentrate on other design choices that may be relevant to the reduction of \ac{SAR} in 5G \ac{UE}. The adopted techniques include: \textit{i}) adjustment of \ac{UE} radiation patterns to reduce the exposure \cite{Brown:19} and \textit{ii}) integration of multiple antenna arrays with dual-mode operation \cite{bang2018sar}. Focusing on the former solution, the authors of \cite{Brown:19} define an antenna array design for 5G \ac{UE} to reduce the body exposure associated with various mobile use cases, e.g., voice-calling, video-calling, and texting. In the analyzed scenarios, a set of smartphone sensors are exploited to infer the \ac{UE} position and orientation. Then, the relative phase between the antenna elements is designed to direct the exposure away from the part of the body currently exposed to the specific \ac{UE} usage. This technique is of particular interest to 5G, since the \ac{UE} will be used for a set of variegate services, which will result in different exposure zones of the body, as well as different \ac{EMF} levels, in contrast to previous studies focused only on head exposure \cite{WanOsa:97,HwaChe:06,IslamAli:11}.

A further improvement towards exposure reduction is then tackled by \cite{bang2018sar}, which targets the \ac{SAR}-aware design of beam-steerable array antenna operating at mm-Waves with dual-mode operation. The main idea is to employ two distinct sub-arrays that are placed in different \ac{UE} positions, and consequently, generate different exposure patterns. More in detail, the first subarray is placed on the back cover, and it is activated only when the user exploits voice services. On the other hand, the second subarray is located at the upper frame, and it is enabled only when the user utilizes video or text services. By alternatively activating the two arrays (based on the type of services employed by the user), a peak \ac{SAR} of 0.88~[W/kg] is achieved, a value much lower compared to other competing solutions \cite{yu2017novel,bang2017mm} that do no employ separate sub-arrays. However, the wide adoption of the proposed approach in commercial devices is still an open issue, again under the light of the lack of space due to the presence of multiple wireless interfaces deployed on the same \ac{UE}.

In summary, different techniques can be exploited to reduce the exposure from 5G \acp{UE}. Differently from approaches adopted for legacy technologies based on voice services, 5G-based solutions have to integrate a variegate set of exposure \ac{UE} types. Also, the co-location of 5G antenna arrays with other wireless interfaces is already an issue, due to the limited available space on the smartphone. Therefore, future work is still needed to tackle the reduction of \ac{SAR} for 5G \ac{UE} at the device level.

    \subsubsection{\ac{EMF}-aware \ac{gNB} Design}
The second approach to reduce the \ac{EMF} is to target the design of \ac{gNB} integrating exposure minimization. In contrast to legacy technologies, the massive adoption of \ac{MIMO} and beamforming in 5G allows to dynamically focus the exposure on territory zones where the 5G service is currently needed. Thus, avoiding to pollute the other zones where the 5G services are not required. Therefore, an \ac{EMF}-aware objective should be naturally targeted during the design of 5G \ac{gNB} implementing \ac{MIMO} and beamforming. Therefore, further research in the field should be devoted, e.g., for designing antenna elements that minimize the exposure outside the main focus of each beam. This last aspect, which is already tackled by 5G \acp{gNB}, is also linked to interference reduction and, consequently,  increased throughput.

A second aspect, often underrated by the population, is that the design of base stations with \ac{EMF} minimization is already in line with the goals of \ac{gNB} manufacturers. In legacy technologies (pre-5G), in fact, a large fraction of the total base station power is used to feed its power amplifiers \cite{improvingwireless, zhang2013overview}. In line with this trend, the definition of \ac{EMF}-aware approaches for 5G \ac{gNB} could lead to a reduction of the radiated power and, consequently, the associated electricity costs. In the literature, different works (see e.g., \cite{lee2016design,chih2020energy}) are tailored to the energy-efficient design of base stations. However, the assessment of the proposed approaches in terms of \ac{EMF} is, in general, not faced, while we advocate the need to integrate it in the context of risk minimization.

Eventually, we point out that different \ac{gNB} types (e.g., small cells, macro cells) are subject to different levels of radiated power, and consequently of \ac{EMF} exposure. For example, the \ac{ITU} guidelines \cite{desset2012flexible} define multiple power classes for the base stations. In the context of 5G, we advocate the need of pursuing different types of \ac{EMF}-aware design approaches, tailored to the \ac{gNB} classes. For example, the classes of \ac{gNB} placed in close proximity to users (e.g., small cells and picocells) should implement the most sophisticated techniques to reduce \ac{EMF} exposure. On the other hand, this goal is less stringent for macro \acp{gNB}, as the (not negligible) distance between the \ac{gNB} and each user already contributes to limit the exposure.

    \subsection{Architectural-based Approaches}
In the following, we focus on solutions that require a change at the architectural level (even going beyond currently available 5G functionalities). To this aim, we analyze: \textit{i}) communications aided by large intelligent surfaces, \textit{ii}) network densification extended at both vertical and horizontal levels, \textit{iii}) network offloading. %In the following, we provide more details about each of the aforementioned solutions.
    
    \subsubsection{Large Intelligent Surfaces Aided Communications}
%\textbf{this section should be reduced in size and should be more focused on the advantaes introduced in terms of exposure reduction.}

    \begin{figure*}[t]
        \centering
        \includegraphics[width=0.99\linewidth,clip]{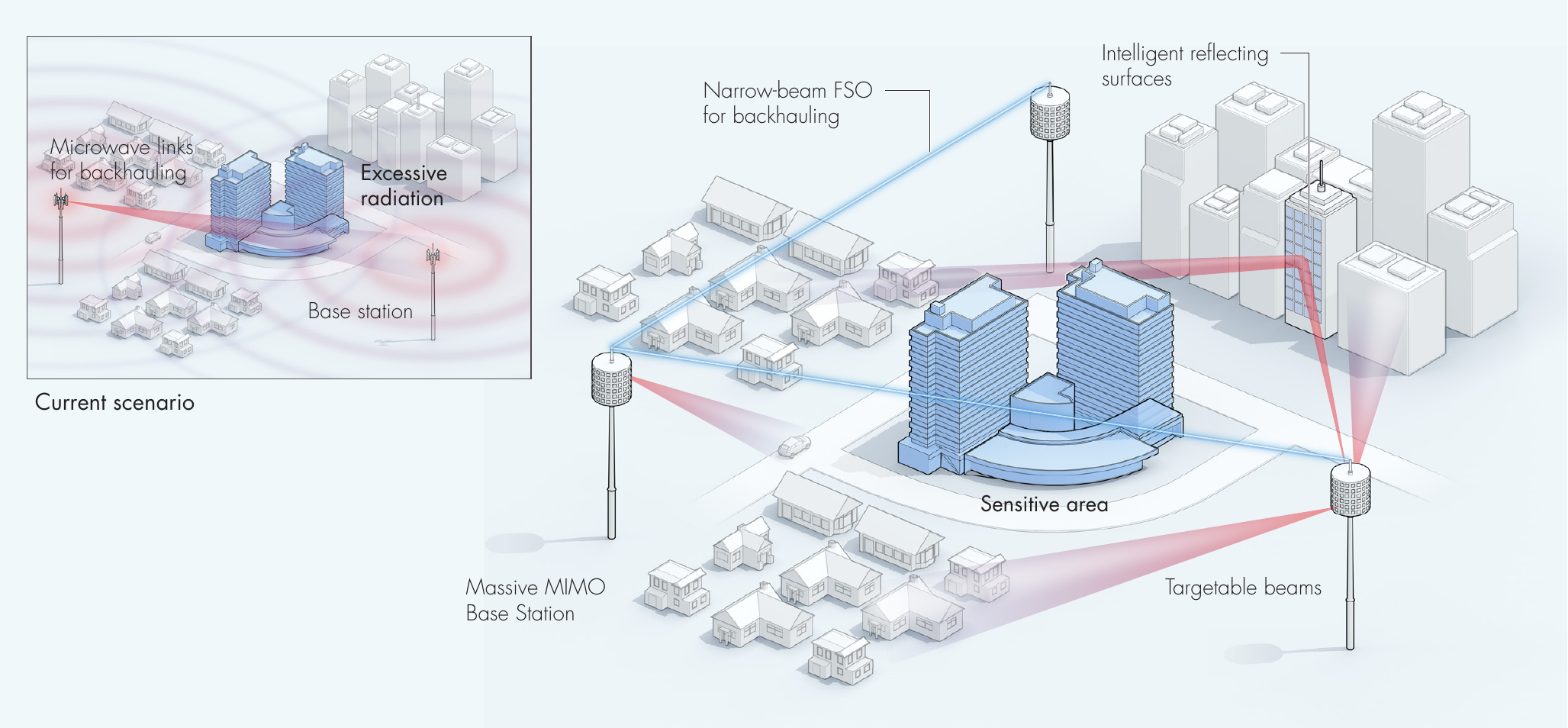}
        \caption{An example of how the adoption of \ac{MIMO}, beamforming, large intelligent surfaces and narrow-beam \ac{FSO} can improve the service level \textit{and} reduce the exposure levels over the territory (including sensitive places) w.r.t. current cellular networks.}
        \label{fig:202006slimfacompt4fig-2}
    \end{figure*}
    
The channel condition between \acp{gNB} and \ac{UE} has a notable impact on human exposure to \ac{EMF}. The unfavorable channel status is a crucial challenging issue for mm-Wave communications, where the \ac{LOS} path can be easily obstructed by large-size and small-size blockages, e.g., buildings and humans. Traditionally, this problem is solved through the introduction of relay stations (see, e.g., \cite{AbrJha:16}). By exploiting relays, in fact, the original long and obstructed path is split into a subset of links, each of them composed by a pair of interfaces in \ac{LOS} conditions. When a single relay is exploited, the \ac{UE}-\ac{gNB} path is divided into two separate links, i.e., one between the \ac{UE} and the relay, and another one between the relay and the \ac{gNB}. From a communications perspective, relays can increase both the coverage and the network throughput \cite{lin2000multihop}. Nevertheless, relays are active transmitters with full \ac{RF} chains and dedicated power sources \cite{RenzoShamai:19}. Therefore, from the health risk perspective, the systematic adoption of relay stations may further increase the \ac{EMF} exposure over the territory.  

In this context, a key question is: Is it possible to exploit the functionalities of relays, without introducing additional \ac{RF} sources? One of the most promising techniques to tackle this question is the adoption of large intelligent surfaces aided communications. According to \cite{BasRenSlim:19}, such devices operate as smart passive controllable scatterers, and they can improve the wireless channel by reflecting the waves into desirable directions to create \ac{LOS}  link for the \ac{UE}. More in detail, the reconfigurable intelligence surfaces can be fabricated from meta-materials that are equipped with programmable electronic circuits to steer the incident wave into customizable ways \cite{RenzoAlouini:19}. Compared to active relays, the scatterers are passive elements, and therefore they do not increase the number of \ac{RF} sources radiating over the territory.

From a communications perspective, the adoption of large intelligent surfaces introduces notable advantages, which include: \textit{i}) coverage probability and \ac{SINR} improvement \cite{KiskSlim:20,NadAlwkamChaDebAlouini:20,NadeemKamChaDebAlouini:2020}, \textit{ii}) high energy efficiency \cite{HuaZapYue:19} and \textit{iii}) low transmission power (also in the uplink direction) \cite{CaoLv:19}.  {\color{black} The impact in terms of \ac{EMF} has not been yet fully analyzed. However, we expect that the exploitation of large intelligent surfaces will be of great help in reducing the exposure (from both \ac{gNB} and \ac{UE}), and consequently, the associated health risks for beyond \ac{5G} networks \cite{LatLep:19,YaaAlouini:20,ShuAminAlouini:20}. Recently, several intial works consider the use of \acp{RIS} to reduce the \ac{EMF} expsoure \cite{ElzanatyLuca:21b,HussamElzanaty:21,ZapDiRenzo:21}.}

To support the previous discussion with a clear example, Fig.~\ref{fig:202006slimfacompt4fig-2} sketches a simple scenario where \ac{MIMO}, beamforming and large intelligent surfaces are exploited. More in detail, the integrated architecture allows us to easily reach the \ac{QoS} requirements of the users over the territory (e.g., the moving car in the figure or the neighborhood on the top left part of the figure), by always guaranteeing \ac{LOS} conditions. On the other hand, the exposure will be diverted from sensitive places (e.g., the central building in the figure). It is important to remark that the current pre-5G network (box on the top left of the figure) introduces an \ac{EMF} exposure that is spread over the territory (including on sensitive buildings). Eventually, the future cellular architecture will also exploit narrow-beam \ac{FSO} for backhauling to further reduce the exposure w.r.t. micro-waves links that are adopted in the current networks \cite{ElzSlimPS:20,AlzShaYaniAlouini:18,Trichili:20}.

    \subsubsection{Vertical and Horizontal Densification}

The goal of cellular densification is to increase the number of \ac{gNB} serving a given portion of the territory. With vertical densification, the number of \ac{gNB} is increased by deploying different cellular layers over the area (e.g., macro cells and small cells). This feature is already exploited in pre-5G networks \cite{brueck2011heterogeneous}, e.g., to provide primary coverage with macro cells and hotspot capacity with small cells. With 5G, the vertical densification will be a pivotal aspect to control the level of exposure. Thanks to the wide exploitation of heterogeneous networks, the (low) emissions from \ac{gNB} will be concentrated on the zones where they are really needed (e.g., to provide capacity in hot spots), and not over the whole territory. In addition, the deployment of multiple layers of \ac{gNB} will be exploited to improve the coverage and service of users. This aspect will be beneficial, especially for devices operating at mm-Waves, which are subject to strong attenuation effects, and hence require in general \ac{LOS} and proximity to the serving \ac{gNB}.  In any case, however, research works tailored to the investigation of the \ac{EMF} levels due to the large adoption of vertical densification in 5G are needed, both from theoretical and practical sides. 

{\color{black} In this regard, an \ac{EMF} aware approach with vertifical denisfication is proposed in \cite{LouElzanaty:21}, where several receive-only tethered \acp{UAV} are deployed to minimize the \ac{EMF} exposure by vertically denisfying the network for assisting the \ac{UL}. Neverthless, research works that analyze the effect of densification not only for mobile users but also for the public with exposure metrics that leverage both uplink and downlink exposures. }

In line with this trend, the exploitation of horizontal densification is another key feature to target the \ac{EMF} reduction at an architectural level. The main goal of this approach is to increase the number of \acp{gNB} over the territory \cite{ge20165g}, in order to reduce the coverage size of each cell and (possibly) the radiated power. Differently from the vertical densification, which considers different types of \ac{gNB}, the horizontal densification is realized by increasing the number of \acp{gNB} of the same type (e.g., only small cells). As already shown in Sec.\ref{sec:densfication_risks}, the horizontal densification is not a threat for the exposure levels, but rather an enabler for a low and uniform \ac{EMF} over the territory. Certainly, strict \ac{EMF} regulations may be a great barrier towards the horizontal densification of 5G networks. For example, in countries imposing minimum distances between \acp{gNB} (of every type) and sensitive places, horizontal densification will be a challenge, especially in densely populated areas that include a multitude of sensitive places. Although some previous research works try to shed light on a preliminary evaluation of \ac{EMF} levels from horizontal densification in 5G networks \cite{ChiaraviglioBiaBleFiore:19,MatMarSer:19,MatDer:18}, future research is still needed, to properly take into account the specific 5G features and the impact from the national exposure regulations.
    
    \subsubsection{Network Offloading} The main idea of this approach is to move the user traffic from cellular macro cells to other wireless stations, e.g., Wi-Fi access points \cite{dimatteo2011cellular}, small cells \cite{rebecchi2014data},  \ac{LiFi} attocells \cite{HaasYinWanChen:16}, and Terahertz access points \cite{ElaAminAlouini:20,FaiSarAlouini:19,HadiAlouiniAlNaffouri:19}. In the following, we provide more details about Wi-Fi, small cell, and \ac{LiFi} offloading techniques. {Other types of offloading, e.g., from users to \ac{MEC} servers} \cite{nguyen2020privacy}{, are intentionally not treated and left for future investigations.}

\textbf{Wi-Fi Offloading.} Nowadays, \ac{Wi-Fi} is undergoing a paradigm shift toward ubiquity, with outdoor/city-wide wireless networks gaining continuous popularity. This trend is fueled by the release of the spectrum in 6~[GHz] band as an unlicensed spectrum \cite{FCCWiFi6E:20}. In this scenario, the \ac{Wi-Fi} 6E networks will make use of the additional spectrum, i.e., $1200$~[MHz] in 6~[GHz] band, leading to higher data rates and lower latency\cite{WiFi6E:20}. To ensure such performance level, the \ac{Wi-Fi} access points have to be deployed in proximity to users. In this scenario, the cellular operator can offload part of its own traffic to the Wi-Fi network. More in-depth, three different offloading strategies can be applied, namely: \textit{i}) cellular network bypass \cite{AijAghAma:15}, \textit{ii}) managed offloading \cite{ZhoWanLeu:18}, and \textit{iii}) integrated \ac{Wi-Fi} core network \cite{3GPPInterworking:17}. With the cellular network bypass, the \ac{UE} bypasses the mobile network by offloading the whole amount of traffic into the \ac{Wi-Fi} network. With managed offloading, the operator manages a data session over the \ac{Wi-Fi} lower layers. Hence, it has more control over the amount of offloaded traffic compared to the network bypass case. Finally, in the integrated Wi-Fi core network, the \ac{Wi-Fi} access points are owned by the operator. Therefore, the traffic always traverses the mobile core network, and the offloading procedure is completely transparent to the user.

From an exposure perspective, it is clear that the three aforementioned categories can greatly contribute in reducing the \ac{EMF} levels, especially in the uplink direction. Although the number of works evaluating the benefits of \ac{Wi-Fi} offloading in terms of exposure for pre-5G networks is overall limited (see, e.g., \cite{brau2016assessing}), we believe that this architectural change could be of great interest in the context of 5G communications. Therefore, future works, tailored at the quantification of the exposure reduction due to \ac{Wi-Fi} offloading in 5G networks, are needed.

\textbf{Small Cell Offloading.} The second approach to realize offloading is to move the user traffic from macro cells to small cells (including pico cells and femto cells). In this context, the offloading is beneficial to the \ac{EMF} exposure perspective for several reasons. First of all, the transmitted power can be greatly reduced \cite{chiaraviglio2018planning}. In addition, differently from \ac{Wi-Fi}, small cells operate in licensed bands, and they are managed by the network operator. Hence, the problems of the reliability of the spectrum and integration issues are not so evident, as in \ac{Wi-Fi} based offloading. Third, the deployment costs of small costs are consistently lower than those of macro cells, and hence, small cells can be beneficial candidates for a pervasive deployment in the context of 5G. 

In the literature, different works \cite{SidAltTal:14,SidAlt:16,de2014emf} demonstrate that small cell offloading introduces several positive effects on the \ac{EMF} levels from pre-5G networks. Thoroughly, a clear reduction in the uplink radiated power is achieved \cite{de2014emf,SidAltTal:14}, which can be coupled with a coordination of the inter-cell interference \cite{SidAlt:16}. However, as shown by \cite{MazloumSamWouJoe:19}, the exposure in the downlink direction may be increased in proximity to the small cells. Therefore, we advocate the need to continue the research of exposure due to the small cell offloading in the context of 5G. Possible avenues of research include the investigation of the impact of traffic-aware offloading strategies in dense 5G networks, where the amount of offloaded data depends on the specific 5G applications run on the user side. Also, the impact of handovers between small cells and macro cells on the exposure should be thoroughly investigated, e.g., by considering exposure-friendly small cell discovery protocols \cite{OniImran:19}.

\textbf{Li-Fi Attocell Offloading.}
A recent technique to perform offloading is to move the user traffic from macro or small cells to what is called \ac{LiFi} attocells.  \ac{LiFi} is an entire networking system, similar in concept to WiFi, but it operates in the visible light frequency band, in contrast to Wi-Fi that uses \ac{RF} \cite{ObeSalAlouini:2019}. Working at such a high-frequency band allows tremendous data rates due to the availability of large bandwidth. Nevertheless, as the frequency increases, the size of the cell decreases, leading to cells with tiny coverage area, i.e.,  attocells \cite{HaasYinWanChen:16}. From the \ac{EMF} perspective, offloading through attocells has more benefits compared to \ac{RF}-based offloading techniques. The main reason is that  \ac{LiFi} technology relies on modulating the light that is already used for the illumination; hence,  no additional  \ac{RF} waves are generated for data offloading, unlike offloading through WiFi and small cells. Nevertheless, the \ac{UE} should be equipped with an additional transceiver consisting of \acp{LED} and avalanche photodetectors \cite{RajChu:15}.
%the existing illumination infrastructure can be used  
% Cellular offloading operating in visible light can significantly increase the achieved rate because of the higher bandwidth available at  
% The tremendous increase in the required data rate derived by bandwidth-hungry services necessities moving to higher frequency bands. Nevertheless, . LiFi is an entire networking system, similar in concept to WiFi. 
% Nevertheless, as the frequency increases 
% The tremendous increase in the required data rate derived by bandwidth-hungry services necessities moving to higher frequency bands. For offloading at indoor environments, Terahertz and visible light communications can achieve the target data rate, where the available  as they are operating in  The spectrum \cite{HadiAminAlouini:20,ObeSalAlouini:2019}. VLC networks provide higher data rates,
% larger EEs, lower battery consumption, and smaller latency.

    \subsection{Network-based Approaches}
The goal of network-based approaches is to tackle the risk minimization of human exposure by devising solutions in which the different 5G devices are jointly considered at a network level in order to reduce the \ac{EMF} exposure over the territory. We divide the related literature into the following categories: \textit{i}) \ac{EMF}-aware 5G cellular network planning, and \textit{ii}) \ac{EMF}-aware resource management and communications protocols. 
    \subsubsection{EMF-aware Cellular Network Planning}
The planning of a cellular network under \ac{EMF} constraints aims at selecting the set of base stations that have to be installed over the territory while ensuring economic feasibility for the operator, \ac{EMF} levels below the maximum limits, and coverage and service constraints. Not surprisingly, this problem has already been faced in the past years to design 2G/3G/4G networks (see, e.g., \cite{amaldi2003planning} for the 3G case). Nevertheless, the planning of 5G cellular networks is a novel and challenging step, as pointed out by \cite{chiaraviglio2018planning,itutksupplement14}. The main reasons are that when considering 5G communications, the set of new radio features are introduced in this technology (e.g., in terms of \ac{MIMO}, beamforming, and mm-Waves),  \ac{5G} planning is coupled with the pervasive deployment of legacy technologies, and stringent \ac{EMF} regulations are adopted. 

More technically, the planning phase of a 5G cellular network requires the following input parameters: \textit{i}) set of candidate \ac{gNB} locations which may host 5G equipment; \textit{ii}) set of possible configurations for each candidate \ac{gNB} in terms of e.g., equipment type, radio parameters (e.g., adopted carrier(s) and bandwidth) and power parameters (e.g., maximum radiated power, radiation pattern for each radiating antenna, duplexing ratio between uplink and downlink communications); \textit{iii}) terrain description in terms of elevation, 3D modeling of buildings (including sensitive places) and obstacles (e.g., trees, lamps, bus shelters), already deployed \ac{RF} sources contributing to the \ac{EMF} (e.g., other base stations and/or TV/radio repeaters and/or civil/military radars); \textit{iv}) spatial-temporal positioning of the users, \textit{v}) minimum service constraints of users (by considering also their trajectories over the territory), \textit{vi}) set of \ac{EMF} limits and procedures to verify the \ac{EMF} limits currently enforced in the territory under consideration. Given the aforementioned parameters, the network planning aims to find the subset of \acp{gNB} that have to be installed over the territory by balancing between the minimization of monetary costs for operators,  maximization of service to users, and minimization of \ac{EMF} levels over the territory. Clearly, a set of constraint has to ensured, and namely: \textit{i}) coverage over the area by the installed \ac{gNB}, \textit{ii}) guaranteed service constraints for users, \textit{iii}) estimated \ac{EMF} levels lower than the maximum limits imposed by law.

%The  {ChiaraviglioGala:19,OugKatEne:19,MatMarSer:19,MatDer:18}
To the best of our knowledge, the closest works targeting the \ac{5G} network planning are \cite{OugKatEne:19,MatMarSer:19,MatDer:18}. Specifically, the work of Oughton \textit{et al.} \cite{OugKatEne:19} is tailored to the assessment of the \ac{5G} planning by designing a new simulator, that can produce as an output the set of \ac{5G} sites and their configurations (e.g., in terms of radiating elements), by taking account multiple features, including the spectrum portfolio and the costs of the assets. However, the work is not tailored to the specific radio features of \ac{5G} networks (e.g., \ac{MIMO}, beamforming, densification) and their evaluation in terms of \ac{EMF}. In addition, irregular coverage layouts are not considered.

A cellular planning problem is also targeted by Matalala \textit{et al.}  \cite{MatMarSer:19}. Specifically, the goal of the authors is to tackle the trade-off between downlink power consumption, exposure from \acp{BS}, and exposure from terminals coverage in a cellular network exploiting \ac{MIMO}. The authors then introduce two distinct objective functions, i.e., by considering downlink and uplink exposure as two separate metrics or as a single one. The problems are then heuristically solved on three scenarios based on a suburban area in Belgium. Results show that the number of users in the scenario strongly affects the exposure from \ac{gNB}. In addition, the increase in the number of antennas elements triggers a decrease in downlink exposure and an increase in the uplink one. Moreover, the selected \ac{5G} planning achieves the same performance in terms of user coverage w.r.t. a 4G planning, coupled with a strong reduction in  downlink exposure. 

Eventually, Matalala \textit{et al.} \cite{MatDer:18} focus on the problem of selecting the subset of \ac{MIMO} \acp{BS} that minimizes the total power consumption, while ensuring coverage and capacity constraints. The considered scenarios include \ac{MIMO} \ac{5G} configurations, as well as a reference one based on \ac{LTE} technology. In addition, the problem is heuristically solved on a custom simulator. Results reveal that the increase in the number of deployed \ac{MIMO} antennas can reduce the total power consumption compared to a 4G reference network while dramatically increasing the capacity offered to users. Moreover, the \ac{MIMO} effectiveness in crowded scenarios with limited mobility emerges.

Although we recognize the importance of \cite{MatMarSer:19,MatDer:18}, we believe that substantial work is still needed to fully investigate the problem of 5G planning in the context of exposure minimization. To this aim, future research may be tailored to: \textit{i}) a precise modelling of the key \ac{5G} features in terms of \ac{EMF} levels, \textit{ii}) the investigation of the \ac{EMF} levels
by considering the deterministic positions of the users over the territory \textit{and} the beam configuration of \acp{gNB} in order to serve the users, \textit{iii}) the evaluation of the impact of strict \ac{EMF} constraints (e.g., exposure limits stricter than \ac{ICNIRP} ones and/or presence of sensitive areas) on the obtained planning, \textit{iv}) the evaluation of the \ac{5G} planning by taking into account the influence of legacy technologies (e.g., 2G/3G/4G) on the combined exposure levels.

Finally, we recognize that the \ac{EMF}-aware \ac{5G} cellular network planning is typically solved by network operators thanks to the exploitation of commercial solutions (see, e.g., \cite{htz}). However, we advocate the need to closely involving the research community (including academia) on this aspect. On one side, in fact, innovative models to estimate exposure from 5G features could be defined. On the other hand, results obtained by organizations without economic ties to the problem may be a winning solution to publicly for demonstrating the benefits introduced by an accurate 5G planning on the exposure levels.

    \subsubsection{EMF-aware Resource Allocation and Communications Protocols}

In general, the level of \ac{EMF} exposure is affected by the amount of radio resources assigned to the user, e.g., time, frequency, and power, along with the considered communication protocols in different layers, e.g., physical, data link, network, and transport layers. Hence, efficient radio resource allocation schemes and communication protocols that aim at minimizing the exposure while preserving a target \ac{QoS} can be interesting and effective solutions for risk minimization (see, e.g, \cite{HocLovJin:14} for the \ac{SAR} case). This problem is similar, albeit not identical, to the well-established research of green communications \cite{BuzKlePoorZapponne:16}. The main difference between \ac{EMF}-aware and energy-efficient approaches is that the first ones mainly focus on the exposure metrics that are closely related to the transmitted power from \acp{BS} and \ac{UE}. On the other hand, the second approaches aim at minimizing the energy efficiency (e.g., in terms of joule/bit), including not only the energy spent in communications but also the energy that is consumed within the hardware components of \ac{gNB} and \ac{UE}. Although we recognize the importance of green communications, we consider henceforth the main works that are explicitly tailored to the \ac{EMF}-aware resource management and communications protocols \cite{SamHelImr:14,SamImaHelImr:17,MatDerTanJoseph:18,BatOniImr:19,PenAgeTes:15,DieAguPen:15,IanDieArg:14}.
%In  \cite{HocLovJin:14}, designing communication systems with low \ac{BER} and SAR constraints is discussed. It is shown how to incorporate the SAR constraint into system performance analysis and code design.   SAR codes use multiple transmit antennas to get good combined far-field error performance and near-field SAR performance,

In this regard, \cite{SamHelImr:14} details a user-scheduling approach to reduce the uplink exposure in \ac{TDMA} systems. The proposed solution manages the scheduling of the user transmissions depending on their total transmitted power in the past frames, leading to a reduction in the user transmitted power and consequently limiting the uplink exposure. Focusing then on \ac{OFDM} based systems, which are typically exploited in \ac{4G} and \ac{5G}, the authors of \cite{SamImaHelImr:17} propose two resource allocation schemes in order to minimize uplink exposure, while guaranteeing a pre-defined throughput for each user. More in-depth, the first approach is an offline algorithm that makes use of the availability of long term \ac{CSI}, while the second one is an online scheme that adopts the current \ac{CSI} estimation. Results demonstrate that the proposed approaches are able to consistently reduce the user transmitted power compared to traditional solutions that solely maximize the spectral and/or power efficiency. The authors' work is further extended in \cite{SabHelImran:17} to the multi-cell scheduling case, confirming the positive outcomes in terms of uplink exposure reduction. 

Focusing then on the downlink exposure, the authors of \cite{MatDerTanJoseph:18} design an algorithm for the exposure-aware association of \ac{UE} to \acp{gNB}. Interestingly, results show that the exposure in massive \ac{MIMO} \ac{5G} networks is almost one order of magnitude lower than the corresponding one from \ac{LTE} systems with the same network coverage. However, the number of deployed \acp{gNB} in the \ac{5G} network is almost double than the one required in the \ac{LTE} networks. This increase is justified by the authors of \cite{MatDerTanJoseph:18} due to the decrease of the downlink transmitted power of each antenna element in 5G w.r.t. 4G. 

An influential aspect of controlling the \ac{EMF} in cellular networks, exploiting beamforming (like 5G), is the design of beams. To this aim, the authors of \cite{WanLinJar:11} propose an algorithm to compute the beamforming vector to reduce uplink exposure. More precisely, the proposed solution can increase the antenna gains of the beams in the direction of the \ac{BS}, while decreasing the localized \ac{SAR} on the head. Eventually, the authors in \cite{YinLovHoc:13} take into account both \ac{SAR} and transmit power in the beamformer optimization process, showing that this approach allows a substantial performance improvement over schemes that are derived from solely power constraints.

\begin{figure*}[t]
        \centering
        \subfigure[Seventies]
        {
            \includegraphics[width=5.3cm]{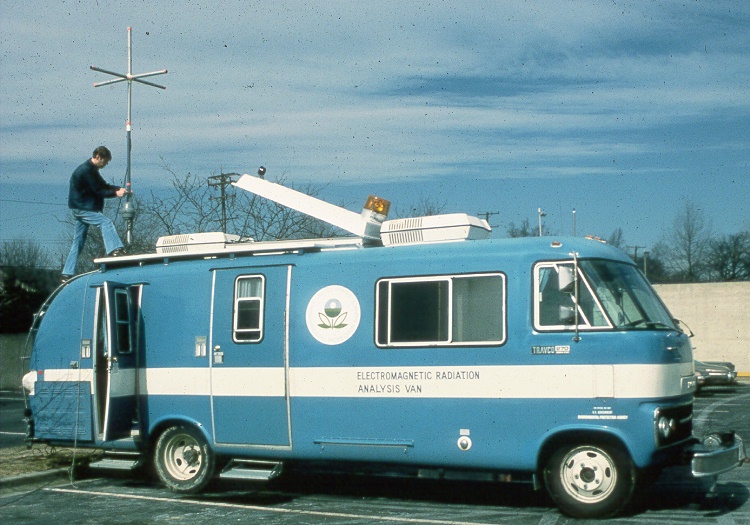}
            \label{fig:seventies}
        }
        \subfigure[Eighties]
        {
            \includegraphics[width=5cm]{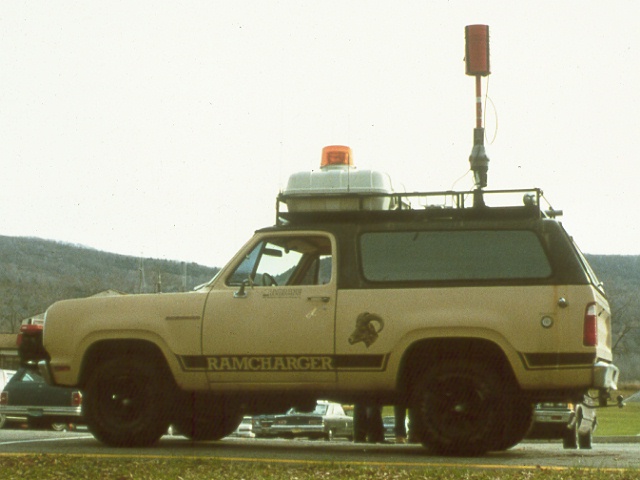}
            \label{fig:eighties}
        }    
        \subfigure[Twenty-Tens]
        {
            \includegraphics[width=5cm]{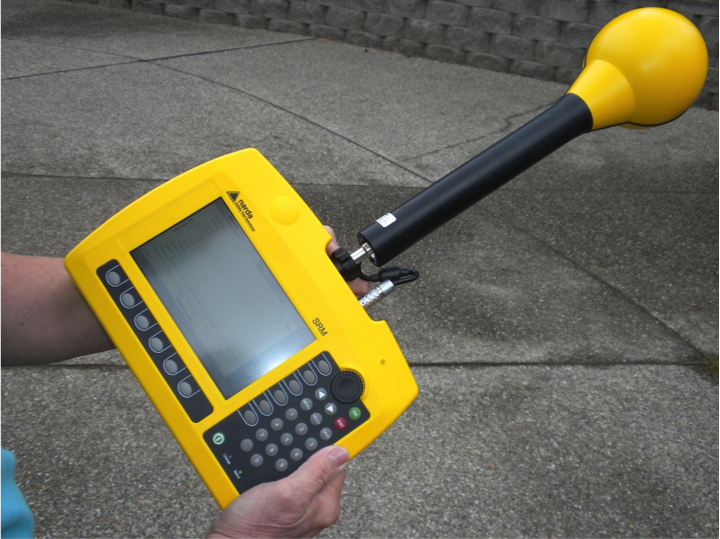}
            \label{fig:portable_spectrum_analuzer}
        }
        \caption{Evolution over the years of the measurement equipment to perform wide-scale \ac{EMF} (photos by Richard A. Tell). The reduction of equipment size is essential to allow extensive \ac{EMF} measurements from 5G \acp{gNB}.}
        \label{fig:evolution}
    \end{figure*}

The \ac{EMF} reduction methods discussed so far are employed in the physical layer. However, the \ac{EMF} exposure can also be minimized by considering higher layers, e.g., \ac{MAC}, link, and transport layers. In this regard, a cross-layer \ac{EMF} reduction approach combining features from physical and link layers is proposed in \cite{BatOniImr:19}. More specifically, an \ac{EMF}-aware  hybrid \ac{ARQ} protocol is designed to minimize the number of re-transmissions, and consequently, the transmitted power, along with the latency. This methodology could be applied to the \ac{URLLC} use case of \ac{5G} with efficient power transmission. On the other hand, \cite{PenAgeTes:15,DieAguPen:15} investigate cross-layer approaches based on link and transport layers to target the decrease of \ac{EMF} exposure in \ac{LTE} networks. The solution proposed in \cite{PenAgeTes:15} prioritizes the radio link control frames according to their significance in terms of \ac{QoS} for video transmission over \ac{LTE}. This approach can limit the number of re-transmissions for the non-critical frames, reducing the transmission power and, consequently, the \ac{EMF} exposure. Eventually, the authors of \cite{DieAguPen:15} show that the cooperation between transport and link layers allows reducing the number of re-transmissions of non-critical data in video transmissions, which in turn decreases the uplink exposure, without jeopardizing the perceived \ac{QoS}.

Although the previous approaches are promising in terms of exposure reduction, future works, tailored to the specific layers that will be implemented in 5G (and consequently to the standardized features in this technology), are needed.

    \subsection{Regulation-based Approaches}
    
The goal of regulation-based approaches is to enforce a change in the current \ac{EMF} regulations to ease the installation of 5G networks while ensuring exposure limitation. In general, these solutions are pursued by decision-makers (e.g., national governments and international organizations), with a significant impact on the exposure levels.

    \subsubsection{Dismission of Legacy 2G/3G/4G Technologies}

The deployment of 5G networks is currently done in parallel to the already deployed pre-5G systems. In a scenario where multiple \ac{RF} sources already radiate over the same territory, and also in the presence of strict \ac{EMF} regulations, the installation of 5G \ac{gNB} is a challenging step, due to the fact there is a small room to install new \ac{gNB} while ensuring the strict \ac{EMF} constraints. To this aim, a possible solution could be the dismission of legacy 2G/3G/4G networks in favor of the adoption of 5G equipment.

Although this approach could be a great driver for the full exploitation of 5G technologies, its actual applicability is not a trivial task. For example, even by assuming the sole dismission of 2G networks, all the services currently in use by this technology will have to shift to 5G. This would include, e.g., most home alarm systems currently communicating through 2G interfaces, as well as voice services, which are still exploiting 2G in many countries. Even by assuming a smooth replacement of \ac{UE} and other terminals with 5G interfaces, the deployed 5G radio access infrastructure should guarantee at least the same level of coverage provided by the 2G network that is dismissed. Despite these constraints, we believe that the disposal of the legacy technologies should be calendared in the activity list of national governments. This step could include, e.g., an incremental and selective dismission of pre-5G networks, where the legacy radio technologies are maintained in parallel to the deployment of the 5G network, for an amount of time defined in the regulations. As a step toward this goal, an operator in Netherlands has recently dismissed its 3G network, where the majority of users utilizes 4G instead of 3G services \cite{3Gdissmessed:20}.\footnote{We would like to note that dismissing 3G cellular systems, and replacing them with \ac{5G} for data services could be easier than dismissing 2G networks.}

    \subsubsection{Harmonization of Exposure Limits and Assessment of Compliance Procedures}

As discussed in Sec.~\ref{sec:national_regulations} and in Sec.~\ref{sec:compliace_assessment}, the {fragmentation} of exposure limits as well as of the methodologies to assess the exposure compliance w.r.t. the exposure limits are a great barrier towards a uniform deployment of 5G networks in the world. Even when considering countries adopting international guidelines, there are clear differences that emerge, e.g., on the maximum limit values, the adopted metrics, and the assessment of compliance methodologies. In this scenario, it is desirable that international organizations will continue to promote harmonization procedures, which should be implemented in the national regulations. For example, in countries adopting strict regulations, the application of international guidelines (and consequently less strict limits), would ease the installation of 5G equipment over the territory. However, we recognize that this choice introduces non-negligible consequences at the political levels, as the risk levels perceived by the population may be increased due to the change of the exposure limits.
    
    \subsubsection{Reduction of Emissions from non-Cellular RF Sources}

The emissions from radio and TV stations represent the largest contributions to human exposure \cite{tell2014survey,tell1980population}, especially for people living in proximity of radio and/or TV towers \cite{chiaraviglio2020safe}. In the context of 5G deployment, it would be advisable to take counter-measures and reduce exposure from such non-cellular \ac{RF} sources. Although the population does not generally associate high health risks to radio and TV towers (due to the fact that these technologies are in use for many decades), the reduction of exposure from these sources would ease the installation of the 5G equipment over the territory. Clearly, the services running on the legacy radio / TV architectures should be shifted to other technologies (e.g., satellites) or be included in 5G. In any case, however, the complete replacement of radio/TV equipment with 5G one is a challenging step.

    \subsubsection{Deployment of Pervasive \ac{EMF} Measurement Campaigns and Data Integration}

The high exposure dynamicity introduced by the novel 5G features (e.g., \ac{MIMO} and beamforming), coupled with the exploitation of relatively new frequencies in the mm-Wave band, require to setup novel methodologies for the measurement and analysis of 5G exposure from \acp{gNB}. In particular, the implementation of continuous and pervasive \ac{EMF} measurements from 5G \acp{gNB} is crucial to face the perceived health risks from the population. Although the \ac{EMF} meters have been continuously decreased in size and usage complexity in the last decades (as shown in Fig.~\ref{fig:evolution}), professional \ac{EMF} meters are not intended to be used by the general public, due to several reasons. On one side, in fact, such devices are subject to high purchase costs, which introduce significant economic barriers against the deployment of pervasive measurement campaigns exploiting a vast number of meters. On the other hand, advanced technical skills are required to perform valid measurements, e.g., to avoid measurement errors and  \ac{EMF} contributions from other \ac{RF} sources apart from \ac{gNB} in the measurement campaign. As a result, the measurement activity is often performed by the technicians of \ac{EMF} protection agencies. Clearly, assuming that these agencies will ensure a pervasive \ac{EMF} monitoring for every location of the territory covered by 5G service is not realistic. In this context, the selection of a meaningful set of sites to concentrate \ac{EMF} measurements will be an engaging and challenging future goal. Again, we believe that this problem can be solved with the help of the communications engineering community. {\color{black} For example, novel techniques for wide spectrum monitoring can be achieved by  using sub-Nyquist \acp{ADC}  exploiting the sparsity and spatio-temporal structures of the measurements, in the context of compressed sensing \cite{ElzGioChi:19,ElzAbdSedGhu:13,MishaliEldar:10,ElzAbdSedGhu:14,ElzAbdSed:14,ElzGioChi:17spl,ElzGioChisyndrome:19,NikaZhaHai:14}.  }

A second aspect, which is often underestimated by the population, is related to the great benefits that could be achieved from the integration of the \ac{EMF} measurements on common platforms at national and international levels. Providing a uniform set of interfaces to store, visualize, and analyze the \ac{EMF} measurements from 5G devices (and especially from 5G \ac{gNB}) would ease the reduction of the health risks perceived by the population. In addition, the sharing of the measurements across different communities would improve the knowledge about 5G exposure by allowing, e.g., the discovery of common exposure patterns and the presence of outliers/anomalies. However, this step requires effective coordination between the \ac{EMF} protection agencies at the national level, as well as the integration of the measured data between the different countries. Eventually, we point out that this goal is being undertaken in some countries (see, e.g., \cite{ecoscienza} for the Italian case).
    \renewcommand{\arraystretch}{1.3}

\section{Summary and Conclusions}
\label{sec:summary_concl}

We have performed an in-depth analysis of the health risks associated with 5G exposure by adopting the perspective of 5G communications engineering. Initially, we have concentrated on the health effects, by analyzing the central allegations of diseases linked to 5G exposure and by {\color{black}investigating} the false claims and hoaxes. Besides, we have applied key concepts of communications engineering to review recent animal-based studies, demonstrating that the claimed health effects about the carcinogenicity of \ac{RF} radiation can not be applied to 5G \acp{gNB} and 5G \ac{UE}. Moreover, we have examined the population-based studies relevant to 5G, showing that their methodologies have to be deeply revised when considering 5G communications.

In the second part of our work, we have analyzed the basic metrics to characterize 5G exposure, in terms of incident \ac{EMF} strength, \ac{PD}, and \ac{SAR}. We have then moved our attention to the \ac{PD}/\ac{EMF}/\ac{SAR} limits that are defined by international organizations (\ac{IEEE}, \ac{ICNIRP}) and federal commissions (\ac{FCC}), by also reporting a timely detailed comparison between the latest guidelines set in 2019-2020 against the previously adopted ones. To this aim, we recognize that the limits are pretty heterogeneous across the different authorities, although a harmonization effort appears for a subset of the considered metrics. In the following part, we have deeply analyzed the national regulations in more than 220 countries in the world, coupled with the actual deployment level of 5G technology. Overall, our picture reveals that there is a massive fragmentation of rules across the different countries (especially for \ac{gNB} deployment), with many of these countries with unknown limits and no plans to deploy 5G, as well as a non-negligible amount of world population subject to strict exposure regulations. Clearly, for countries that adopt limits more stringent than \ac{ICNIRP}/\ac{FCC} ones, deploying the 5G networks and minimizing the perceived risks are two conflicting goals. Finally, we have analyzed in detail the different procedures defined by \ac{IEEE}, \ac{IEC}, and \ac{ITU} to assess compliance of 5G exposure against the limits. Overall, we have found that the definition level of these approaches is already mature to be implemented in practice, although some guidelines have to be officially finalized.

In the third part of the paper, we have faced the main concerns associated with key 5G features, including: \textit{i}) extensive adoption of \ac{MIMO} and beamforming, \textit{ii}) densification of 5G sites over the territory, \textit{iii}) adoption of frequencies in the mm-Wave bands, \textit{iv}) connection of millions of \ac{IoT} devices and \textit{v}) coexistence of 5G with legacy technologies. By applying sounds concepts of communications engineering to review the related literature, we have shown that such features do not represent in general a threat to the population health.

Finally, the last part of our work has been devoted to the review of the main approaches that can be targeted to reduce the exposure from 5G \acp{gNB} and 5G \ac{UE},  thus minimizing the perceived health risks. We have analyzed solutions working at the device, architectural, network, and regulation levels in-depth. Although some efforts have already been considered in the literature to reduce the 5G exposure, we have pointed out different avenues that could be followed in the future to achieve this goal fully. In particular, the role of the national governments in defining regulation-based solutions appears fundamental at this stage.

In conclusion, our work suggests that the health concerns about the deployment of 5G \acp{gNB}  of 5G \ac{UE} are not supported by communications engineering evidence. Therefore, there is no compelling motivation to stop the deployment of 5G networks, especially when precautionary principles are applied. However, we point out the importance of continuing to research possible health effects (not proven at the present time), associated with the realistic exposure (i.e., below maximum limits) of 5G devices. Clearly, we advocate further research works that aim to design exposure-aware cellular networks for 5G and beyond systems properly.

%\cite{ThoCol:16}
\section{Acknowledgment}
Fig.~\ref{fig:202006slimfacompt4fig-1} and Fig.~\ref{fig:202006slimfacompt4fig-2} were produced by Xavier Pita, scientific illustrator at King Abdullah University of Science and Technology (KAUST). We also thank Richard A. Tell for having granted permission to include the pictures reported in Fig.~\ref{fig:fmproximity} and Fig.~\ref{fig:evolution}. {Finally, we thank Prof. Kenneth R. Foster, Prof. Martin R{\"o}{\"o}sli, Dr. Muhammad Ali Jamshed and the anonymous reviewers for their fruitful suggestions on how to improve our work.}
%%%%%%%%%%%%%%%%%%%%%%%%%%%%%%%%%%%%%%%%%%%%%%%%%%%%%%%%%%%%%%%%%%%%%%%%%%%%%%%%%%%%%%%%%%%%%%%%%%%%%%%%%%%%%%%%%%%%%%%%%%%%%%%%%%%%%%%%%%%%%%%%%%%%%%%%%%%%%%%%%%%%%%%%%%%%%%%%%%%%%%%%%%%%%%%%%%%%%%%%%%%%%%%%%%%%%%%%%%%%%%%
\bibliographystyle{IEEEtran}
%\enlargethispage{-10.4cm}
\bibliography{IEEEabrv,emf_ref}
\begin{IEEEbiographynophoto}{Luca Chiaraviglio}(M’09-SM’16)
is Associate Professor at the University of Rome Tor Vergata
(Italy). He holds a M.Sc. in Computer Engineering and Ph.D. in Telecommunication and Electronics Engineering, both obtained from Polytechnic of Turin (Italy). In the past years, he held research/visiting positions at Boston University (USA), INRIA (France), Auckland University of Technology (New Zealand) and University of Rome Sapienza (Italy). Luca has co-authored 140+ papers published in international journals, books and conferences, resulting from the collaboration with 200+ co-authors. He participates in the TPC of IEEE INFOCOM, IEEE GLOBECOM, IEEE ICC, IEEE VTC and IEEE GlobalSIP. He has received the Best Paper Award in different conferences, including IEEE VTC and ICIN. Some of his papers are listed as Best Readings on Green Communications by IEEE. Moreover, he has been recognized as an author in the top 1\% most highly cited papers in the ICT field worldwide. He serves in the Editorial Board of IEEE Communications Magazine and IEEE Transactions on Green Communications and Networking. He is also the founding Specialty Chief Editor for the Networks Section of Frontiers in Communications and Networking. Luca is an IEEE Senior Member and a founding member of the IEEE Communications Society Technical Subcommittee on Green Communications and Computing. His current research topics cover 5G networks, electromagnetic fields, optimization applied to telecommunication networks, and new architectures to reduce the digital divide in rural and low-income areas.
\end{IEEEbiographynophoto}
\vspace{10.0 em}
\begin{IEEEbiographynophoto}{Ahmed Elzanaty}(S’13-M’19) received the Ph.D. degree (excellent cum laude) in Electronics, Telecommunications, and Information technology from the University of Bologna, Italy, in 2018.  He was a research fellow at the University of Bologna from 2017 to 2019. Currently, he is a post-doctoral fellow at King Abdullah University of Science and Technology (KAUST), Saudi Arabia. He has participated in several national and European projects, such as GRETA and EuroCPS. His research interests include cellular network design with EMF constraints,  coded modulation, compressive sensing, and distributed training of neural networks. %He is the recipient of the best paper award at the IEEE Int. Conf. on Ubiquitous Wireless Broadband (ICUWB 2017). 
\end{IEEEbiographynophoto}
%
%\vspace{-10.0 em}
%
\begin{IEEEbiographynophoto}{Mohamed-Slim Alouini}
(S'94-M'98-SM'03-F'09)  was born in Tunis, Tunisia. He received the Ph.D. degree in Electrical Engineering
from the California Institute of Technology (Caltech), Pasadena,
CA, USA, in 1998. He served as a faculty member in the University of Minnesota,
Minneapolis, MN, USA, then in the Texas A\&M University at Qatar,
Education City, Doha, Qatar before joining King Abdullah University of
Science and Technology (KAUST), Thuwal, Makkah Province, Saudi
Arabia as a Professor of Electrical Engineering in 2009. His current
research interests include the modeling, design, and
performance analysis of wireless communication systems.
\end{IEEEbiographynophoto}

\end{document}